\documentclass[twocolumn]{aa}

\usepackage{graphicx}
\usepackage{amsmath,amsfonts,amssymb}
\usepackage{txfonts}
\usepackage{color}
\usepackage{natbib}
\usepackage{float}
\usepackage{dblfloatfix}
\usepackage{booktabs, makecell, multirow, rotating}
\usepackage{afterpage}
\usepackage{ifthen}
\usepackage[morefloats=12]{morefloats}
\usepackage{placeins}
\usepackage{multicol}

\bibpunct{(}{)}{;}{a}{}{,}
\usepackage[switch]{lineno}
\definecolor{linkcolor}{rgb}{0.6,0,0}
\definecolor{citecolor}{rgb}{0,0,0.75}
\definecolor{urlcolor}{rgb}{0.12,0.46,0.7}
\usepackage[breaklinks, colorlinks, urlcolor=urlcolor,
    linkcolor=linkcolor,citecolor=citecolor,pdfencoding=auto]{hyperref}
\hypersetup{linktocpage}

\def\setsymbol#1#2{\expandafter\def\csname #1\endcsname{#2}}
\def\getsymbol#1{\csname #1\endcsname}

\def\Planck{\textit{Planck}}

\newbox\tablebox    \newdimen\tablewidth
\def\leaderfil{\leaders\hbox to 5pt{\hss.\hss}\hfil}
\def\tablenote#1 #2\par{\begingroup \parindent=0.8em
    \abovedisplayshortskip=0pt\belowdisplayshortskip=0pt
    \noindent
    $$\hss\vbox{\hsize\tablewidth \hangindent=\parindent \hangafter=1 \noindent
    \hbox to \parindent{$^#1$\hss}\strut#2\strut\par}\hss$$
    \endgroup}

\def\L2{\ifmmode L_2\else $L_2$\fi}

\def\DeltaT{\ifmmode \Delta T\else $\Delta T$\fi}
\def\deltat{\ifmmode \Delta t\else $\Delta t$\fi}
\def\fknee{\ifmmode f_{\rm knee}\else $f_{\rm knee}$\fi}
\def\Fmax{\ifmmode F_{\rm max}\else $F_{\rm max}$\fi}
\def\solar{\ifmmode{\rm M}_{\mathord\odot}\else${\rm M}_{\mathord\odot}$\fi}
\def\Msolar{\ifmmode{\rm M}_{\mathord\odot}\else${\rm M}_{\mathord\odot}$\fi}
\def\Lsolar{\ifmmode{\rm L}_{\mathord\odot}\else${\rm L}_{\mathord\odot}$\fi}
\def\inv{\ifmmode^{-1}\else$^{-1}$\fi}
\def\mo{\ifmmode^{-1}\else$^{-1}$\fi}
\def\sup#1{\ifmmode ^{\rm #1}\else $^{\rm #1}$\fi}
\def\expo#1{\ifmmode \times 10^{#1}\else $\times 10^{#1}$\fi}
\def\,{\thinspace}
\def\lsim{\mathrel{\raise .4ex\hbox{\rlap{$<$}\lower 1.2ex\hbox{$\sim$}}}}
\def\gsim{\mathrel{\raise .4ex\hbox{\rlap{$>$}\lower 1.2ex\hbox{$\sim$}}}}

\def\simprop{\mathrel{\raise .4ex\hbox{\rlap{$\propto$}\lower 1.2ex\hbox{$\sim$}}}}
\def\deg{\ifmmode^\circ\else$^\circ$\fi}
\def\pdeg{\ifmmode $\setbox0=\hbox{$^{\circ}$}\rlap{\hskip.11\wd0 .}$^{\circ}
          \else \setbox0=\hbox{$^{\circ}$}\rlap{\hskip.11\wd0 .}$^{\circ}$\fi}
\def\arcs{\ifmmode {^{\scriptstyle\prime\prime}}
          \else $^{\scriptstyle\prime\prime}$\fi}
\def\arcm{\ifmmode {^{\scriptstyle\prime}}
          \else $^{\scriptstyle\prime}$\fi}
\newdimen\sa  \newdimen\sb
\def\parcs{\sa=.07em \sb=.03em
     \ifmmode \hbox{\rlap{.}}^{\scriptstyle\prime\kern -\sb\prime}\hbox{\kern -\sa}
     \else \rlap{.}$^{\scriptstyle\prime\kern -\sb\prime}$\kern -\sa\fi}
\def\parcm{\sa=.08em \sb=.03em
     \ifmmode \hbox{\rlap{.}\kern\sa}^{\scriptstyle\prime}\hbox{\kern-\sb}
     \else \rlap{.}\kern\sa$^{\scriptstyle\prime}$\kern-\sb\fi}
\def\ra[#1 #2 #3.#4]{#1\sup{h}#2\sup{m}#3\sup{s}\llap.#4}
\def\dec[#1 #2 #3.#4]{#1\deg#2\arcm#3\arcs\llap.#4}
\def\deco[#1 #2 #3]{#1\deg#2\arcm#3\arcs}
\def\rra[#1 #2]{#1\sup{h}#2\sup{m}}

\def\dots{\relax\ifmmode \ldots\else $\ldots$\fi}
\def\WHzsr{\ifmmode $W\,Hz\mo\,sr\mo$\else W\,Hz\mo\,sr\mo\fi}
\def\mHz{\ifmmode $\,mHz$\else \,mHz\fi}
\def\GHz{\ifmmode $\,GHz$\else \,GHz\fi}
\def\mKs{\ifmmode $\,mK\,s$^{1/2}\else \,mK\,s$^{1/2}$\fi}
\def\muKs{\ifmmode \,\mu$K\,s$^{1/2}\else \,$\mu$K\,s$^{1/2}$\fi}
\def\muKRJs{\ifmmode \,\mu$K$_{\rm RJ}$\,s$^{1/2}\else \,$\mu$K$_{\rm RJ}$\,s$^{1/2}$\fi}
\def\muKHz{\ifmmode \,\mu$K\,Hz$^{-1/2}\else \,$\mu$K\,Hz$^{-1/2}$\fi}
\def\MJysr{\ifmmode \,$MJy\,sr\mo$\else \,MJy\,sr\mo\fi}
\def\MJysrmK{\ifmmode \,$MJy\,sr\mo$\,mK$_{\rm CMB}\mo\else \,MJy\,sr\mo\,mK$_{\rm CMB}\mo$\fi}
\def\microns{\ifmmode \,\mu$m$\else \,$\mu$m\fi}

\def\muK{\ifmmode \,\mu$K$\else \,$\mu$\hbox{K}\fi}
\def\microK{\ifmmode \,\mu$K$\else \,$\mu$\hbox{K}\fi}
\def\muW{\ifmmode \,\mu$W$\else \,$\mu$\hbox{W}\fi}
\def\kms{\ifmmode $\,km\,s$^{-1}\else \,km\,s$^{-1}$\fi}
\def\kmsMpc{\ifmmode $\,\kms\,Mpc\mo$\else \,\kms\,Mpc\mo\fi}
\providecommand{\sorthelp}[1]{}

\def\WMAP{\textit{WMAP}}

\renewcommand{\d}[0]{\vec{d}}

\newcommand{\n}[0]{\vec{n}}

\newcommand{\s}[0]{\vec{s}}
\renewcommand{\a}[0]{\vec{a}}

\newcommand{\f}[0]{\vec{f}}

\newcommand{\B}[0]{\tens{B}}

\renewcommand{\L}[0]{\tens{L}}
\newcommand{\g}[0]{\vec{g}}
\newcommand{\N}[0]{\tens{N}}
\newcommand{\M}[0]{\tens{M}}

\renewcommand{\r}[0]{\vec{r}}

\renewcommand{\P}[0]{\tens{P}}

\newcommand{\BP}{\textsc{BeyondPlanck}}

\def\inv{^{-1}}

\begin{document}

\title{\BP\ VI. Noise characterization and modelling}

\newcommand{\nersc}[0]{11}
\newcommand{\princeton}[0]{7}
\newcommand{\helsinkiA}[0]{9}
\newcommand{\milanoA}[0]{2}
\newcommand{\triesteA}[0]{14}
\newcommand{\haverford}[0]{12}
\newcommand{\helsinkiB}[0]{10}
\newcommand{\triesteB}[0]{5}
\newcommand{\milanoB}[0]{3}
\newcommand{\milanoC}[0]{4}
\newcommand{\oslo}[0]{1}
\newcommand{\jpl}[0]{8}
\newcommand{\mpa}[0]{13}
\newcommand{\planetek}[0]{6}
\author{\small
\textcolor{black}{H.~T.~Ihle}\inst{\oslo}\thanks{Corresponding author: H.~T.~Ihle; \url{h.t.ihle@astro.uio.no}}
\and
M.~Bersanelli\inst{\milanoA, \milanoB, \milanoC}
\and
C.~Franceschet\inst{\milanoA,\milanoC}
\and
E.~Gjerl{\o}w\inst{\oslo}
\and
K.~J.~Andersen\inst{\oslo}
\and
\textcolor{black}{R.~Aurlien}\inst{\oslo}
\and
\textcolor{black}{R.~Banerji}\inst{\oslo}
\and
S.~Bertocco\inst{\triesteB}
\and
M.~Brilenkov\inst{\oslo}
\and
M.~Carbone\inst{\planetek}
\and
L.~P.~L.~Colombo\inst{\milanoA}
\and
H.~K.~Eriksen\inst{\oslo}
\and
J.~R.~Eskilt\inst{\oslo}
\and
\textcolor{black}{M.~K.~Foss}\inst{\oslo}
\and
\textcolor{black}{U.~Fuskeland}\inst{\oslo}
\and
S.~Galeotta\inst{\triesteB}
\and
M.~Galloway\inst{\oslo}
\and
S.~Gerakakis\inst{\planetek}
\and
\textcolor{black}{B.~Hensley}\inst{\princeton}
\and
\textcolor{black}{D.~Herman}\inst{\oslo}
\and
M.~Iacobellis\inst{\planetek}
\and
M.~Ieronymaki\inst{\planetek}
\and
J.~B.~Jewell\inst{\jpl}
\and
\textcolor{black}{A.~Karakci}\inst{\oslo}
\and
E.~Keih\"{a}nen\inst{\helsinkiA, \helsinkiB}
\and
R.~Keskitalo\inst{\nersc}
\and
G.~Maggio\inst{\triesteB}
\and
D.~Maino\inst{\milanoA, \milanoB, \milanoC}
\and
M.~Maris\inst{\triesteB}
\and
A.~Mennella\inst{\milanoA, \milanoB, \milanoC}
\and
S.~Paradiso\inst{\milanoA, \milanoC}
\and
B.~Partridge\inst{\haverford}
\and
M.~Reinecke\inst{\mpa}
\and
M.~San\inst{\oslo}
\and
A.-S.~Suur-Uski\inst{\helsinkiA, \helsinkiB}
\and
T.~L.~Svalheim\inst{\oslo}
\and
D.~Tavagnacco\inst{\triesteB, \triesteA}
\and
H.~Thommesen\inst{\oslo}
\and
D.~J.~Watts\inst{\oslo}
\and
I.~K.~Wehus\inst{\oslo}
\and
A.~Zacchei\inst{\triesteB}
}
\institute{\small
Institute of Theoretical Astrophysics, University of Oslo, Blindern, Oslo, Norway\goodbreak
\and
Dipartimento di Fisica, Universit\`{a} degli Studi di Milano, Via Celoria, 16, Milano, Italy\goodbreak
\and
INAF-IASF Milano, Via E. Bassini 15, Milano, Italy\goodbreak
\and
INFN, Sezione di Milano, Via Celoria 16, Milano, Italy\goodbreak
\and
INAF - Osservatorio Astronomico di Trieste, Via G.B. Tiepolo 11, Trieste, Italy\goodbreak
\and
Planetek Hellas, Leoforos Kifisias 44, Marousi 151 25, Greece\goodbreak
\goodbreak
\and
Department of Astrophysical Sciences, Princeton University, Princeton, NJ 08544,
U.S.A.\goodbreak
\and
Jet Propulsion Laboratory, California Institute of Technology, 4800 Oak Grove Drive, Pasadena, California, U.S.A.\goodbreak
\and
Department of Physics, Gustaf H\"{a}llstr\"{o}min katu 2, University of Helsinki, Helsinki, Finland\goodbreak
\and
Helsinki Institute of Physics, Gustaf H\"{a}llstr\"{o}min katu 2, University of Helsinki, Helsinki, Finland\goodbreak
\and
Computational Cosmology Center, Lawrence Berkeley National Laboratory, Berkeley, California, U.S.A.\goodbreak
\and
Haverford College Astronomy Department, 370 Lancaster Avenue,
Haverford, Pennsylvania, U.S.A.\goodbreak
\and
Max-Planck-Institut f\"{u}r Astrophysik, Karl-Schwarzschild-Str. 1, 85741 Garching, Germany\goodbreak
\and
Dipartimento di Fisica, Universit\`{a} degli Studi di Trieste, via A. Valerio 2, Trieste, Italy}

\authorrunning{Ihle et al.}
\titlerunning{\BP\ noise modelling}

\abstract{We present a Bayesian method for estimating instrumental
  noise parameters and propagating noise uncertainties within the
  global \textsc{BeyondPlanck} Gibbs sampling framework, and apply this to
  \textit{Planck} Low Frequency Instrument (LFI)
  time-ordered data.  Following previous literature, we initially adopt a
  $1/f$ model for the noise power spectral density (PSD), but find the need for an additional lognormal component in the noise model for the 30 and 44\,GHz bands. We implement an optimal Wiener-filter (or constrained realization)
  gap-filling procedure to account for masked data.  We then use this
  procedure to both estimate the gapless correlated noise in the
  time-domain, $\vec{n}_\mathrm{corr}$, and to sample the noise PSD
  parameters, $\xi^n = \{\sigma_0, f_\mathrm{knee},
  \alpha, A_\mathrm{p}\}$. In contrast to previous \textit{Planck} analyses, we assume
  piecewise stationary noise only within each pointing period (PID), not
  throughout the full mission, but we adopt the LFI Data 
  Processing Center (DPC) results as
  priors on $\alpha$ and $f_\mathrm{knee}$. On average, we find
  best-fit correlated noise parameters that are mostly consistent with
  previous results, with a few notable exceptions. However, a detailed
  inspection of the time-dependent results reveals many important
  findings. First and foremost, we find strong evidence for
  statistically significant temporal variations in all noise PSD
  parameters, many of which are directly correlated with satellite
  housekeeping data. Second, while the simple $1/f$ model appears to
  be an excellent fit for the LFI 70\,GHz channel, there is 
  evidence for additional correlated noise not described by a $1/f$
  model in the 30 and 44\,GHz channels, including within the primary
  science frequency range of 0.1--1\,Hz. In general, most 30 and
  44\,GHz channels exhibit deviations from $1/f$ at the 2--$3\,\sigma$ level in each one hour pointing period, motivating the addition of the lognormal noise component for these bands. For some periods of time, we also
  find evidence of strong common mode noise fluctuations across the
  entire focal plane. Overall, we conclude that a simple
  $1/f$ profile is not adequate to fully characterize the \textit{Planck} LFI
  noise, even when fitted hour-by-hour, and a more general model is
  required. These findings have important implications for large-scale
  CMB polarization reconstruction with the \textit{Planck} LFI data, and
  the current work is a first attempt at understanding and mitigating these issues.}


\maketitle


\section{Introduction}
\label{sec:introduction}

One of the main algorithmic achievements made within the field of Cosmic Microwave Bacground (CMB)
analysis during the last few decades is accurate and nearly lossless
data compression. Starting from data sets that typically comprise
$\mathcal{O}(10^8-10^{11})$ time-ordered measurements, we are now able
to routinely produce sky maps that contain $\mathcal{O}(10^3-10^{7})$
pixels \citep[e.g.,][]{tegmark_mapmaking,ashdown:2007}. From these, we may constrain
the angular CMB power spectrum, which spans $\mathcal{O}(10^3)$
multipoles \citep[e.g.,][]{gorski:1994,hivon:2002,wandelt2004}. Finally, from
these we may derive tight constraints on a small set of cosmological
parameters \citep[e.g.,][]{bond:2000,cosmomc,planck2016-l05,planck2016-l06},
which typically is the ultimate goal of any CMB experiment.

Two fundamental assumptions underlying this radical compression process
are that the instrumental noise may be modelled to a sufficient
precision, and that the corresponding induced uncertainties may be
propagated faithfully to higher-order data products. The starting
point for this process is typically to assume that the noise is
Gaussian and random in time, and does not correlate with the true sky
signal at any given time. Under the Gaussian hypothesis, the net noise
contribution therefore decreases as $1/\sqrt{N_{\mathrm{obs}}}$, where
$N_{\mathrm{obs}}$ is the number of observations of the given sky pixel,
while the signal contribution is independent of $N_{\mathrm{obs}}$.

However, it is not sufficient to assume that the noise is simply
Gaussian and random. One must also assume something about its
statistical properties, both in terms of its correlation structure in
time and its stationarity period. Regarding the correlation structure,
the single most common assumption in the CMB literature is that the
temporal noise power spectrum density (PSD) can be modelled as a sum
of a so-called $1/f$ term and a white noise term
\citep[e.g.,][]{bennett2012,planck2016-l02,planck2016-l03}. The white
noise term arises from intrinsic detector and amplifiers' thermal
noise, and is substantially reduced by cooling the instrument to
cryogenic temperatures, typically to $\sim$$20$\,K for coherent
receivers (as in the case of \Planck\ Low Frequency Instrument (LFI)) and to 0.1--0.3\,K for
bolometric detectors. Traditionally, the white noise of coherent
radiometers is expressed in terms of system noise temperature, $T_{\rm
  sys}$, per unit integration time (measured in K\,Hz$^{-1/2}$), while
for bolometers it is expressed as noise equivalent power,
$\mathrm{NEP}$ (W\,Hz$^{-1/2}$). The sources of the $1/f$ noise
component include intrinsic instabilities in the detectors, amplifiers
and readout electronics, as well as environmental effects, and,
notably, atmospheric fluctuations for sub-orbital experiments. In the
case of \Planck\ LFI, the $1/f$ noise was dominated by gain and noise
temperature fluctuations and thermal instabilities
\citep{planck2016-l02}, and was minimized by introducing the 4\,K
reference loads and gain modulation factor to optimize the receiver
balance; see, e.g., \citet{planck2013-p02,planck2014-a03} and
\citet{bp01} for further details.

Regarding stationarity, the two most common assumptions are either
that the statistical properties remain constant throughout the entire
observation period \citep[e.g.,][]{planck2016-l02}, or that it may at
least be modelled as piecewise stationary within for instance one
hour of observations \citep[e.g.,][]{quiet:2011}. Given such basic
assumptions, the effect of the instrumental noise on higher-order data
products has then traditionally been assessed, and propagated, either
through the use of detailed end-to-end simulations
\citep[e.g.,][]{planck2014-a14} or in the form of explicit noise
covariance matrices
\citep[e.g.,][]{tegmark1997,page2007,planck2016-l05}.

The importance of accurate noise modelling is intimately tied to the
overall signal-to-noise ratio of the science target in question. For
applications with very high signal-to-noise ratios, detailed noise
modelling is essentially irrelevant, since other sources of systematic
errors dominate the total error budget. One prominent example of this
is the CMB temperature power spectrum as measured by \Planck\ on large
angular scales \citep{planck2016-l04,planck2016-l05}. The white noise
contribution to the power spectrum can be misestimated by orders of magnitude without making
any difference in terms of cosmological parameters, because the full
error budget is vastly dominated by cosmic variance.

The cases for which accurate noise modelling is critically important
are those with signal-to-noise ratios of order unity. For these, noise
misestimation may be the difference between obtaining a tantalizing,
but ultimately unsatisfying, $2\,\sigma$ result, and claiming a
ground-breaking and decisive $5\,\sigma$ discovery or, the
worst-case scenario, erroneously reporting a baseless positive
detection.

This regime is precisely where the CMB field is expected to find
itself in only a few years from now, as the next-generation CMB
experiments (e.g., CMB-S4, \textit{LiteBIRD}, \textit{PICO}, Simons Observatory, and
many others; \citealp{cmbS4,litebird2018,litebird2020,litebird2022,pico2019,SO2019}) are
currently being planned, built and commissioned in the search for
primordial gravitational waves imprinted in $B$-mode polarization. The
predicted magnitude of this signal is expected to be at most a few
tens of nanokelvins on angular scales larger than a degree,
corresponding to a relative precision of $\mathcal{O}(10^{-8})$, and
extreme precision is required for a robust detection. It will
therefore become critically important to take into account all sources
of systematic uncertainties, including correlated noise, and propagate 
these into the final results.

The \BP\ project \citep{bp01} is an initiative that aims to meet
this challenge by implementing the first global Bayesian CMB analysis
pipeline that supports faithful end-to-end error propagation from raw
time-ordered data to final cosmological parameters. One fundamental
aspect of this approach is a fully parametric data model that is
fitted to the raw measurements through standard posterior sampling
techniques, simultaneously constraining both instrumental and
astrophysical parameters. Within this framework, the instrumental
noise is just one among many different sources of uncertainty, all of
which are treated on the same statistical basis. The sample-based
approach introduced by \BP\ therefore represents a novel and third way
of propagating noise uncertainties \citep{bp02,bp10}, complementary to the existing
simulation and covariance matrix based approaches used by traditional
pipelines.

As a real-world demonstration of this novel framework, the
\BP\ collaboration has chosen the \Planck\ LFI measurements
\citep{planck2016-l01,planck2016-l02} as its main scientific target
\citep{bp01}. These data represent an important and realistic testbed
in terms of overall data volume and complexity, and they also have
fairly well-understood properties after more than a decade of detailed
scrutiny by the \Planck\ team \citep[see][and references
  therein]{planck2013-p02,planck2014-a03,planck2016-l02}. However, as
reported in this paper, there are still a number of subtle unresolved
and unexplored issues relating even to this important and
well-studied data set that potentially may have an impact on
higher-level science results. Furthermore, as demonstrated by the
current analysis, the detailed low-level Bayesian modelling approach
is ideally suited to identify, study and, eventually, mitigate these
effects.

Thus, the present paper has two main goals. The first is to describe
the general algorithmic framework implemented in the \BP\ pipeline for
modelling instrumental noise in CMB experiments. The second goal is
then to apply these methods to the \Planck\ LFI observations, and
characterize the performance and systematic effects of the instrument
as a function of time and detector.

The rest of the paper is organized as follows. First, in
Sect.~\ref{sec:bp} we briefly review the \BP\ analysis framework and
data model, with a particular emphasis on noise modelling aspects. In
Sect.~\ref{sec:methods}, we present the individual sampling steps
required for noise modelling, as well as some statistics that are useful
for efficient data monitoring. In Sect.~\ref{sec:degeneracies} we
discuss various important degeneracies relevant for noise modelling, 
and how to minimize the impact of modelling errors. Next, in
Sect.~\ref{sec:results} we give a high-level overview of the various
noise posterior distributions and their correlation
properties, as well as detailed specifications for each detector. In
Sect.~\ref{sec:systematics} we discuss anomalies found in the data,
and interpret these in terms of the instrument and the thermal
environment. Finally, we summarize in
Sect.~\ref{sec:conclusions}.

\section{The \BP\ data model and framework}
\label{sec:bp}
The \BP\ project is an attempt to build up an end-to-end data analysis
pipeline for CMB experiments going all the way from raw time-ordered
data to cosmological parameters in a consistent Bayesian
framework. This allows us to characterize degeneracies between
instrumental and astrophysical parameters in a statistically
well-defined framework, from low-level instrumental quantities such as
gain \citep{bp07}, bandpasses \citep{bp09}, far sidelobes \citep{bp08}, and correlated noise via Galactic parameters such as
the synchrotron amplitude or spectral index \citep{bp13,bp14}, to the angular
CMB power spectrum and cosmological parameters \citep{bp11,bp12}.

The LFI dataset consists of three bands, at frequencies of
roughly 30, 44, and 70\,GHz. These bands have two, three, and six radiometer
pairs each, respectively, which for historical reasons are numbered
from 18 to 28. The two radiometers in each pair are labeled by M and S\footnote{Each pair of radiometers are connected to the same feedhorn through the main arm (M) or side arm (S) of an orthomode transducer, and are therefore sensitive to orthogonal linear polarizations.}
\citep{planck2013-p02}. In \BP, the raw uncalibrated data, $\d$, produced by
each of these radiometers is modelled in time-domain as follows,
\begin{equation}
	\begin{split}
		d_{j,t} = g_{j,t}&\P_{tp,j}\left[\B^{\mathrm{symm}}_{pp',j}\sum_{c}
		\M_{cj}(\beta_{p'}, \Delta_\mathrm{bp}^{j})a^c_{p'}  + \B^{\mathrm{asymm}}_{pp',j}\left(s^{\mathrm{orb}}_{j,t}  
		+ s^{\mathrm{fsl}}_{j,t}\right)\right] + \\
		+ &s^{\mathrm{1hz}}_{j,t} + n^{\mathrm{corr}}_{j,t} + n^{\mathrm{wn}}_{j,t}.
	\end{split}
	\label{eq:todmodel}
\end{equation}
Here the subscript $t$ denotes a sample index in time domain; $j$ denotes
radiometer; $p$ denotes the pixel number; $c$ is an index denoting the different 
signal components; $\g$
denotes the gain; $\P$ denotes the pointing matrix; $\B^{\mathrm{symm}}$ and
$\B^{\mathrm{asymm}}$ denote the symmetric and asymmetric beam matrices,
respectively; $\a$ are the astrophysical signal amplitudes; $\vec{\beta}$ are
the corresponding spectral parameters; $\Delta_\mathrm{bp}$ are the bandpass
corrections; $\M_{cj}$ is the bandpass-dependent component mixing matrix;
$\s^{\mathrm{orb}}$ is the orbital dipole; $\s^{\mathrm{fsl}}$ are the far
sidelobe corrections; $s^{\mathrm{1hz}}$ represents electronic 1\,Hz spike corrections; $\n^{\mathrm{corr}}$ is the correlated noise; and
$\n^{\mathrm{wn}}$ is the white noise. For more details on each of these
parameters see \citet{bp01} and the other companion papers. 

Let us denote the combined set of all free parameters by
 $\omega\equiv\{g, \n^{\mathrm{corr}}, \xi^n,\Delta_\mathrm{bp},\n^{\mathrm{corr}}, \a,
\beta, C_{\ell}\}$. Here we have defined the noise power spectral density (PSD) parameters, $\xi^n \equiv \{\sigma_0, f_\mathrm{knee},
\alpha, A_\mathrm{p}\}$. 
The goal of the Bayesian approach is now to sample from the joint posterior distribution,
\begin{equation}\label{eq:full_distribution}
	P(\g, \n^{\mathrm{corr}}, \xi^n,\Delta_\mathrm{bp}, \a, \beta, C_{\ell}\mid\d),
\end{equation}
where the notation $P(A\mid B)$ denotes the conditional probability of $A$ given $B$ (i.e. keeping
$B$ fixed). 
This is a large and complicated distribution, with many
degeneracies. However, using Gibbs sampling we can divide the sampling
process into a set of managable steps. Gibbs sampling is a simple
algorithm in which samples from a joint multi-dimensional
distribution are generated by iterating through all corresponding
conditional distributions. Using this method, the \BP\ sampling scheme may be summarized as follows \citep{bp01},
\begin{alignat}{11}
\label{eq:gain_samp_dist}\g &\,\leftarrow P(\g&\,|&\,\d,&\, & &\,\xi^n, 
&\,\Delta_\mathrm{bp}, &\,\a, &\,\beta, &\,C_{\ell})\\
\label{eq:ncorr_samp_dist} \n^{\mathrm{corr}} &\,\leftarrow P(\n^{\mathrm{corr}}&\,|&\,\d, &\,\g, &\,&\,\xi^n, 
&\,\Delta_\mathrm{bp}, &\,\a, &\,\beta, &\,C_{\ell})\\ 
\label{eq:xi_samp_dist} \xi^n &\,\leftarrow P(\xi^n&\,|&\,\d, &\,\g, &\,\n^{\mathrm{corr}}, &\,
&\,\Delta_\mathrm{bp}, &\,\a, &\,\beta, &\,C_{\ell})\\
\Delta_\mathrm{bp} &\,\leftarrow P(\Delta_\mathrm{bp}&\,|&\,\d, &\,\g, &\,\n^{\mathrm{corr}}, &\,\xi^n, 
&\,&\,\a, &\,\beta, &\,C_{\ell})\\
\beta &\,\leftarrow P(\beta&\,|&\,\d, &\,\g, &\,\n^{\mathrm{corr}}, &\,\xi^n, 
&\,\Delta_\mathrm{bp}, & &\,&\,C_{\ell})\\
\a &\,\leftarrow P(\a&\,|&\,\d, &\,\g, &\,\n^{\mathrm{corr}}, &\,\xi^n, 
&\,\Delta_\mathrm{bp}, &\,&\,\beta, &\,C_{\ell})\\
C_{\ell} &\,\leftarrow P(C_{\ell}&\,\mid &\,\d, &\,\g, &\,\n^{\mathrm{corr}}, &\,\xi^n, &\,\Delta_\mathrm{bp}, &\,\a, &\,\beta&\,\phantom{C_{\ell}})&.
\end{alignat}
Here, $\leftarrow$ indicates sampling from the distribution on the
right-hand side. 

Note that for some of these steps we are not following the strict
Gibbs approach of conditioning on all but one variable. Most notably
for us, this is the case for the gain sampling step in
Eq.~(\ref{eq:gain_samp_dist}), where we do not condition on
$\n_\mathrm{corr}$. In effect, we instead sample the gain and
correlated noise jointly by exploiting the definition of a conditional distribution,
\begin{equation}
	P(\g, \n^{\mathrm{corr}}\mid\d, \cdots) = P(\g\mid\d, \cdots)P(\n^{\mathrm{corr}}\mid\d,\g, \cdots).
\end{equation}
This equation implies that a joint sample $\{\g,\n^{\mathrm{corr}}\}$
may be produced by first sampling the gain from the marginal
distribution with respect to $\n_\mathrm{corr}$, and then sampling
$\n_\mathrm{corr}$ from the usual conditional distribution with
respect to $\g$. The advantage of this joint sampling procedure is a
much shorter Markov correlation length as compared to standard Gibbs
sampling, as discussed by \citet{bp07}.

A convenient property of Gibbs sampling is its modular nature, as the
various parameters are sampled independently within each conditional
distribution, but joint dependencies are still explored through the
iterative scheme. In this paper, we are therefore only concerned with
two of the above steps, namely Eqs.~(\ref{eq:ncorr_samp_dist}) and
(\ref{eq:xi_samp_dist}). For details on the complete Gibbs chain and the
other sampling steps, see \citet{bp01} and the companion papers.

When sampling the correlated noise we want to allow for the possibility that the noise properties change over time. Perhaps the simplest way to allow for this, and the approach we will use in this paper, is to chunk the data in time and assume that the noise properties are constant within each chunk and independent between chunks. The next step is then to choose the timescale, $t_\mathrm{chunk}$, over which we assume that the noise properties are stationary. A number of considerations are relevant for deciding this timescale. In general, there are at least three important timescales to consider, namely the scanning period, $t_\mathrm{scan}$ (i.e., how long it takes between the telescope comes back to roughy the same position on the sky, which is about 60 seconds for Planck), the correlated noise timescale $t_\mathrm{corr} \sim 1/f_\mathrm{knee}$ (i.e., the timescale at which the correlated noise becomes relevant compared to the white noise, typically 10--100\,s for LFI), and the timescale at which the noise properties change, $t_\mathrm{change}$ (which will depend strongly on what gives rise to the change). Ideally we would want to chunk the data at a timescale that is much longer than the two former and much shorter than the latter timescale, i.e., $t_\mathrm{scan}, t_\mathrm{corr} \ll t_\mathrm{chunk} \ll t_\mathrm{change}$. However, as changes to the noise properties can occur both suddenly, due to glitches or sudden changes in the thermal environment (as is discussed in Sect.~\ref{sec:systematics}), or drift slowly over time, the last of the three timescales is not always well defined.

Another factor to consider is that if we are using Fourier transforms of the timestreams, or other analyses which scale superlinearly with the chunk size in time, a larger chunk size would be more computationally expensive. In general, the details of the specific data are very relevant, even if we do not have a completely well defined way to determine an optimal chunk size. The sample rate of the LFI time-ordered data is 32.5079\,Hz, 46.5455\,Hz and 78.7692\,Hz for the 30, 44 and 70\,GHz bands respectively.

The LFI time-ordered data are divided into roughly 45\,000 pointing
periods, denoted PIDs (pointing ID), during which the satellite spin 
axis direction was maintained fixed (see \citet{planck2011-1.3} 
for detailes on the Planck scanning strategy). Most PIDs have a duration of
30--60 minutes, which is at least an order of magnitude longer than both the scanning period and the correlated noise timescale for LFI, so they are a natural and convenient choice for chunking the data. Therefore when sampling the correlated noise and the corresponding PSD parameters, we assume that the noise is stationary within each PID, but independent between PIDs. The gain is also
assumed to be constant within each PID, however, this is not fit
independently for each PID, but rather sampled smoothly on longer
timescales \citep{bp07}.

Following previous literature
\citep{mennella2010,planck2013-p02,planck2016-l02}, we start by assuming that the noise PSD may be described by a so-called $1/f$ model,
\begin{equation}
	P(f) = \sigma_0^2\left[1 +
          \left(\frac{f}{f_\mathrm{knee}}\right)^\alpha\right].
        \label{eq:1fmodel}
\end{equation} 
Here $f$ denotes a temporal frequency; $\sigma_0$ quantifies the white
noise level of the time-ordered data\footnote{The white noise levels $\sigma_0$ are specified for an integration time corresponding to three samples per FWHM beam crossing (see \citet{maris2009a} for details). $\sigma_0$ has different units if we are talking about the uncalibrated data, $\sigma_0$\,[V], calibrated data, $\sigma_0$\,[K] $\equiv \sigma_0$\,[V] $/ g$, or the white noise PSD, $\sigma_0^2$\,[K${}^2$ Hz${}^{-1}$] $\equiv \left(\sigma_0\right.$\,[K]$\left. \right)^2 \frac{2}{R_\mathrm{samp}}$, where $R_\mathrm{samp}$ is the sample rate (in Hz) of the time ordered data. Where this distinction is important, we include the units explicitly.}; $\alpha$ is the slope (typically negative) of the correlated
noise spectrum; and the knee frequency, $f_\mathrm{knee}$, denotes the
(temporal) frequency at which the variance of the correlated noise is equal to the white noise variance. 
 
While 70 GHz noise properties are well described with the $1/f$ model, we show in the following that this model is not sufficient for representing the noise properties of the 30 and 44 GHz bands. These detectors often show a small amount of excess power at intermediate temporal frequencies (0.01--1 Hz), which is not well fit by the $1/f$ model. In order to address this, we allow for an additional lognormal component in the noise PSD for these bands
\begin{equation}
	P(f) = \sigma_0^2\left[1 +
	\left(\frac{f}{f_\mathrm{knee}}\right)^\alpha\right] + A_\mathrm{p} \exp\left[-\frac{1}{2}\left(\frac{\log_{10}f - \log_{10} f_\mathrm{p}}{\sigma_\mathrm{dex}}\right)^2\right],
	\label{eq:1fmodel_lognorm}
\end{equation} 
where $A_\mathrm{p}$ and $f_\mathrm{p}$ are the amplitude and the frequency of the peak of the lognormal component, and $\sigma_\mathrm{dex}$ denotes the width of the component, in dex, around this peak. As the deviations from the $1/f$ model are small, and not very significant in a single PID, we will not be able to estimate the three parameters of the lognormal, $A_\mathrm{p}$, $f_\mathrm{p}$  and $\sigma_\mathrm{dex}$, for each PID. For this reason, we will keep $f_\mathrm{p}$  and $\sigma_\mathrm{dex}$ fixed, and only sample $A_\mathrm{p}$ for each PID. We use a fairly wide lognormal centered on $f_\mathrm{p} = 0.011$ Hz with a width of $\sigma_\mathrm{dex} = 1$ in order to accomodate excess power in a wide range of frequencies. A more detailed analysis would fit individual values for $f_\mathrm{p}$  and $\sigma_\mathrm{dex}$ for each radiometer, allowing an even better model fit, however we find that the current model leads to acceptable model fits, as quantified in the time stream $\chi^2$ values, for all radiometers. The four PSD parameters that we will fit are then
collectively denoted $\xi^n = \{ \sigma_0, f_\mathrm{knee},\alpha, A_\mathrm{p}\}$. 

The choice of using a lognormal component to paramatrize the deviations from the $1/f$ model, as opposed to other ways of extending the noise model, is not motivated by any fundamental considerations. Rather, we simply want to add an extra degree of freedom to the noise model in the relevant frequency range, and a lognormal component was a practical way to do this. By using a lognormal, which decays both at high and low frequencies, we can use the high frequencies to determine the white noise level, $\sigma_0$, and the low frequencies to measure $f_\mathrm{knee}$ and $\alpha$. This way we reduce the degeneracy between the added noise component and the $1/f$ parameters, although some level of degeneracy between them is inevitable. 

\section{Methods}
\label{sec:methods}

As outlined above, noise estimation in the Bayesian \BP\ framework
amounts essentially to being able to sample from two conditional
distributions, namely $P(\n^{\mathrm{corr}}\mid\d, \omega\setminus
\n^{\mathrm{corr}})$ and $P(\xi^n\mid \d, \omega\setminus\xi^n)$, 
where $A \setminus B$ denotes all the elements of the set $A$ except 
the ones also present in $B$. The 
first presentation of Bayesian noise estimation for time-ordered CMB
data that was applicable to the current problem was presented by
\citet{wehus:2012}, and the main novel feature presented in the
current paper is simply the integration of these methods into the
larger end-to-end analysis framework outlined above. In addition, the
current analysis also employs important numerical improvements as
introduced by \citet{bp02}, in which optimal mapmaking is re-phrased
into an efficient Bayesian language.

The starting point for both conditional distributions is the following
parametric data model,
\begin{equation}
 \d = g \s^{\mathrm{tot}} + \s^{\mathrm{1hz}} + \n^{\mathrm{corr}} + \n^{\mathrm{wn}},
\end{equation}
where $\d$ denotes the raw time ordered data (TOD) organized into a
column vector; $g$ is the gain; 
\begin{equation}
 \s^\mathrm{tot} \equiv \P\left[\B^{\mathrm{symm}}
 \M a  + \B^{\mathrm{asymm}}\left(s^{\mathrm{orb}} 
 + s^{\mathrm{fsl}}\right)\right],
\end{equation}
describes the total sky signal, comprising both CMB and foregrounds,
projected into time-domain; $\s^{\mathrm{1hz}}$ represents electronic 1\,Hz spike corrections; $\n^\mathrm{corr}$ represents the
correlated noise in time domain; and $\n^\mathrm{wn}$ is white
noise. The two noise terms are both assumed to be Gaussian distributed
with covariance matrices $\tens{N}_\mathrm{corr} \equiv \langle
\vec{n}_\mathrm{corr} \vec{n}_\mathrm{corr}^{T}  \rangle$ and
$\tens{N}_\mathrm{wn} \equiv \langle \vec{n}_\mathrm{wn}
\vec{n}_\mathrm{wn}^{T}  \rangle$, respectively. The complete noise PSD
is then (in Fourier space) given by $P(f) = \tens{N}_\mathrm{wn} + \tens{N}_\mathrm{corr} =
\sigma_0^2 + \sigma_0^2\left(\frac{f}{f_\mathrm{knee}}\right)^\alpha + A_\mathrm{p} \exp\left[-\frac{1}{2}\left(\frac{\log_{10}f - \log_{10} f_\mathrm{p}}{\sigma_\mathrm{dex}}\right)^2\right]$. 

\subsection{Sampling correlated noise, $P(\vec{n}^\mathrm{corr}\mid\vec{d}, \xi^n, \vec{s}_\mathrm{tot}, g)$}
\label{sec:ncorr_samp}

Our first goal is to derive an appropriate sampling prescription for
the time-domain correlated noise conditional distribution,
${P(\vec{n}^\mathrm{corr}\mid\vec{d}, \xi^n, \vec{s}^\mathrm{tot}, \s^{\mathrm{1hz}},
  g)}$. To this end, we start by defining the signal-subtracted data,
$\vec{d}'$, directly exploiting the fact that $g$, $\s^{\mathrm{1hz}}$ and $\s^{\mathrm{tot}}$ are currently conditioned upon,\footnote{When a parameter appears on the right-hand side of a conditioning bar in a probability distribution, it is assumed known to infinite precision. It is therefore for the moment a constant quantity, and not associated with any stochastic degrees of freedom or uncertainties.}
\begin{equation}
 \d' \equiv \d - g \s^\mathrm{tot}  - s^{\mathrm{1hz}} = \n^\mathrm{corr} + \n^\mathrm{wn}. 
\end{equation}
Since both $\n^\mathrm{corr}$ and $\n^\mathrm{wn}$ are assumed
Gaussian with known covariance matrices, the appropriate sampling
equation for $\n^{\mathrm{corr}}$ is also that of a multivariate
Gaussian distribution, which is standard textbook material; for a
brief review, see Appendix~A in \citet{bp01}. In particular, the
maximum likelihood (ML) solution for $n_t^\mathrm{corr}$ is given by
the so-called Wiener-filter equation,
\begin{equation}\label{eq:ncorr}
 \left( \tens{N}_\mathrm{corr}^{-1} + \tens{N}_\mathrm{wn}^{-1}\right) \vec{n}^\mathrm{corr} = \tens{N}_\mathrm{wn}^{-1} \vec{d}', 
\end{equation}
while a random sample of $\vec{n}^\mathrm{corr}$ may be found by solving the following equation,
 \begin{equation}\label{eq:ncorr_samp}
 \left( \tens{N}_\mathrm{corr}^{-1} + \tens{N}_\mathrm{wn}^{-1}\right) \vec{n}^\mathrm{corr} = \tens{N}_\mathrm{wn}^{-1} \vec{d}' + \tens{N}_\mathrm{wn}^{-1 /2} \vec{\eta}_1 + \tens{N}_\mathrm{corr}^{- 1/2} \vec{\eta}_2,
 \end{equation}
where $\vec{\eta}_1$ and $\vec{\eta}_2$ are two independent vectors of
random variates drawn from a standard Gaussian distribution,
${\vec{\eta}_{1,2} \sim \mathcal{N}(\mu = 0, \sigma^2 = 1)}$.

\subsubsection{Ideal data}

Assuming for the moment that both $\tens{N}_\mathrm{corr}$ and
$\tens{N}_\mathrm{wn}$ are diagonal in Fourier space, we note that
Eq.~(\ref{eq:ncorr_samp}) may be solved in a closed form in Fourier
space,
\begin{equation}
  n_f^\mathrm{corr} = \frac{d'_f +
    C\left(\tens{N}_\mathrm{wn}^{1/2}(f) w_1 + \tens{N}_\mathrm{wn}(f)
    \tens{N}_\mathrm{corr}^{-1/2}(f) w_2\right)}{1 +
    \tens{N}_\mathrm{wn}(f)/\tens{N}_\mathrm{corr}(f)},
  \label{eq:ncorr_fspace}
\end{equation}
for any non-negative frequency $f$, where the correlated noise TOD has
been decomposed as $n^{\mathrm{corr}}_f = \sum_{t} n^{\mathrm{corr}}_t \mathrm
e^{-2\pi ift}$. For
completeness, $C$ is a constant factor that depends on the Fourier
convention of the numerical library of choice,\footnote{We use the
  \texttt{FFTW} library, in which case $C=\sqrt{n_\mathrm{samples}}$, where
  $n_\mathrm{samples}$ is the number of time samples.} and $w_{1,2}$ are two
independent random complex samples from a Gaussian distribution,
\begin{equation}
w_{1,2} \equiv \frac{\eta_R + i \eta_I}{\sqrt{2}},
\end{equation}
where $\eta_{R,I} \sim \mathcal{N}(\mu = 0, \sigma^2 = 1)$.

\begin{figure}
  \begin{center}
    \includegraphics[width=\linewidth]{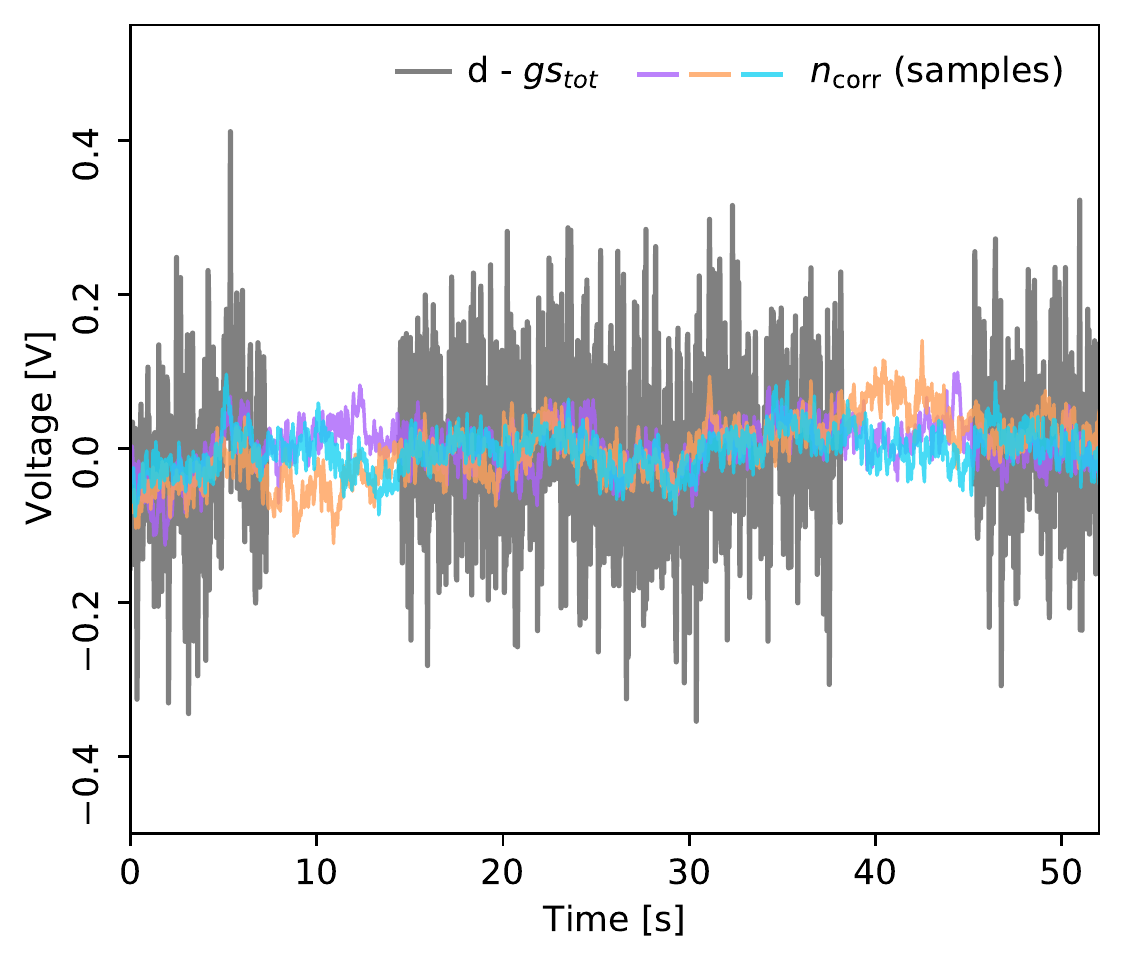}
  \end{center}
  \caption{Illustration of three constrained correlated noise
    realizations (colored curves) drawn from
    $P(\vec{n}^\mathrm{corr}\mid\vec{d}, \xi^n, \vec{s}_\mathrm{tot}, g)$
    for the \Planck\ 30\,GHz data (grey curve). Regions for which parts
    of the data have been masked, either due to a processing mask or
    flagged data, are marked as white gaps. }
  \label{fig:ncorr}
\end{figure}

Figure~\ref{fig:ncorr} shows three independent
realizations of $\vec{n}^\mathrm{corr}$ that all correspond to the
same signal-subtracted \Planck\ 30\,GHz TOD segment. Each correlated
noise sample is essentially a Wiener-filtered version of the original
data, and traces as such the slow variations in the data, with minor
variations corresponding the two random fluctuation terms in
Eq.~(\ref{eq:ncorr_fspace}), as allowed by the white noise level
present in the data. We can also see that there are gaps in the data, 
which we will need to deal with. 

\subsubsection{Handling masking through a conjugate gradient solver}
When writing down an explicit solution of Eq.~(\ref{eq:ncorr_samp}) in
Eq.~(\ref{eq:ncorr_fspace}), we assumed that both
$\tens{N}_\mathrm{corr}$ and $\tens{N}_\mathrm{wn}$ were diagonal in
Fourier space. However, as illustrated in Fig.~\ref{fig:ncorr}, real
observations have gaps, either because of missing or flagged data. The
most typical example of missing data is the application of a
processing mask that removes all samples with too strong foreground
contamination, either from Galactic diffuse sources or from
extragalactic point sources.

We can represent these gaps in our statistical model by setting the
white noise level for masked samples to infinity. This ensures that
Eqs.~(\ref{eq:ncorr}) and (\ref{eq:ncorr_samp}) are still well
defined, albeit somewhat harder to solve. The new difficulty lies in
the fact that while $\tens{N}_\mathrm{wn}$ is still diagonal in
the time domain, it is no longer diagonal in the Fourier domain. This problem
may be addressed in two ways. Specifically, we can either solve
Eqs.~(\ref{eq:ncorr}) and (\ref{eq:ncorr_samp}) directly, using an
iterative method such as the conjugate gradient (CG) method
\citep{wehus:2012,bp02}, or we can fill any gap in $\vec{d}'$ with a
simpler interpolation scheme, for instance a polynomial plus white
noise, and then use Eq.~(\ref{eq:ncorr_fspace}) directly. The former method is mathematically superior, as it results in a
statistically exact result. However, the CG method is in general not
guaranteed to converge due to numerical round-off errors, and since the 
current algorithm is to be
applied millions of times in a Monte Carlo environment, the second
approach is useful as a fallback solution for the few cases for which the
exact CG approach fails. 

As shown by \citet{bp02}, Eq.~(\ref{eq:ncorr_samp}) may be recast into
a compressed form using the Sherman-Morrison-Woodbury formula,
effectively separating the masked from the unmasked degrees of
freedom, and the latter may then be handled with the direct formula in
Eq.~(\ref{eq:ncorr_fspace}). This approach, in addition to having a
lower computational cost per CG iteration, also needs fewer iterations
to converge compared to the untransformed equation. We adopt this
approach without modifications for the main \BP\ pipeline. 

Returning to Fig.~\ref{fig:ncorr}, we note that the correlated noise has significantly larger variance between the samples within the gaps than in the data-dominated regime. As a result, one should expect to see a
slightly higher conditional $\chi^2$ inside the processing mask in a
full analysis than outside, since $\n^{\mathrm{corr}}$ will
necessarily trace the real data less accurately in that range. This is
in fact seen in the main \BP\ analysis, as reported by
\citet{bp01} and \citet{bp10}. However, when marginalizing over all allowed
correlated noise realizations, the final uncertainties will be
statistically appropriate, due to the fluctuation terms in
Eq.~(\ref{eq:ncorr_samp}).

\subsubsection{Gap-filling by polynomial interpolation}

\begin{figure}
	\begin{center}
		\includegraphics[width=\linewidth]{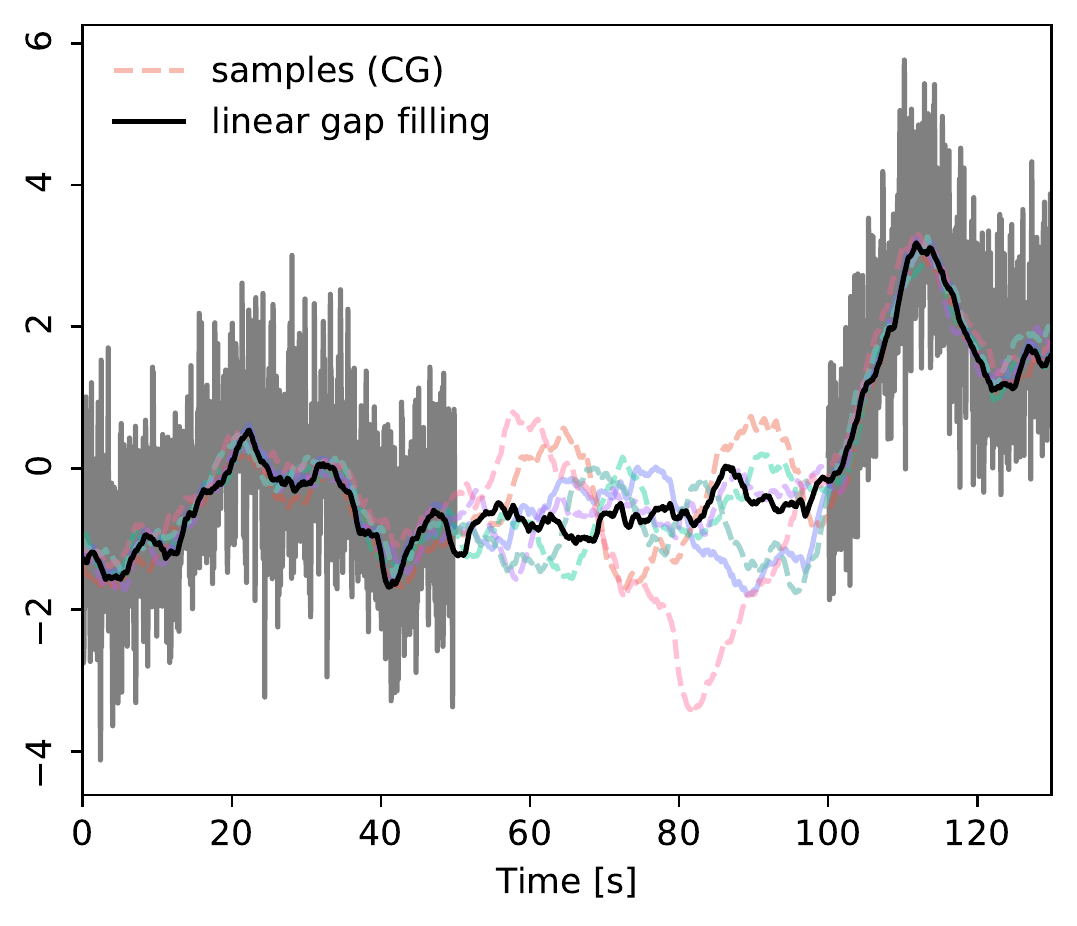}
	\end{center}
	\caption{Correlated noise
		realizations using linear gap-filling (black) and full constrained realizations using CG
		(colored dashed lines) for simulated data with extreme noise
		properties and a large gap. This illustrates the limitation of the linear
          gap-filling procedure. In general, the linear
          gap-filling procedure tends to underestimate the
          fluctuations in $n^\mathrm{corr}$ on long timescales within the gaps.}
	\label{fig:bias}
\end{figure}

As mentioned above, the CG algorithm does not always converge, and for
Monte Carlo applications that will run millions of times without human
supervision, it is useful to establish a robust fallback solution. For
this purpose, we adopt the basic approach of simply interpolating
between the values on each side of a gap. Specifically, we compute the
average of the non-masked points among the 20 points on each side of
the gap, and interpolate linearly between these two values. The choice of exactly 20 points is somewhat arbitrary, but it is a compromise between integrating down the white noise and making sure that the correlated noise stays roughly constant in the interval. In
addition, we add a white noise component to $\vec{d}'$, based on
$\N^{\mathrm{wn}}$, to each masked sample.

An important limitation of the linear gap-filling procedure is
associated with estimation of the noise PSD parameters, $\xi^n$. As
described in Sect.~\ref{sec:noise_psd}, these parameters are estimated
directly from $\vec{n}^\mathrm{corr}$ by Gibbs sampling. A
statistically suboptimal sample of $\vec{n}^\mathrm{corr}$ may
therefore also bias $\xi^n$, which in turn may skew
$\vec{n}^\mathrm{corr}$ even further. If the gaps are short, then this
bias is usually negligible, but for large gaps it can be
problematic. This situation is illustrated in Fig.~\ref{fig:bias},
which compares the linear gap filling procedure with the exact CG
approach. In general, the linear method tends to underestimate the
fluctuations on large timescales within the gap.

Because of the close relative alignment of the \Planck\ scanning
strategy with the Galactic plane that takes place every six
months \citep{planck2011-1.1}, some pointing periods happen to have larger gaps than
others. For these, two long masked regions occur every minute, when
the telescope points toward the Galactic plane. Any systematic bias
introduced by the gap-filling procedure itself will then not be
randomly distributed in the TOD, but rather systematically contribute
to the same modes, with a specific period equal to the satellite spin
rate. For these, the statistical precision of the CG algorithm is
particularly important to avoid biased noise parameters.

Overall, the linear gap filling procedure should only be used when
strictly necessary. In practice, we use it only when the CG solver
fails to converge within 30 iterations, which happens in less
than 0.03\,\% of all cases.

Another simpler and more accurate gap filling procedure is suggested
by \citet{bp02}: We may simply fill the gaps in $\d'$ with the
previous sample of the correlated noise, and then add white noise
fluctuations.  This corresponds to Gibbs sampling over the white noise
as a stochastic parameter, which is statistically fully valid. However,
this approach requires us to store the correlated noise TOD in memory
between consecutive Gibbs iterations. Since memory use is already at
its limit \citep{bp03}, this method is not used for the main
\BP\ analysis. However, for systems with more available RAM, this
method is certainly preferable over simple linear interpolation.

\subsection{Sampling noise PSD parameters, $P(\xi^n\mid\vec{n}^\mathrm{corr})$}
\label{sec:noise_psd}

\begin{figure}
	\begin{center}
		\includegraphics[width=\linewidth]{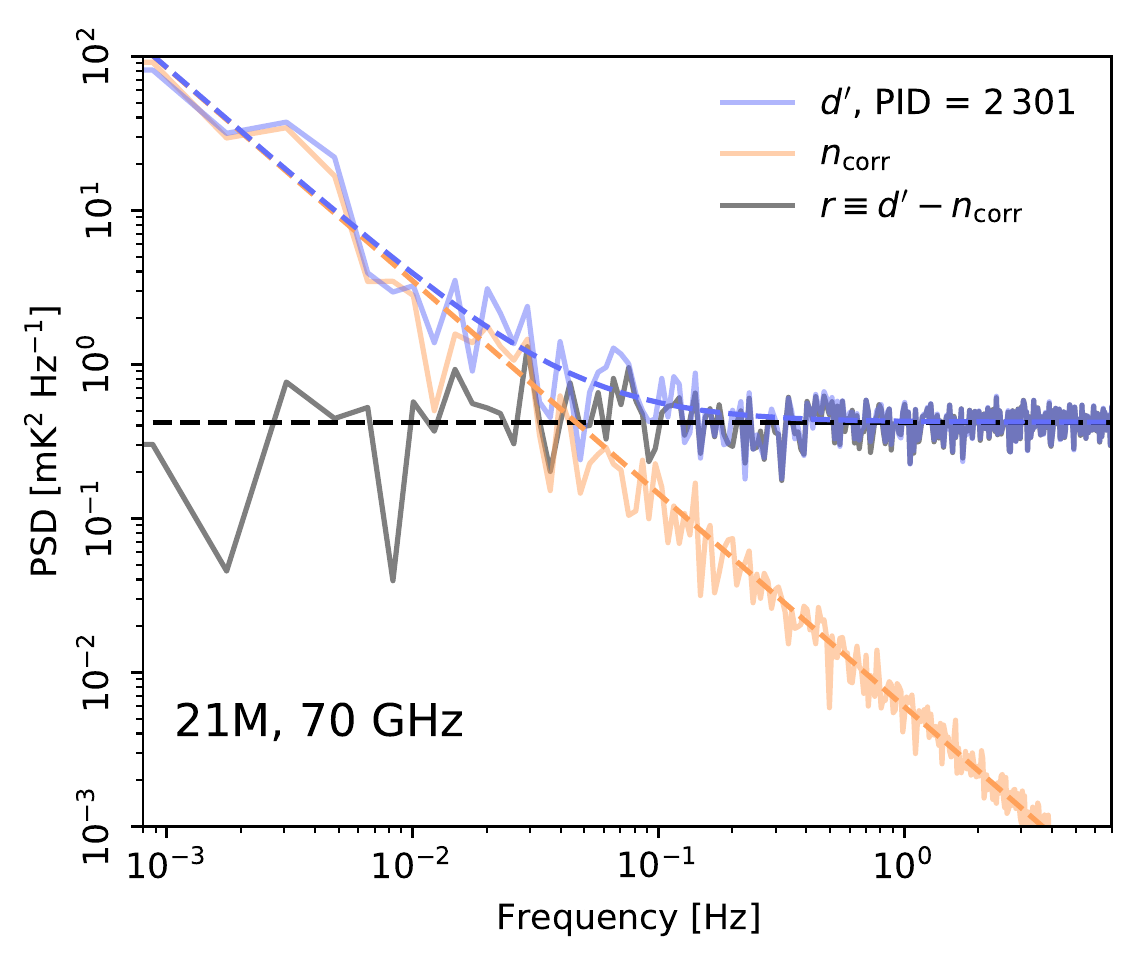}
	\end{center}
	\caption{Comparison of temporal PSDs for different
          components. The blue curve shows the PSD of the
          signal-subtracted data; the orange curve shows the fitted
          correlated noise PSD; and the gray line shows the PSD of the
          residual TOD. The dashed curves correspond to the best fit
          $1/f$-noise model, with (blue) and without (orange) white
          noise. Note that the 70 GHz channels do not have a 
          lognormal component.}
	\label{fig:ps_1f}
\end{figure}

The second noise-related conditional distribution in the \BP\ Gibbs
chain is $P(\xi^n\mid\vec{n}^\mathrm{corr})$, which describes the
noise PSD. As discussed in Sect.~\ref{sec:bp}, in this paper we model
this function in terms of a $1/f$ spectrum as defined by
Eq.~(\ref{eq:1fmodel}), with an added lognormal component in the 30 and 44 GHz bands Eq.~(\ref{eq:1fmodel_lognorm}). 
We emphasize, however, that any functional form for $P(f)$ may be
fitted using the methods described below. Figure~\ref{fig:ps_1f}
illustrates the PSD of the different components for a 70 GHz radiometer, and our task is now
to sample each of the noise PSD parameters $\xi^n = \{\sigma_0,
f_\mathrm{knee}, \alpha\}$, corresponding to the dashed blue line in
this figure.

\subsubsection{Sampling the white noise level, $\sigma_0$}

We start with the white noise level, which by far is the most
important noise PSD parameter in the system. We first note from
Eq.~(\ref{eq:1fmodel}) that if $\alpha$ is close to zero, the
correlated and white noise terms are perfectly degenerate. Even for
$\alpha\approx-1$ there is a significant degeneracy between the two
components for a finite-length TOD.

Of course, for other parameters in the full Gibbs chain, only the
combined $P(f)$ function is relevant, and not each component
separately. At the same time, and as described by \citet{bp01},
marginalization over the two terms within other sampling steps happens
using two fundamentally different methods: while white noise
marginalization is performed analytically through a diagonal
covariance matrix, marginalization over correlated noise is done by
Monte Carlo sampling of $\n^{\mathrm{corr}}$. It is therefore
algorithmically advantageous to make sure that the white noise term
accounts for as much as possible of the full noise variance, as this
will lead to an overall shorter Markov chain correlation length.

For this reason, we employ a commonly used trick in radio astronomy
for estimating the white noise level, and define this to be
\begin{equation}
  \sigma_0^2 \equiv \frac{\mathrm{Var}(r_{i+1} - r_i)}{2},
  \label{eq:sigma0}
\end{equation}
where $\vec{r} \equiv \vec{d}' - \vec{n}^\mathrm{corr}$. By differencing consecutive samples, any residual temporal correlations
are effectively eliminated, and will therefore not bias the
determination of $\sigma_0$. This is very important, because the comparison of this white noise level to the full variance of the residual is what defines the $\chi^2$ goodness-of-fit statistic (see Sect.~\ref{sec:time_variability}) that we use to check if our model fits the data.

This method is equivalent to fixing the white noise level to the
highest frequencies in Fig.~\ref{fig:ps_1f}. Formally speaking, this
means that $\sigma_0$ should not be considered a free parameter within
the Gibbs chain, but rather a derived quantity fixed by the data,
$\vec{d}$, the gain, $g$, the signal model, $\vec{s}^\mathrm{tot}$,
and the correlated noise, $\vec{n}^\mathrm{corr}$. However, this
distinction does not carry any particular statistical significance
with respect to other parameters, and we will in the following
therefore discuss $\sigma_0$ on the same footing as any of the other
noise parameters.

\subsubsection{Sampling correlated noise parameters, $f_\mathrm{knee}$, $\alpha$, and $A_\mathrm{p}$}

With $\sigma_0^2$ fixed by Eq.~(\ref{eq:sigma0}), the other noise
parameters, $f_\mathrm{knee}$, $\alpha$ and $A_\mathrm{p}$, are sampled from their
exact conditional distributions. Since we assume that also the
correlated noise component is Gaussian distributed, the appropriate
functional form is that of a multivariate Gaussian,
\begin{equation}
	P(f_\mathrm{knee},\alpha, A_\mathrm{p}\mid\sigma_0, \vec{n}^\mathrm{corr})
        \propto \frac{\mathrm e^{-\frac{1}{2}(\vec{n}^\mathrm{corr})^T
            \tens{N}_\mathrm{corr}^{-1}\vec{n}^\mathrm{corr}}}{\sqrt{|\tens{N}_\mathrm{corr}|}}
        P(f_\mathrm{knee},\alpha, A_\mathrm{p}),
\end{equation}
where $\tens{N}_{\mathrm{corr}} = \tens{N}_\mathrm{corr}(f_\mathrm{knee},\alpha, A_\mathrm{p})$, and
$P(f_\mathrm{knee},\alpha, A_\mathrm{p})$ is an optional prior. This may be
efficiently evaluated in Fourier space as
\begin{equation}\label{eq:logP_S}
	-\ln P = \sum_{f=f_{\mathrm{min}}}^{f_{\mathrm{max}}}
        \left[\frac{|n_f^\mathrm{corr}|^2}{\tens{N}_\mathrm{corr}(f)}
          + \ln{\tens{N}_\mathrm{corr}(f)}\right] - \ln P(f_\mathrm{knee},\alpha, A_\mathrm{p}),
\end{equation}
where $\tens{N}_\mathrm{corr}(f) = \sigma_0^2 \left(\frac{f}{f_\mathrm{knee}}\right)^\alpha + A_\mathrm{p} \exp\left[-\frac{1}{2}\left(\frac{\log_{10}f - \log_{10} f_\mathrm{p}}{\sigma_\mathrm{dex}}\right)^2\right]$.

To explore this joint distribution, we iteratively Gibbs sample over
$f_\mathrm{knee}$, $\alpha$ and $A_\mathrm{p}$, using an inversion sampler for
each of the three conditional distributions, $P(f_\mathrm{knee}\mid\alpha, A_\mathrm{p},\sigma_0, n^\mathrm{corr})$, $P(\alpha\mid f_\mathrm{knee}, A_\mathrm{p}, \sigma_0, n^\mathrm{corr})$ and $P(A_\mathrm{p}\mid \alpha, f_\mathrm{knee}, \sigma_0, n^\mathrm{corr})$; see
Appendix~A in \citet{bp01} for details regarding the inversion
sampler.

If we naively apply our statistical model, all frequencies should in
principle be included in the sum in Eq.~(\ref{eq:logP_S}).  At the
same time, we note that frequencies well above $f_{\mathrm{knee}}$
ideally should carry very little statistical weight, since the
correlated noise variance then by definition is smaller than the white
noise variance.  This means that the sampled $\n^\mathrm{corr}$ is
almost completely determined by the prior (i.e., the previous values of
$\sigma_0, f_\mathrm{knee}, \alpha, A_\mathrm{p}$), at these high frequencies. The
sum in Eq. (\ref{eq:logP_S}), on the other hand, is completely
dominated by those high frequencies. The result of this is an
excessively long Markov chain correlation length when including all
frequencies in Eq.~(\ref{eq:logP_S}); the inferred values of $\alpha$, $f_\mathrm{knee}$, and $A_\mathrm{p}$ will always be extremely close to the previous values.

One way to avoid these long correlation lengths would be not to 
condition on $\n^\mathrm{corr}$ at all, but rather use the likelihood for 
$\d'$ to sample $\alpha$, $f_\mathrm{knee}$ and $A_\mathrm{p}$ (and sample 
$\n^\mathrm{corr}$ afterwards). This is equivalent to sampling $\xi^n$
from the marginal distribution with respect to $\n^{\mathrm{corr}}$,
and fully analogous to how the degeneracy between $g$ and
$\n^{\mathrm{corr}}$ is broken through joint sampling. However, 
for real world data, residual signals or
systematics may leak into $\d'$, in particular at frequencies around
and above the satellite scanning frequency. While some of these 
systematics may also leak into $\n^\mathrm{corr}$, in general 
$\n^\mathrm{corr}$ is cleaner, especially at frequencies 
below $f_\mathrm{knee}$, where $\n^\mathrm{corr}$ is dominated
by the random sampling terms. 

A useful solution that both makes the correlated noise
parameters robust against modelling errors and results in a short
Markov chain correlation length is therefore to condition on
$\n^\mathrm{corr}$ above some pre-specified frequency. In practice, we
therefore choose to only include frequencies below
$f_\mathrm{max} = 3.7, 3.0$ or 0.14 Hz for the 30, 44, and 70 GHz bands, 
respectively.
That is, we only use the part of
$\n^\mathrm{corr}$ where we are able to measure the $1/f$ slope with
an appreciable signal to noise ratio. For the lower frequency cutoff
in Eq.~(\ref{eq:logP_S}), we adopt $f_{\mathrm{min}}>0$, and only
exclude the overall mean per PID.

\subsubsection{Priors on $\alpha$ and $\f_\mathrm{knee}$}

\begin{figure}
  \begin{center}
    \includegraphics[width=\linewidth]{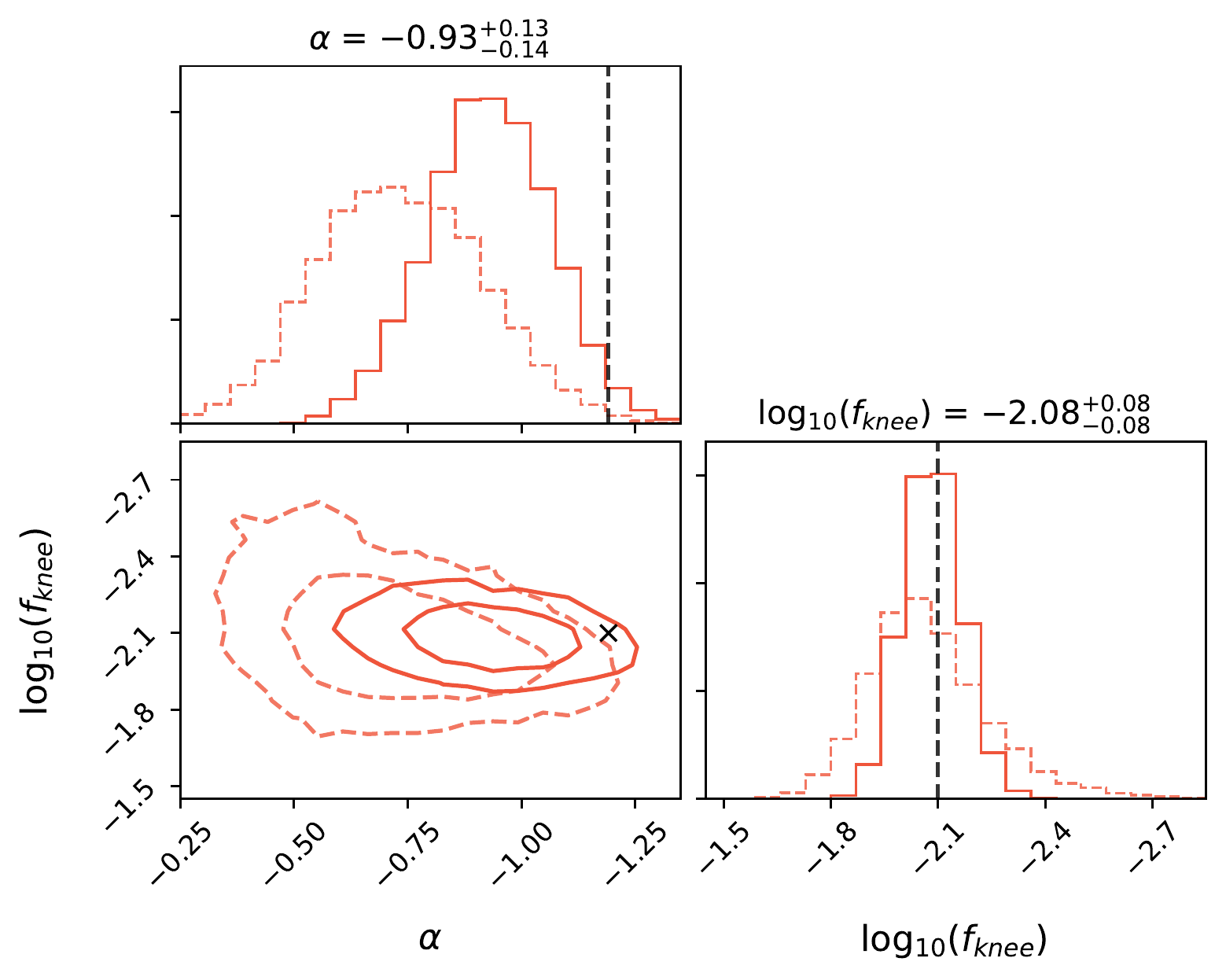}
  \end{center}
  \caption{Distribution of noise parameters for PID 3401 of
    radiometer 20M, one of the 70 GHz channels, for fixed
    $\vec{s}_\mathrm{tot}$ and $g$. Dashed red lines correspond to
    results obtained without an active prior, while the solid line
    corresponds to results after including the active priors on
    $f_\mathrm{knee}$ and $\alpha$ from Eqs.~(\ref{eq:fknee_prior}) and
    (\ref{eq:alpha_prior}). The black cross indicates the best fit values
    derived by the DPC pipeline for this radiometer.
    \label{fig:conditional}}
\end{figure}

As described by \citet{planck2016-l02}, the official \Planck\ LFI Data 
Processing Center (DPC)
analyses assume the noise PSD to be stationary throughout the
mission. Here we allow these parameters to
vary from PID to PID in order to accommodate possible changes in the
thermal environment of the satellite. However, since the duration of a
single PID is typically one hour or shorter, there is only a limited
number of large-scale frequencies available to estimate the correlated
noise parameters, and this may in some cases lead to significant
degeneracies between $\alpha$, $f_\mathrm{knee}$ and 
$A_\mathrm{p}$. In particular, if
$f_\mathrm{knee}$ is low (which of course is the ideal case), $\alpha$
is essentially unconstrained. To avoid pathological cases, it is
therefore useful to impose priors on these parameters, under
the assumption that the system should be relatively stable as a
function of time.

Specifically, we adopt a log-normal prior for $f_\mathrm{knee}$,
\begin{equation}\label{eq:fknee_prior}
	-\ln P(f_\mathrm{knee}) =  \frac{1}{2}\left(\frac{\log_{10} f_\mathrm{knee}  - \log_{10} f^\mathrm{DPC}_\mathrm{knee} }{\sigma_{f_\mathrm{knee}}}\right)^2 + \ln f_\mathrm{knee},
\end{equation}
where $f^\mathrm{DPC}_\mathrm{knee}$ is the DPC result for a given
radiometer \citep{planck2016-l02} and $\sigma_{f_\mathrm{knee}} = 0.1$. For $\alpha$, we
adopt a Gaussian prior of the form
\begin{equation}\label{eq:alpha_prior}
	-\ln P(\alpha) =  \frac{1}{2}\left(\frac{\alpha  - \alpha^\mathrm{DPC}}{\sigma_\alpha}\right)^2,
\end{equation}
where $\alpha^\mathrm{DPC}$ again is the DPC result for the given
radiometer and $\sigma_\alpha = 0.2$. 
Figure \ref{fig:conditional} shows a comparison of the posterior
distributions with (solid lines) and without (dashed lines) active
priors for a typical example.

The prior widths have been chosen to be
sufficiently loose that the overall impact of the priors is moderate
for most cases. The priors are in practice only used to exclude
pathological cases. Technically speaking, we also
impose absolute upper and lower limits for each parameter, as this is
needed for gridding the conditional distribution within the inversion
sampler. However, the limits are chosen to be sufficiently wide so that they have no
significant impact on final results. We do not use an active prior on the amplitude $A_\mathrm{p}$.

\section{Mitigation of modelling errors and degeneracies}
\label{sec:degeneracies}

When applying the methods described above to real-world data as part
of a larger Gibbs chain, several other degeneracies and artifacts may
emerge beyond those discussed above. In this section, we discuss some
of the main challenges for the current setup, and we also describe
solutions to break or mitigate these issues.

\begin{figure}
	\begin{center}
		\includegraphics[width=\linewidth]{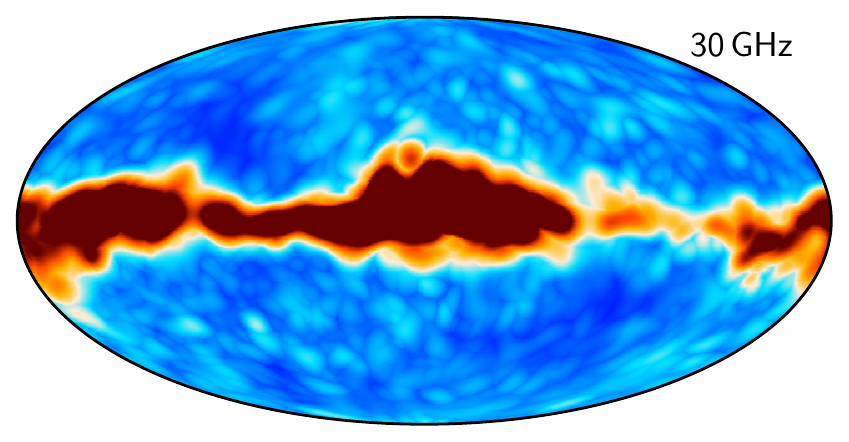}\\
		\includegraphics[width=\linewidth]{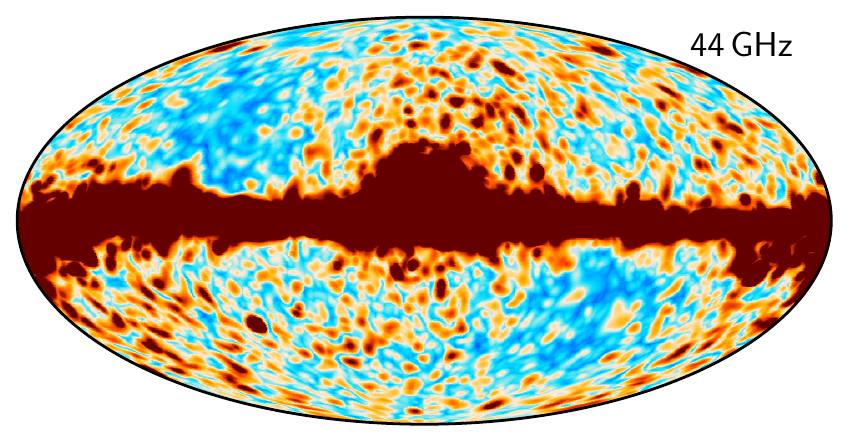}\\
		\includegraphics[width=\linewidth]{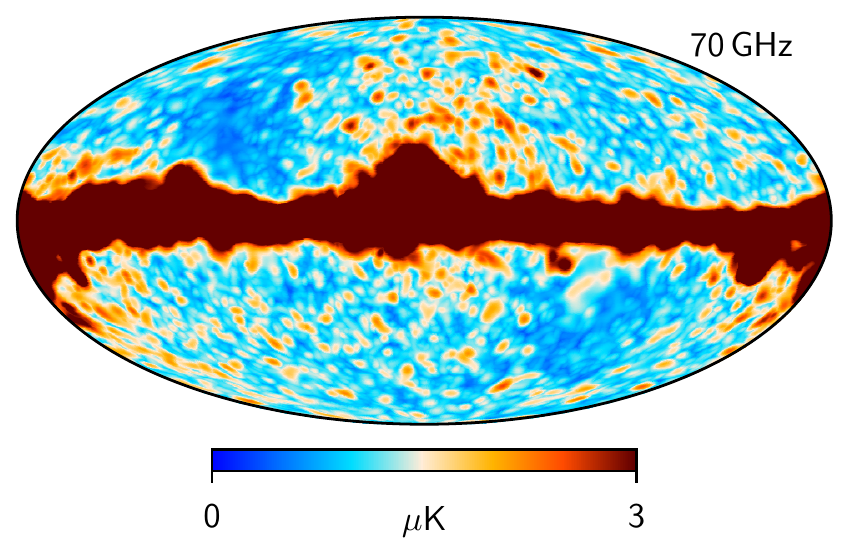}\\
	\end{center}
	\caption{Residual maps, $\r_{\nu}$, for each of the three
		\Planck\ LFI frequencies, smoothed to a common angular resolution
		of $10^{\circ}$ FWHM. 
		\label{fig:todres}}
\end{figure}

\begin{figure}
	\begin{center}
		\includegraphics[width=\linewidth]{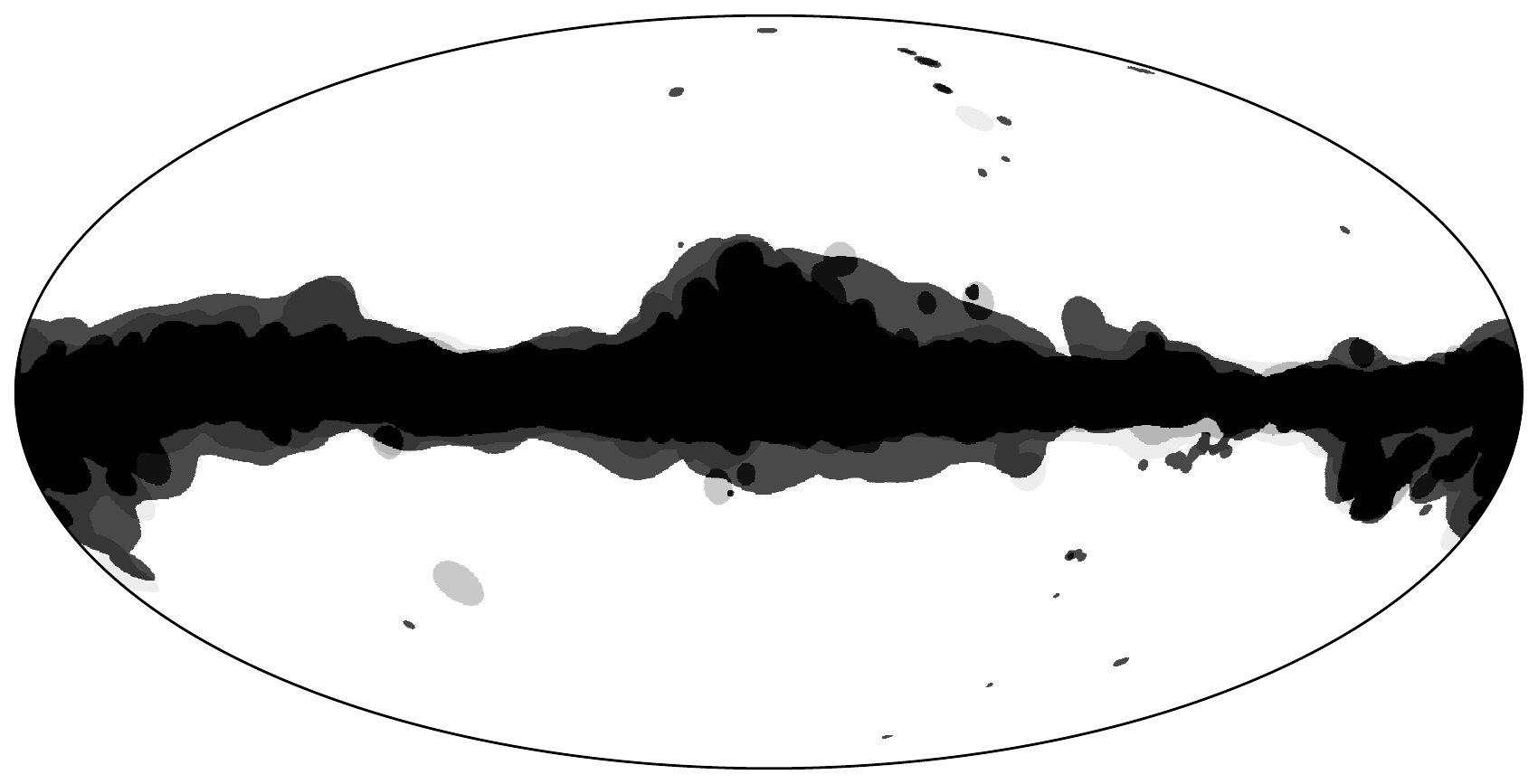}    
	\end{center}
	\caption{Processing masks used for correlated noise
		sampling. Different shades of gray indicate different frequency
		masks. The allowed 30\,GHz sky fraction (light)
		is $f_{\mathrm{sky}}=0.73$; the 44\,GHz sky fraction
		(intermediate) is $f_{\mathrm{sky}}=0.81$; and the 70\,GHz sky fraction
		(dark) is $f_{\mathrm{sky}}=0.77$.
		\label{fig:procmask}}
\end{figure}

\subsection{Signal modelling errors and processing masks}

First, we note that the correlated noise component is by nature
entirely instrument specific, and depends on the intrinsic 
behavior of the individual radiometers (mainly fluctuations in 
gain and noise temperature of the amplifiers) and on the stability 
of the thermal environment. It is therefore difficult to impose any
strong spatial priors on $\n^{\mathrm{corr}}$, beyond the loose PSD
priors described above, and these provide only very weak
constraints in the map-domain. The correlated noise is from first
principles the least known parameter in the
entire model, and its allowed parameter space is able to describe a
wide range of different TOD combinations, without inducing a
significant likelihood penalty relative to the noise PSD model. As a
result, a wide range of systematic errors or model mismatches may be
described quite accurately by modifying $\vec{n}^\mathrm{corr}$,
rather than ending up in the residual,
\begin{equation}
  \r \equiv \vec{d} - \vec{n}^\mathrm{corr}
  - g\vec{s}_\mathrm{tot} - s^{\mathrm{1hz}}.
  \label{eq:residual}
\end{equation}

Colloquially speaking, the correlated noise component may in many
respects be considered the ``trash can'' of CMB time-ordered analysis,
capturing anything that does not fit elsewhere in the model. This is
both a strength and a weakness. On the one hand, the flexibility of
$\n^{\mathrm{corr}}$ protects against modelling errors for other (and
far more important) parameters in the model, including the CMB
parameters. On the other hand, in many cases it is preferable that
modelling errors show up as $\chi^2$ excesses, so that they can
be identified and mitigated, rather than leaking into the correlated
noise. To check for different types of modelling errors, it is
therefore extremely useful to inspect both $\chi^2$s and binned sky
maps of $\r_\nu$ and $\n^{\mathrm{corr}}$ for artifacts. For an
explicit example of this, see the discussion of data selection for
\BP\ in \citet{bp10}, where these statistics are used as efficient
tools to identify bad observations.

In general, the most problematic regions of the sky are those with
bright foregrounds, either in the form of diffuse Galactic emission or
strong compact sources. If residuals from such foregrounds are present
in the signal-subtracted data, $\d'$, while estimating the correlated
noise TOD, the correlated noise Wiener filter in
Eq.~(\ref{eq:ncorr_samp}) will attempt to fit these in
$\n^{\mathrm{corr}}$, and this typically results in stripes along the
scanning path with a correlation length defined by the ratio between
$f_\mathrm{knee}$ and the scanning frequency.

To suppress such artifacts, we impose a processing mask for each
frequency, as discussed in Sect.~\ref{sec:ncorr_samp}. In the current
analysis, we define these masks as follows:
\begin{enumerate}
\item We bin the time-domain residual\footnote{Note that the residual used here is based on a preliminary test run, since these masks are used internally in the final analysis.} in Eq.~(\ref{eq:residual}) into
  an $IQU$ pixelized sky map for each frequency (as defined by Eq.~(77) in
  \citealp{bp01}), and smooth this map to an angular resolution of
  $10^{\circ}$ FWHM.
\item We take the absolute value of the smoothed map, and then smooth
  again with a $30\arcm$ beam to account for pixels which the raw
  residual map changes sign.
\item We then compute the maximum absolute value for each pixel over
  each of the three Stokes parameters. The resulting maps are shown in
  Fig.~\ref{fig:todres} for each of the three \Planck\ LFI frequencies.
\item These maps are then thresholded at values well above the
  noise level, and these thresholded maps form the main input to the
  processing masks. 
\item To remove particularly bright compact objects that may not be
  picked up by the smooth residual maps described above, we
  additionally remove all pixels with high free-free and/or AME
  levels, as estimated in an earlier analysis.
\end{enumerate}
The final processing masks are shown in Fig.~\ref{fig:procmask}, and
allow 73, 81, and 77\,\% of the sky to be included while fitting
correlated noise at 30, 44, and 70\,GHz, respectively.

\begin{figure}
	\begin{center}
		\includegraphics[width=\linewidth]{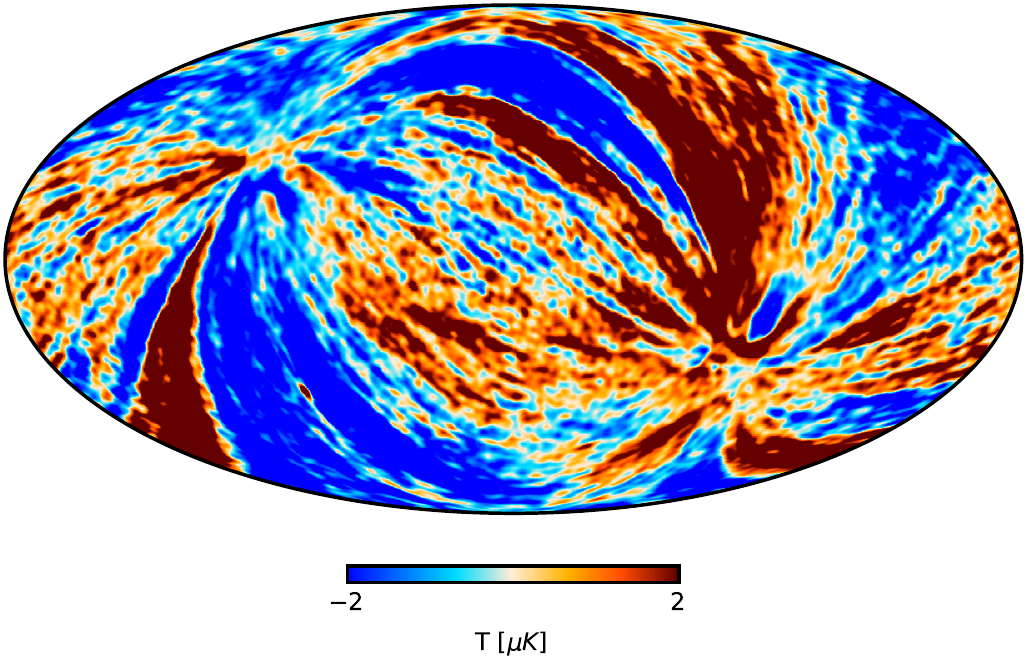}
	\end{center}
	\caption{Correlated noise intensity sample for the 30\,GHz
          band when fitting a model that assumes constant gains throughout
          the mission. This map has been smoothed to an angular resolution
          of $2.5^{\circ}$ FWHM. 
		\label{fig:gain_trouble}}
\end{figure}

\begin{figure}
	\begin{center}
		\includegraphics[width=\linewidth]{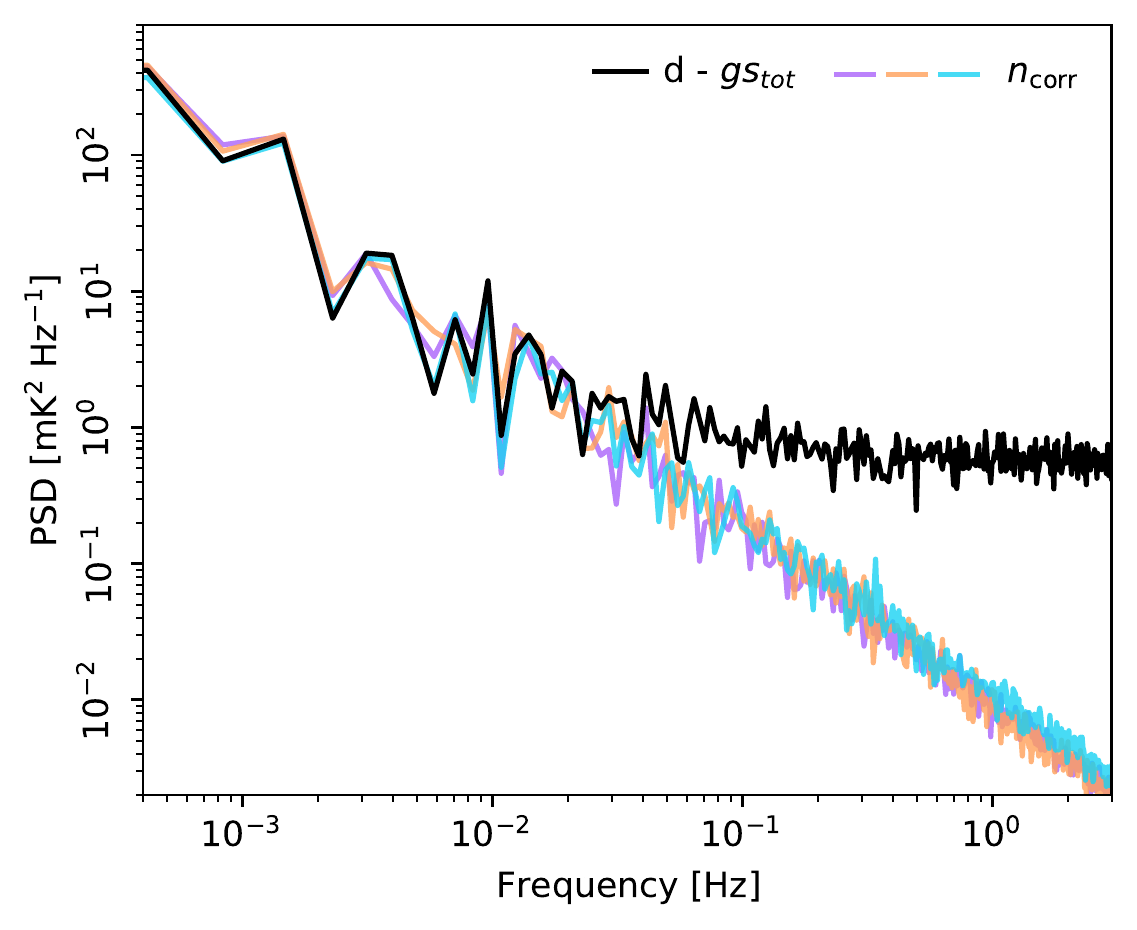}
	\end{center}
	\caption{Three subsequent samples (colored curves) of the
          correlated noise PSD for 23S, one of the 70 GHz
          radiometers. The black line shows the PSD of the
          signal-subtracted data.
		\label{fig:ps_samples}}
\end{figure}

Note that while we use a processing mask when estimating the correlated noise, masking out data from high foreground signal regions, the data from these high signal regions is still used in the other parts of the \BP\, pipeline (e.g. mapmaking, component separation etc.). The difference is only that we have a more uncertain noise model in these regions (since the noise is extrapolated from outside, and will have a high sample variance). 

\subsection{Degeneracies with the gain}

The brightest component of the entire \BP\ signal model is the Solar
CMB dipole, which has an amplitude of 3\,mK. This component plays a
critical role in terms of gain estimation \citep{bp07}, and serves as
the main tool to determine relative calibration differences between
detectors. Both the gain and CMB dipole parameters are of course
intrinsically unknown quantities, and must be fitted jointly. Any
error in the determination of these will therefore necessarily result
in a nonzero residual, in the same manner as Galactic foregrounds
described above, and this may potentially also bias
$\n^{\mathrm{corr}}$. Unlike the Galactic residuals, however, it is
not possible to mask the CMB dipole, since it covers the full sky. The
correlated noise component is therefore particularly susceptible to
errors in either the gain or CMB dipole parameters, and residual
large-scale dipole features in the binned $\n^{\mathrm{corr}}$ map is
a classic indication of calibration errors. To illustrate the effect
of an incorrect gain model, Fig.~\ref{fig:gain_trouble} shows a
30\,GHz correlated noise sample when assuming that the gain is
constant throughout the entire \Planck\ mission.

The gain also has a direct connection with the white noise level,
$\sigma_0$. This manifests itself in different ways, depending on the
choice of units adopted for $\sigma_0$. When expressed in units of
volts, the white noise level is simply given by the radiometer
equation,
\begin{equation}
	\sigma_0 [\mathrm{V}] \propto g_\mathrm{phys} T_\mathrm{sys},
\end{equation}
where $g_\mathrm{phys}$ is the actual physical gain of the radiometer,
and $T_\mathrm{sys}$ is the system temperature \citep{bp01}. In
calibrated units of $\mathrm{K}_{\mathrm{CMB}}$, however, the white
noise level is
\begin{equation}
	\sigma_0 [\mathrm{K}] \propto \frac{g_\mathrm{phys} }{g_\mathrm{model}}T_\mathrm{sys},
\end{equation}
where $g_\mathrm{model}$ is the gain estimate in our model. When
considering the evolution of the noise parameters as a function of
time, we then note that $\sigma_0 [\mathrm{V}]$ will correlate with
the physical gain, which depends strongly on the thermal environment
at any given time. 
On the other hand, if our gain model is correct, i.e.,
$g_{\mathrm{model}}\approx g_{\mathrm{phys}}$, these fluctuations
will cancel in temperature units, and $\sigma_0 [\mathrm{K}]$ should
instead correlate with
the system temperature, $T_\mathrm{sys}$. The system temperature also
depends on the physical temperature, $T_{\mathrm{phys}}$, as the amplifiers' noise and waveguide losses increase with temperature. These were measured in pre-flight tests to be at a level
${dT_{\mathrm{sys}} / dT_{\mathrm{phys}} } \approx 0.2$--0.5\,K/K, depending on
the radiometer \citep{terenzi2009b}.
In conclusion, if we observe a sudden change in $\sigma_0 [\mathrm{K}]$ that is not present in $\sigma_0 [\mathrm{V}]$, this might indicate a problem in the gain model. We also expect that changes in $\sigma_0 [\mathrm{K}]$ reflect genuine variations of the white noise level, mainly driven by changes in the 20\,K stage.
In the following, we will plot $\sigma_0$ as a function
of time in both units of volts and kelvins, and use these to
disentangle gain and system temperature variations.

\section{Results}
\label{sec:results}

\begin{figure*}
	\begin{center}
		\includegraphics[width=0.33\linewidth]{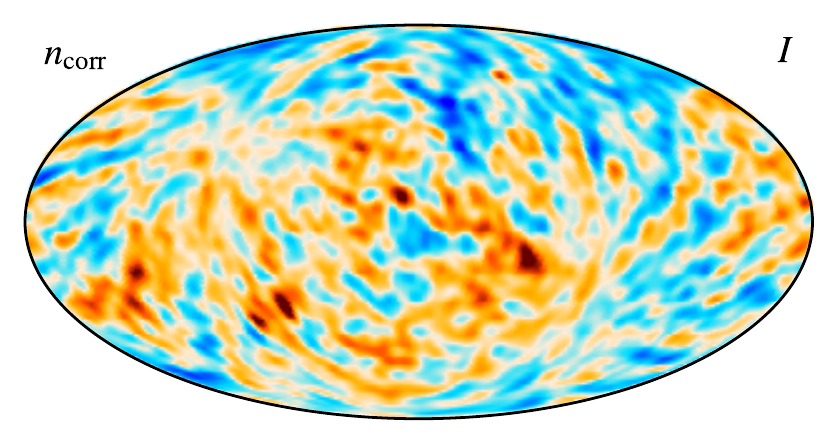}
		\includegraphics[width=0.33\linewidth]{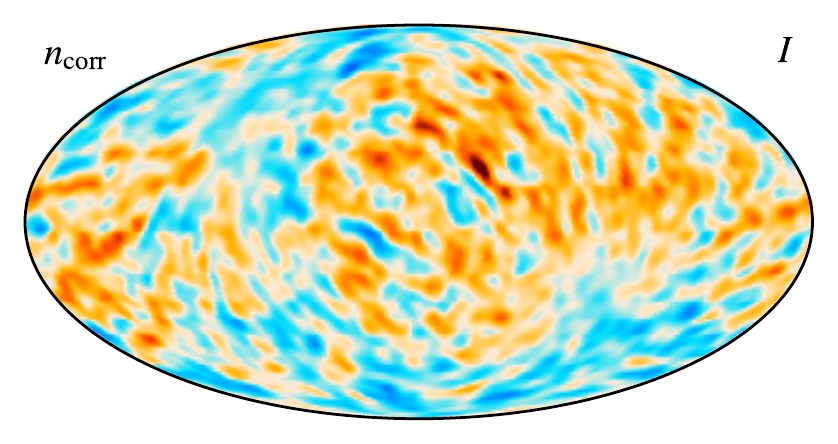}
		\includegraphics[width=0.33\linewidth]{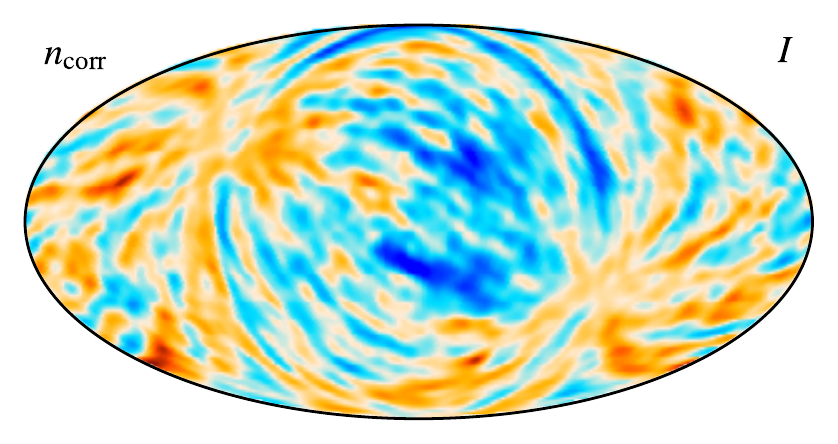}
	\end{center}
	\begin{center}
		\includegraphics[width=0.33\linewidth]{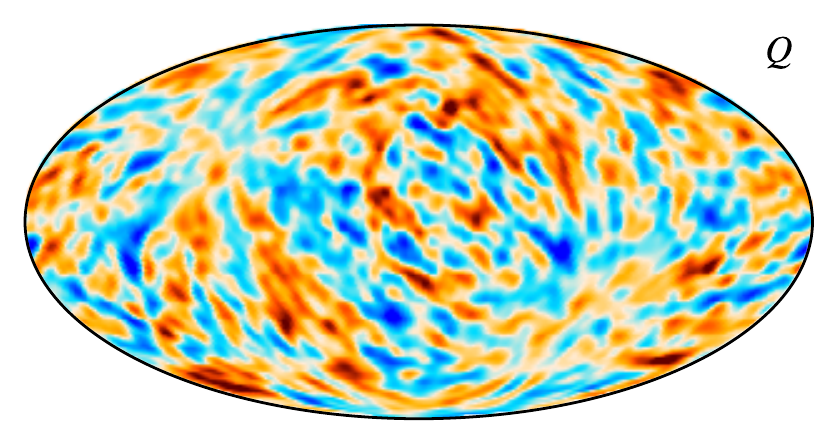}
		\includegraphics[width=0.33\linewidth]{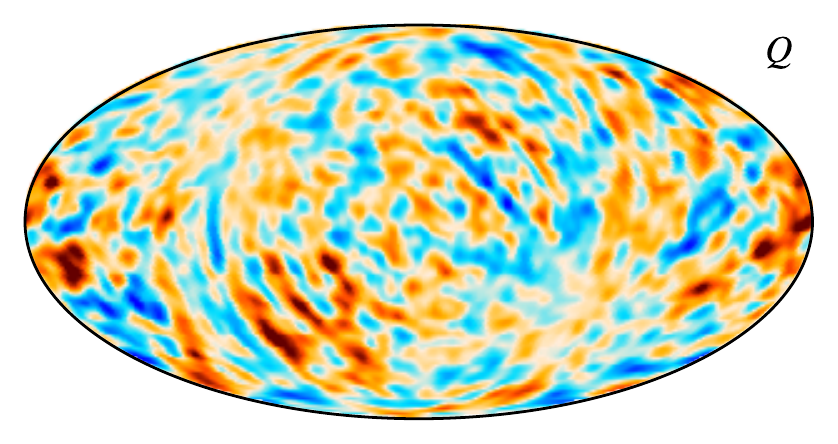}
		\includegraphics[width=0.33\linewidth]{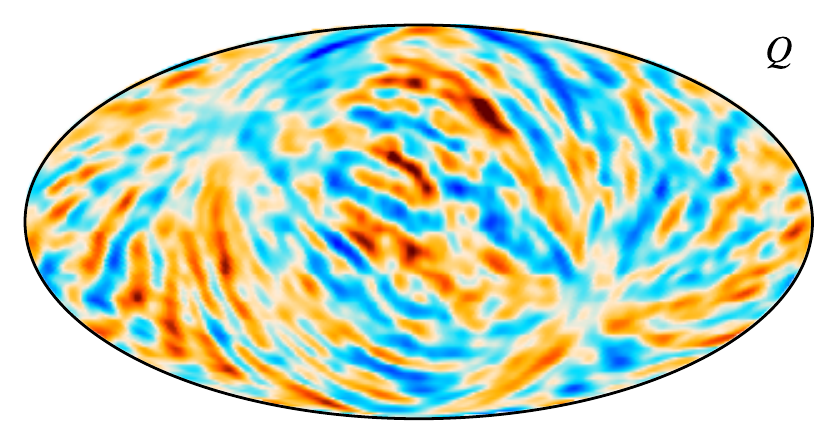}
	\end{center}
	\begin{center}
		\includegraphics[width=0.33\linewidth]{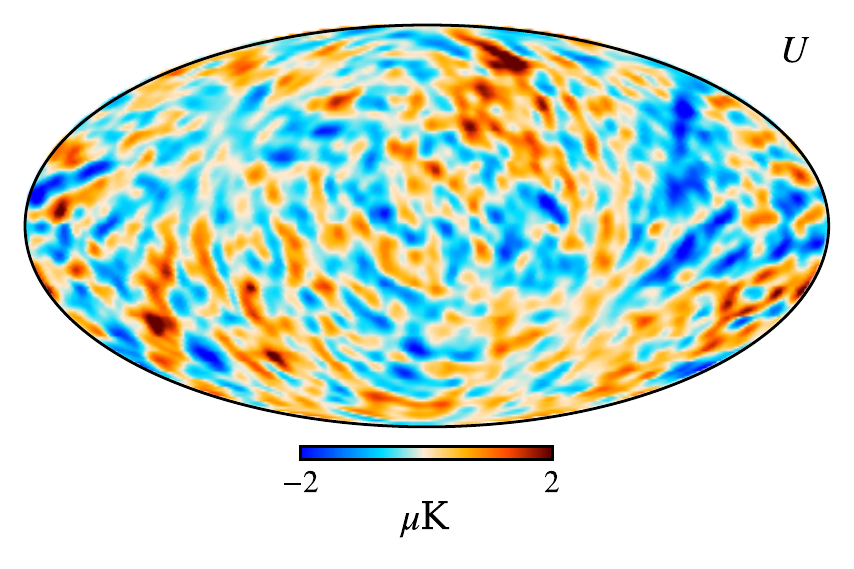}
		\includegraphics[width=0.33\linewidth]{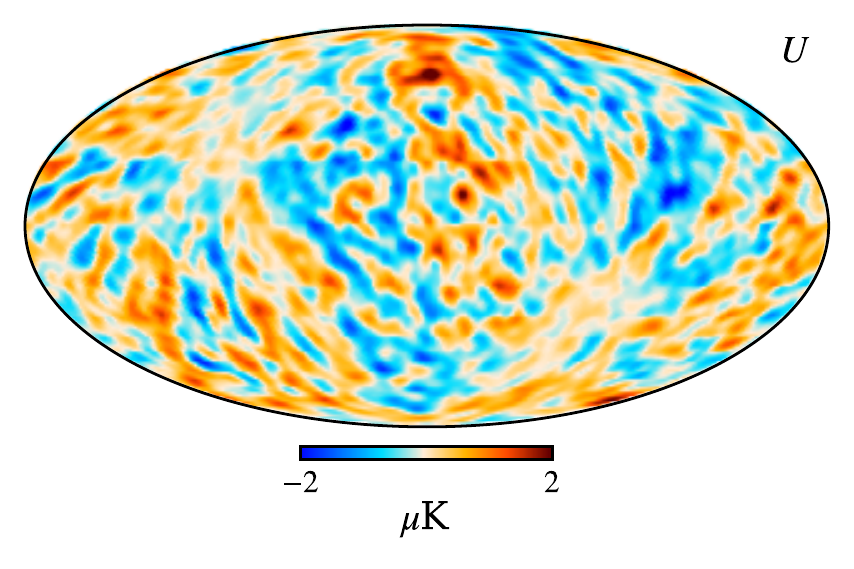}
		\includegraphics[width=0.33\linewidth]{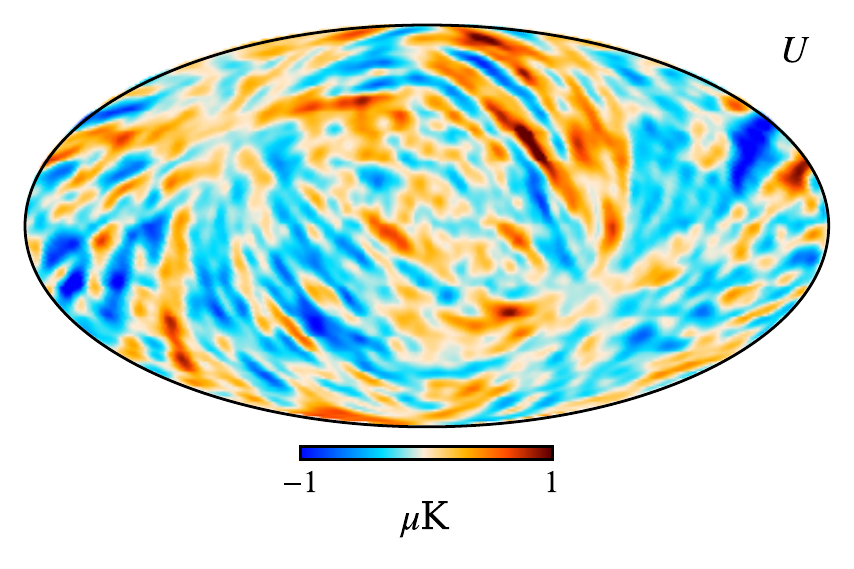}
	\end{center}
	\caption{Maps of a single Gibbs sample of the correlated noise
		added over all radiometers in the 30 GHz (left), 44 GHz
		(middle) and 70 GHz (right) bands. From top to bottom, rows show Stokes $I$, $Q$
		and $U$, respectively. Maps have been smoothed to a common angular resolution
		of $5^{\circ}$ FWHM. 
		\label{fig:ncorr_map}}
\end{figure*}

\begin{figure}
	\begin{center}
		\includegraphics[width=\linewidth]{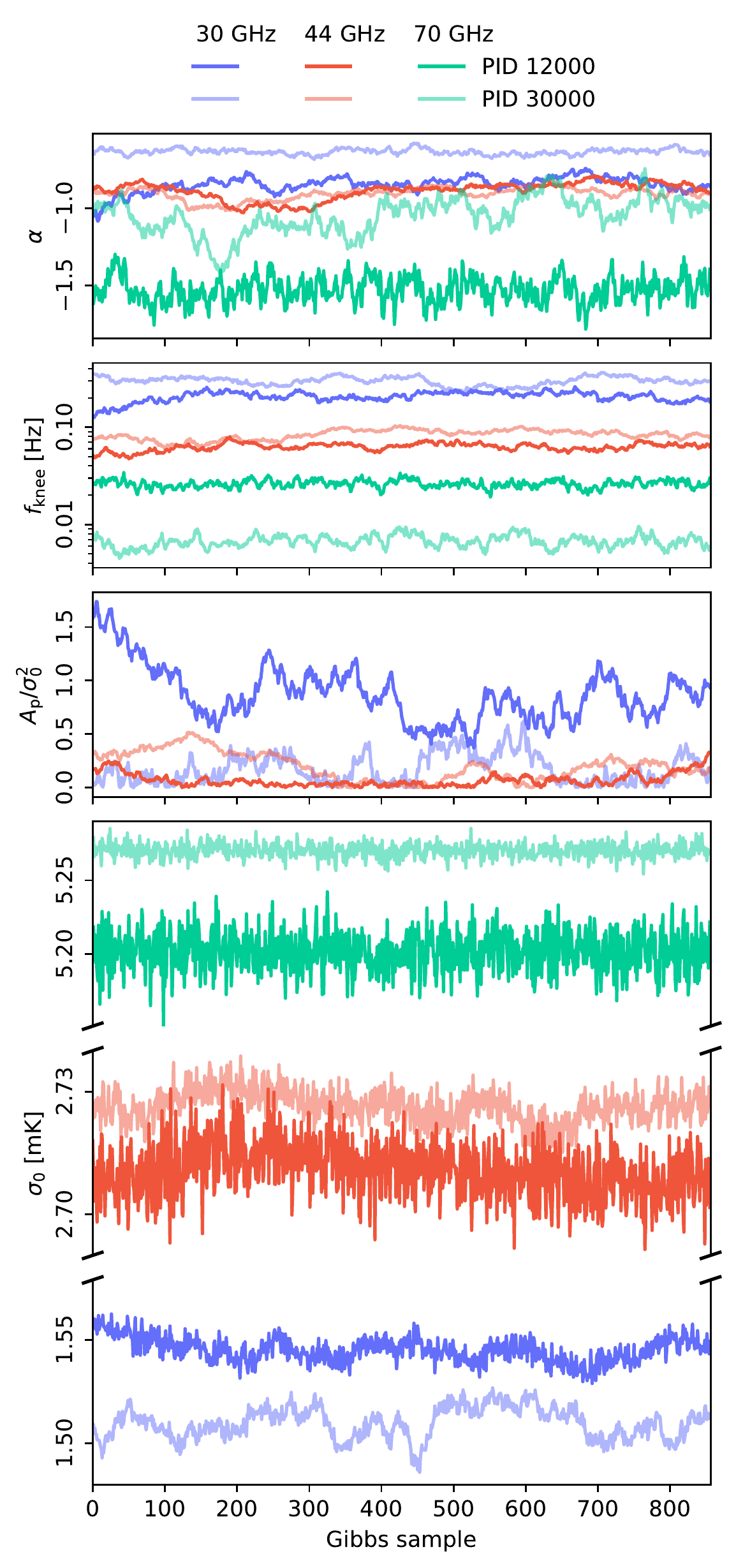}
	\end{center}
	\caption{Gibbs samples of noise parameters for two different PIDs for detectors 27M, 24S and 20M of the 30, 44 and 70\,GHz bands respectively.
	\label{fig:samples}}
\end{figure}

{\renewcommand{\arraystretch}{1.3}%
	\begin{table}
		\centering
		\caption{Distribution of posterior mean noise parameters for each radiometer. Error bars represent variation over time of the posterior mean values, and not the width of the posterior distribution for any given PID.}
		\label{tab:mean_values}
		\begin{tabular}{p{0.4em}|l|c|c|c|c}
		\hline
& Det & $\alpha$ & $f_\mathrm{knee}\,$[mHz] & $\sigma_0$ [mK] & $A_\mathrm{p} / \sigma_0^2$ \\ 
\hline
\multirow{4}{1pt}{\rotatebox[origin=c]{90}{\large 30$\ $GHz}} & 27M & $-0.71^{+0.06}_{-0.08}$ & $254^{+29}_{-66}$ & $1.535^{+0.019}_{-0.016}$ & $0.40^{+0.54}_{-0.24}$ \\ 
&27S & $-0.83^{+0.07}_{-0.09}$ & $102^{+16}_{-13}$ & $ 1.706^{+0.010}_{-0.015} $ & $0.57^{+0.24}_{-0.23}$ \\ 
&28M & $-0.85^{+0.12}_{-0.17}$ & $122^{+28}_{-36}$ & $1.781^{+0.014}_{-0.013}$ & $0.7^{+0.5}_{-0.4}$ \\ 
&28S & $-0.98^{+0.18}_{-0.25}$ & $37^{+13}_{-13}$ & $1.632^{+0.009}_{-0.012} $ & $0.45^{+0.27}_{-0.31}$ \\ 
\hline
\multirow{6}{1pt}{\rotatebox[origin=c]{90}{\large 44$\ $GHz}} & 24M & $-1.05^{+0.16}_{-0.13}$ & $23.3^{+3.3}_{-2.6}$ & $3.149^{+0.015}_{-0.013}$ & $0.22^{-0.11}_{+0.11}$ \\ 
&24S & $-0.90^{+0.05}_{-0.05}$ & $73^{+10}_{-8}$ & $2.714^{+0.014}_{-0.013}$ & $0.20^{+0.33}_{-0.09}$ \\ 
&25M & $-1.04^{+0.24}_{-0.13}$ & $17.7^{+2.7}_{-2.1}$ & $2.823^{+0.012}_{-0.017}$ & $0.28^{+0.11}_{-0.16}$ \\ 
&25S & $-1.07^{+0.15}_{-0.12}$ & $38^{+6}_{-6}$ & $2.681^{+0.020}_{-0.017}$ & $0.45^{+0.13}_{-0.20}$ \\ 
&26M & $-0.94^{+0.12}_{-0.13}$ & $57^{+8}_{-8}$ & $3.263^{+0.019}_{-0.021}$ & $0.25^{+0.25}_{-0.10}$ \\ 
&26S & $-0.68^{+0.21}_{-0.19}$ & $67^{+36}_{-21} $ & $2.829^{+0.026}_{-0.065}$ & $0.25^{+0.39}_{-0.15}$ \\ 
\hline
\multirow{12}{1pt}{\rotatebox[origin=c]{90}{\large 70$\ $GHz} }&18M & $-1.02^{+0.13}_{-0.15}$ & $15^{+4}_{-3}$ & $4.568^{+0.031}_{-0.026}$ & -  \\ 
&18S & $-1.11^{+0.13}_{-0.13}$ & $19^{+4}_{-4}$ & $4.174^{+0.023} _{-0.031}$& -\\ 
&19M & $-1.13^{+0.13}_{-0.13}$ & $12.2^{+3.8}_{-2.5}$ & $5.198^{+0.012}_{-0.039}$ &  -\\ 
&19S & $-1.06^{+0.12}_{-0.12}$ & $14.2^{+3.4}_{-2.5}$ & $4.960^{+0.019}_{-0.034}$ & - \\ 
&20M & $-1.07^{+0.13}_{-0.14}$ & $8.2^{+3.7}_{-2.9}$ & $5.258^{+0.022}_{-0.027}$ &-  \\
&20S & $-1.11^{+0.13}_{-0.13}$ & $6.1^{+2.9}_{-2.1}$ & $5.563^{+0.020}_{-0.045}$ & -\\ 
&21M & $-1.31^{+0.15}_{-0.12}$ & $39^{+9}_{-7}$ & $4.029^{+0.016}_{-0.015}$ & - \\ 
&21S & $-1.10^{+0.11}_{-0.14}$ & $13.5^{+3.5}_{-2.6}$ & $5.016^{+0.026}_{-0.023}$ & - \\ 
&22M & $-1.26^{+0.14}_{-0.16}$ & $11^{+8}_{-4}$ & $4.377^{+0.020}_{-0.020}$ & - \\ 
&22S & $-1.15^{+0.15}_{-0.22}$ & $14^{+10}_{-5}$ & $4.745^{+0.022}_{-0.025}$ & - \\ 
&23M & $-1.03^{+0.09}_{-0.11}$ & $32^{+5}_{-4}$ & $4.494^{+0.020}_{-0.021}$ &  -\\ 
& 23S & $-1.19^{+0.08}_{-0.07}$ & $60^{+7}_{-5}$ & $4.813^{+0.020}_{-0.028}$ & - \\
\hline
		\end{tabular}
	\end{table}
}

\begin{figure*}[p]
  \begin{center}
    \includegraphics[width=0.495\linewidth]{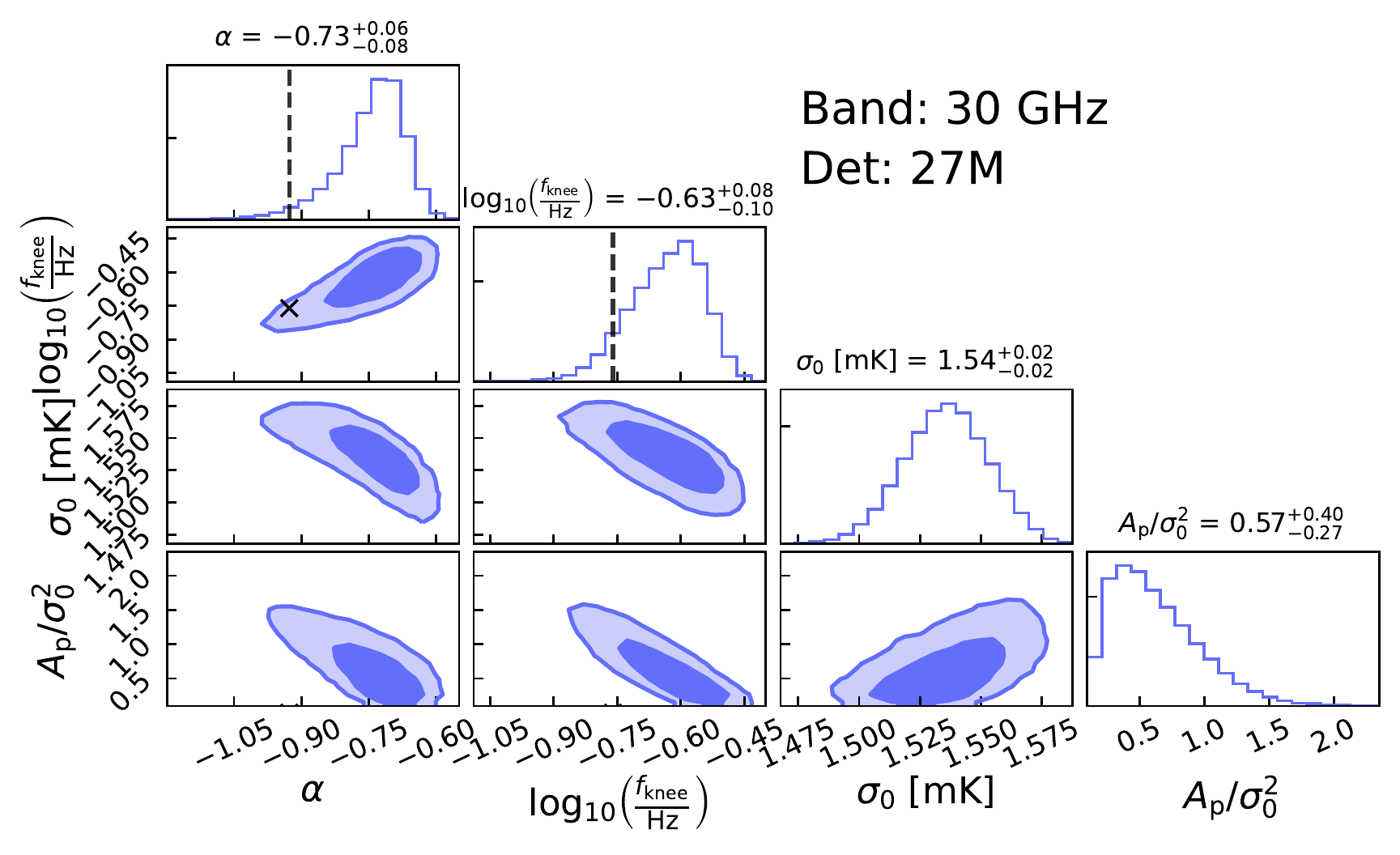}
    \includegraphics[width=0.495\linewidth]{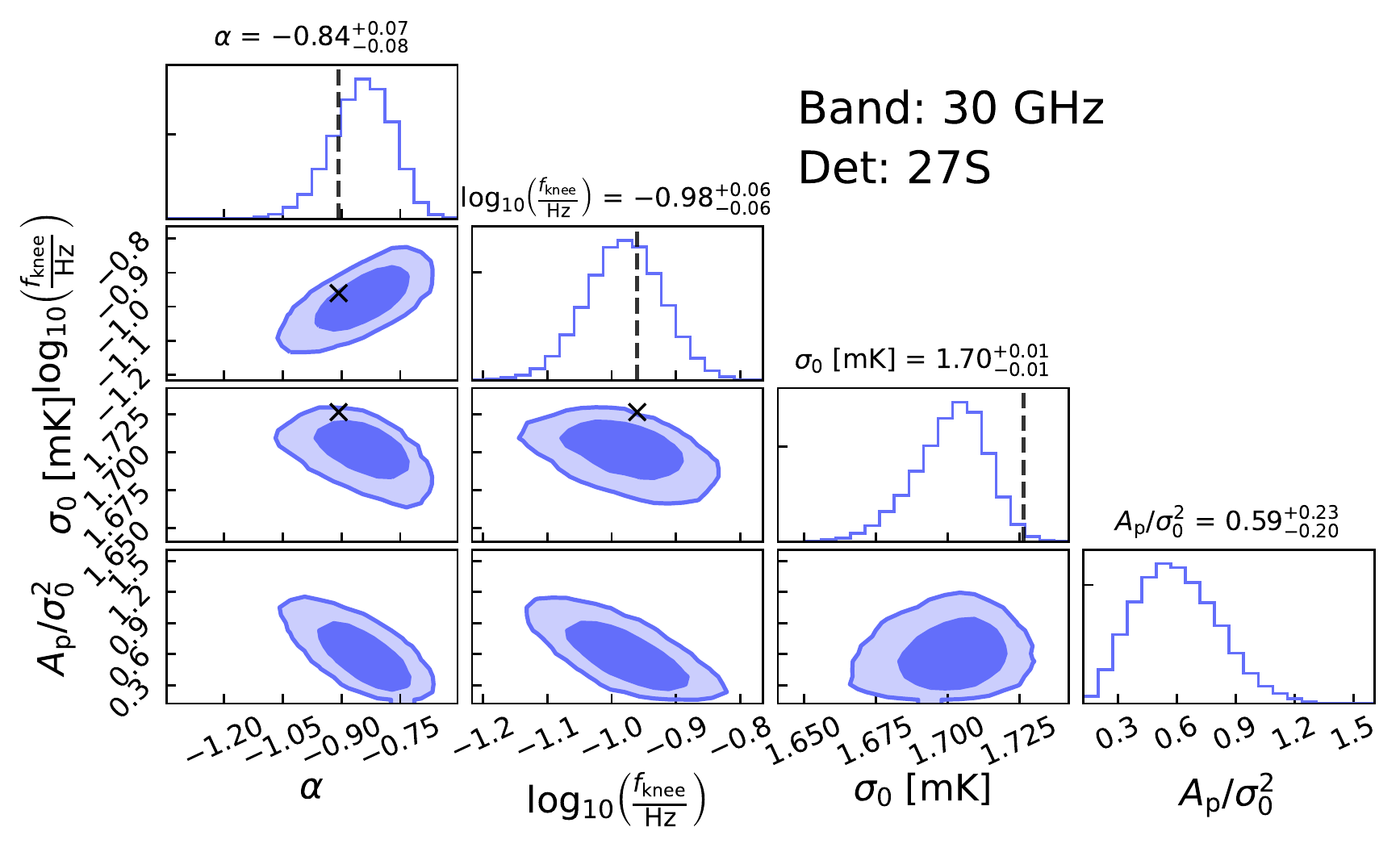}\\
    \includegraphics[width=0.495\linewidth]{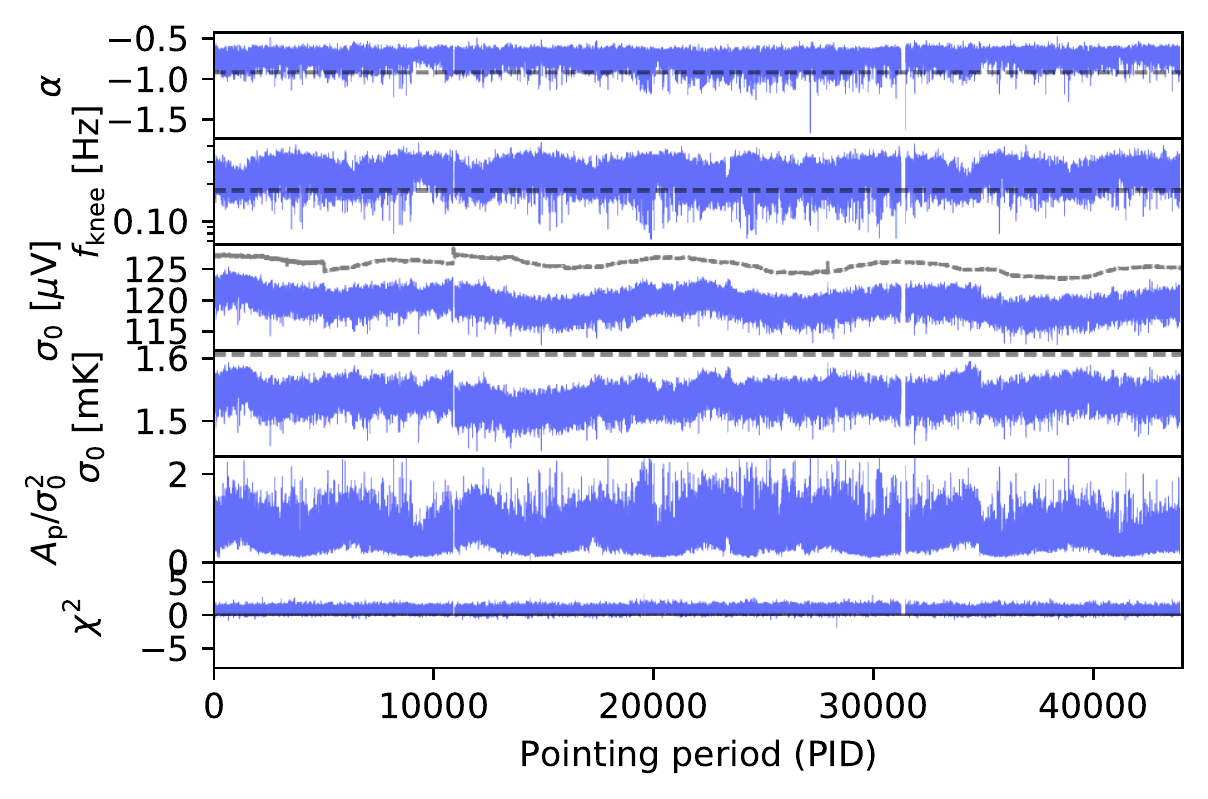}
    \includegraphics[width=0.495\linewidth]{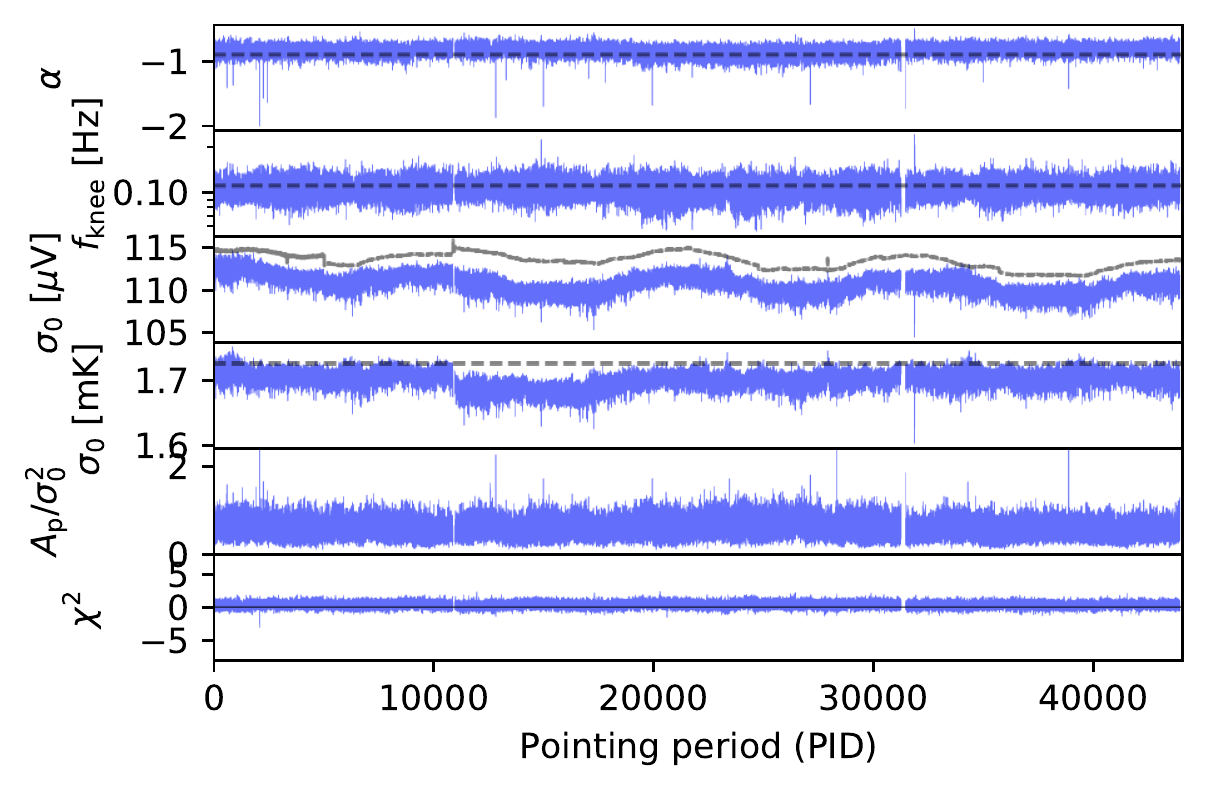}\\
    \vspace*{1mm}
    \includegraphics[width=0.495\linewidth]{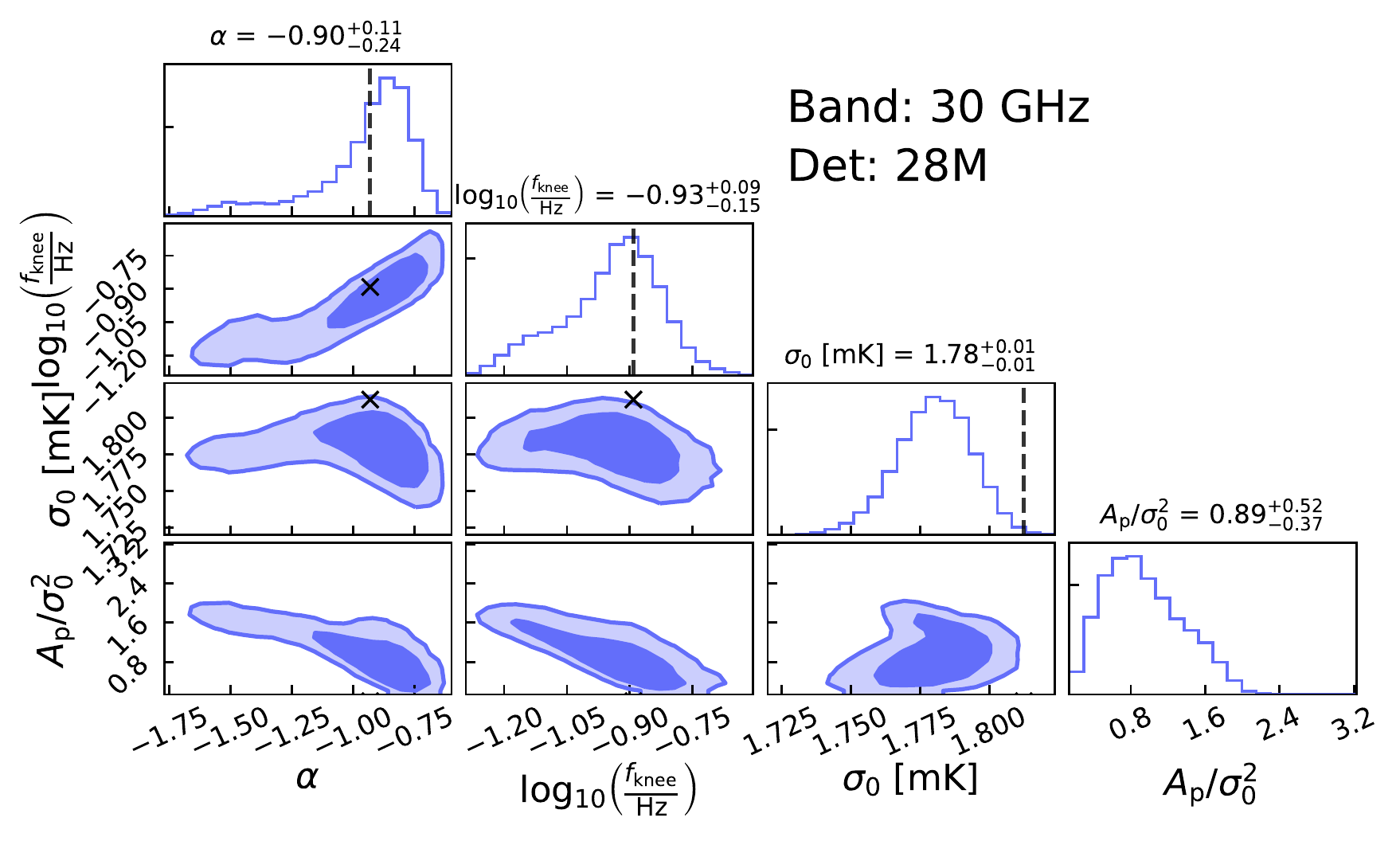}
    \includegraphics[width=0.495\linewidth]{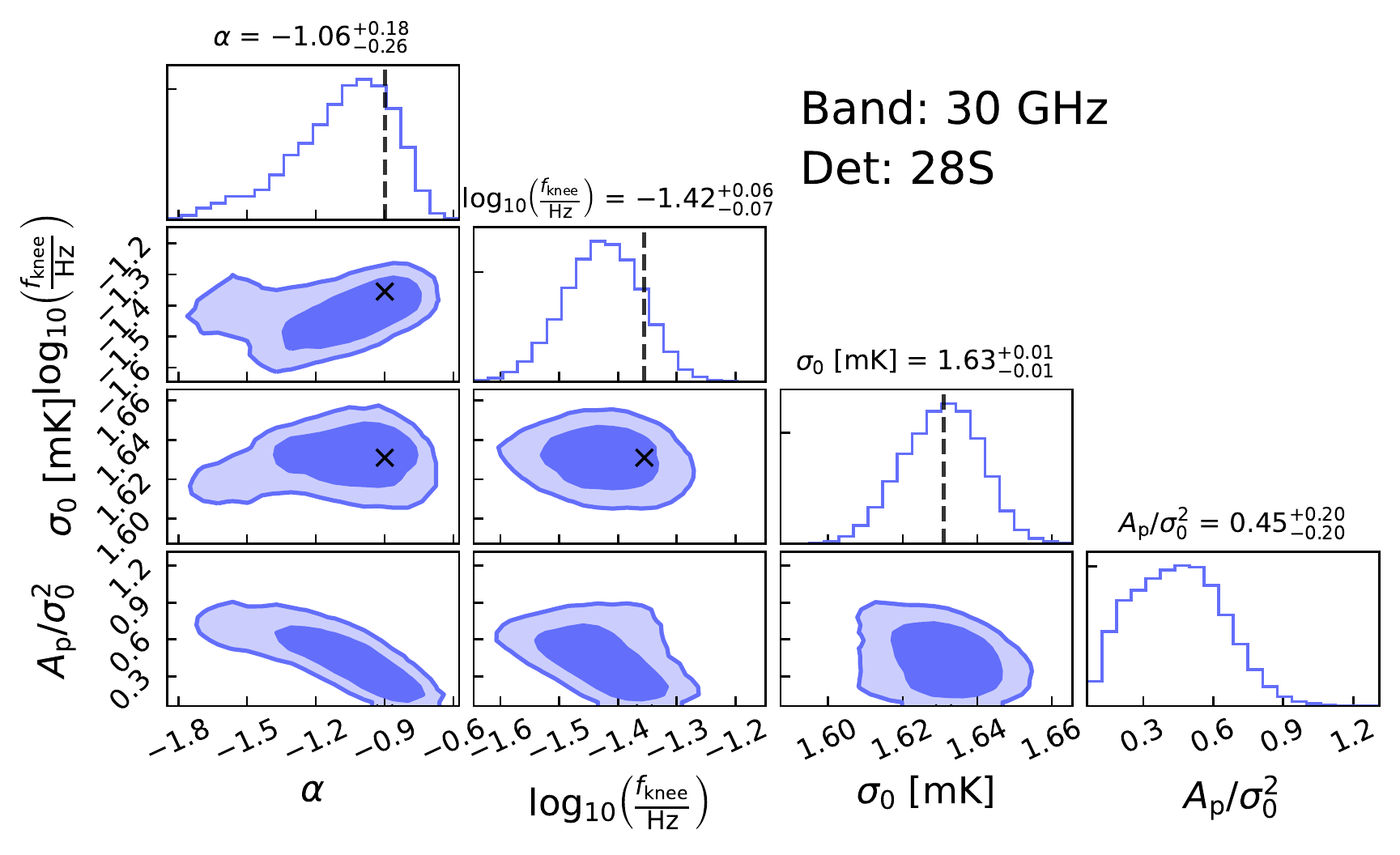}\\
    \includegraphics[width=0.495\linewidth]{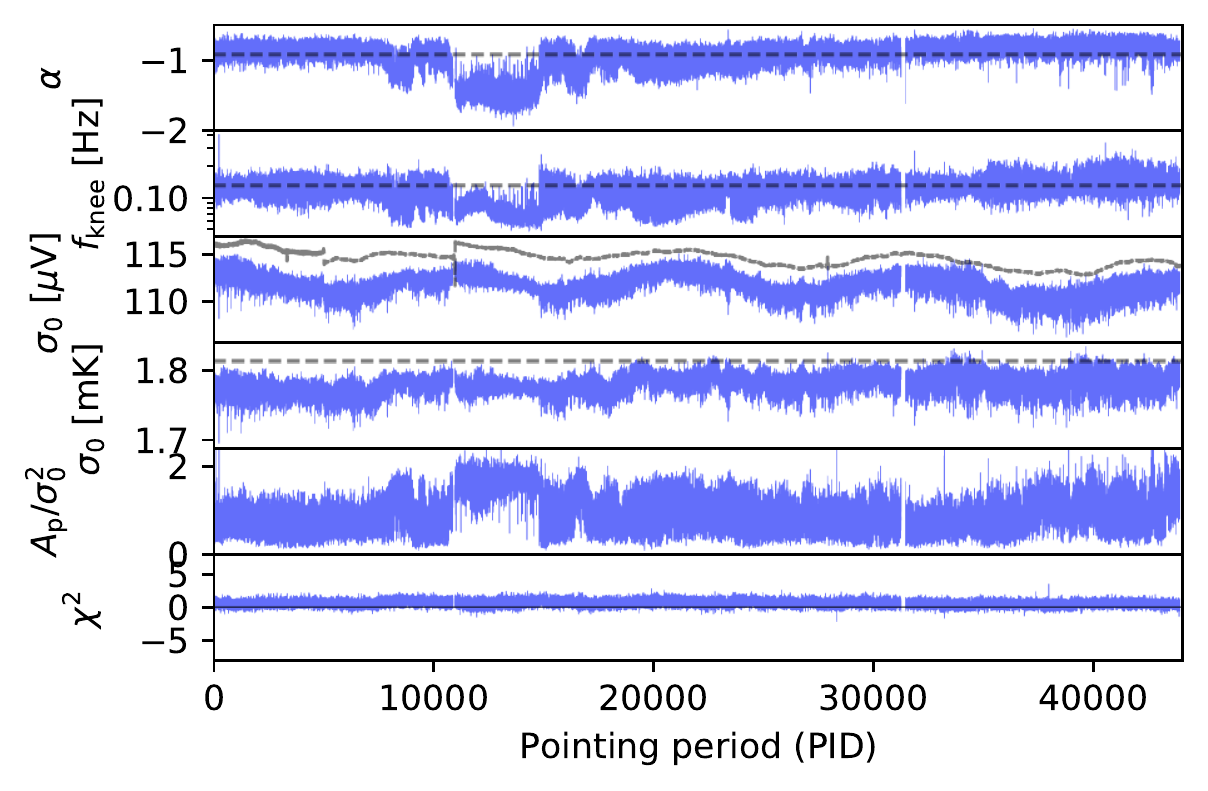}
    \includegraphics[width=0.495\linewidth]{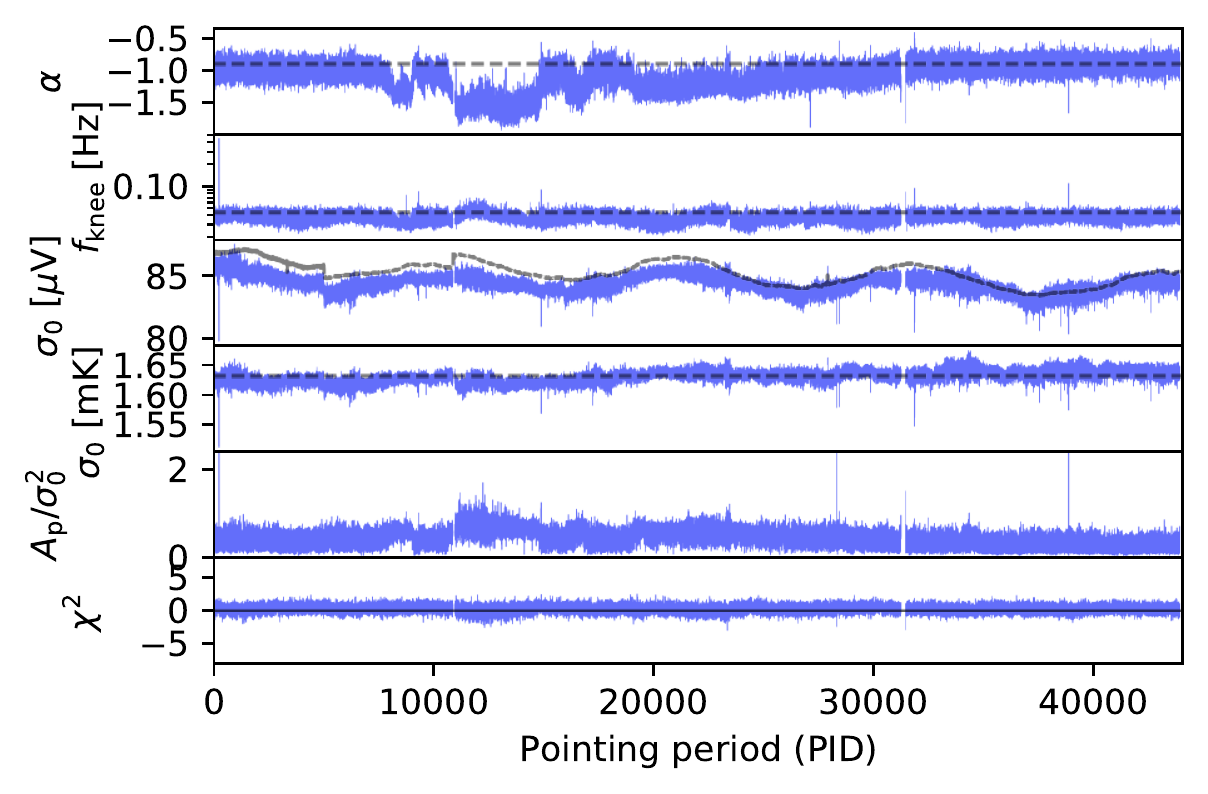}
     \vspace*{-5.5mm}    
  \end{center}
  
  \caption{Noise characterization of the \Planck\ LFI 30~GHz
    radiometers; 27M (\emph{top left}), 27S (\emph{top right}); 28M
    (\emph{bottom left}), and 28S (\emph{bottom right}). For each
    radiometer, the top figure shows distributions of noise parameters
    PSD, $\xi^n = \{\sigma_0, f_\mathrm{knee}, \alpha, A_\mathrm{p}\}$, averaged
    over all Gibbs samples for the full mission. The bottom figure
    shows the time evolution of the posterior mean of the noise
    parameters, and the bottom panel shows the evolution in reduced
    normalized $\chi^2$ in units of $\sigma$. Black dashed curves and crosses show corresponding values as derived by, and used in, the
    official \Planck\ LFI DPC pipeline.
    \label{fig:xi_prop_30}}
\end{figure*}

\begin{figure*}[p]
	\begin{center}
		\includegraphics[width=0.495\linewidth]{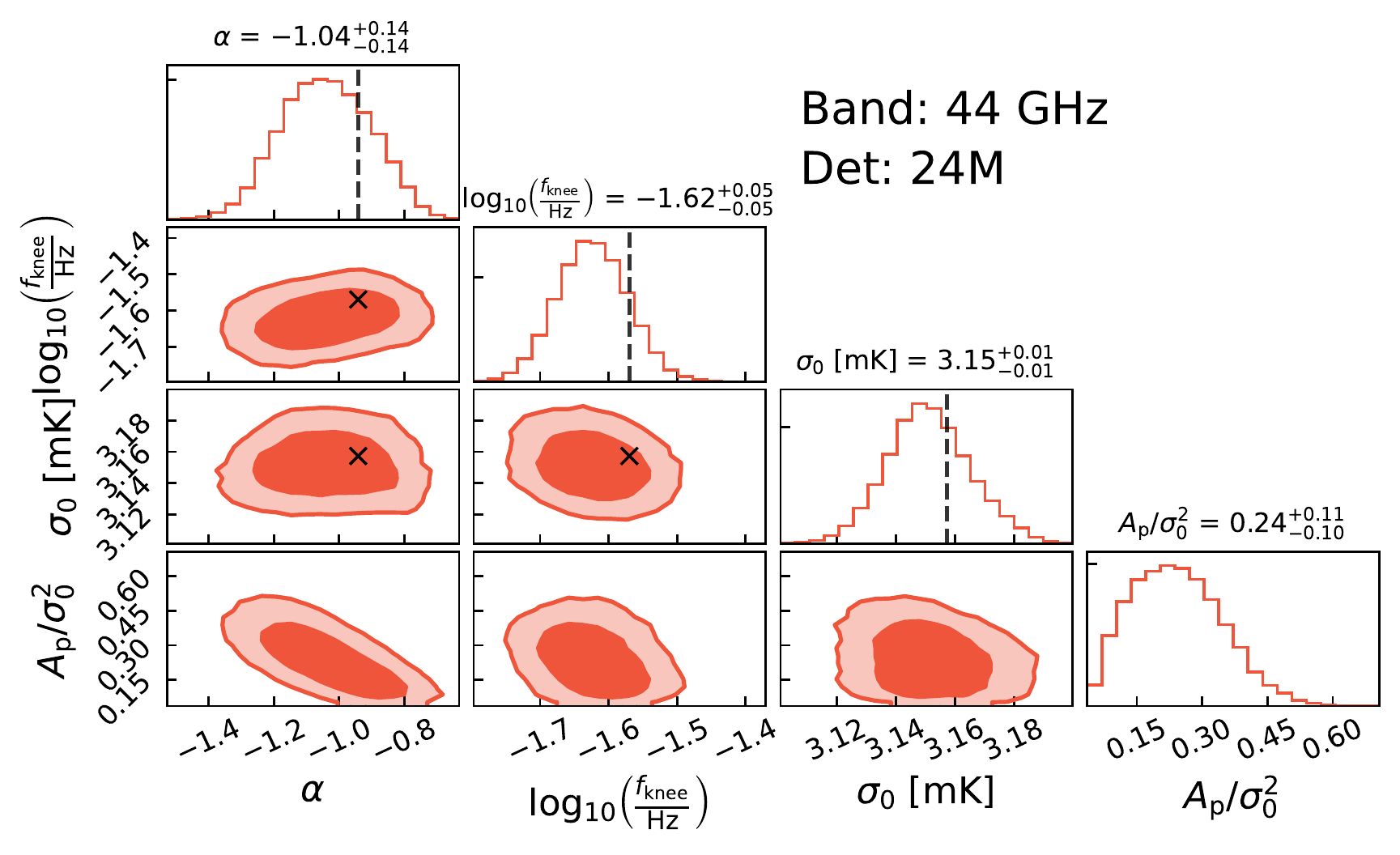}
		\includegraphics[width=0.495\linewidth]{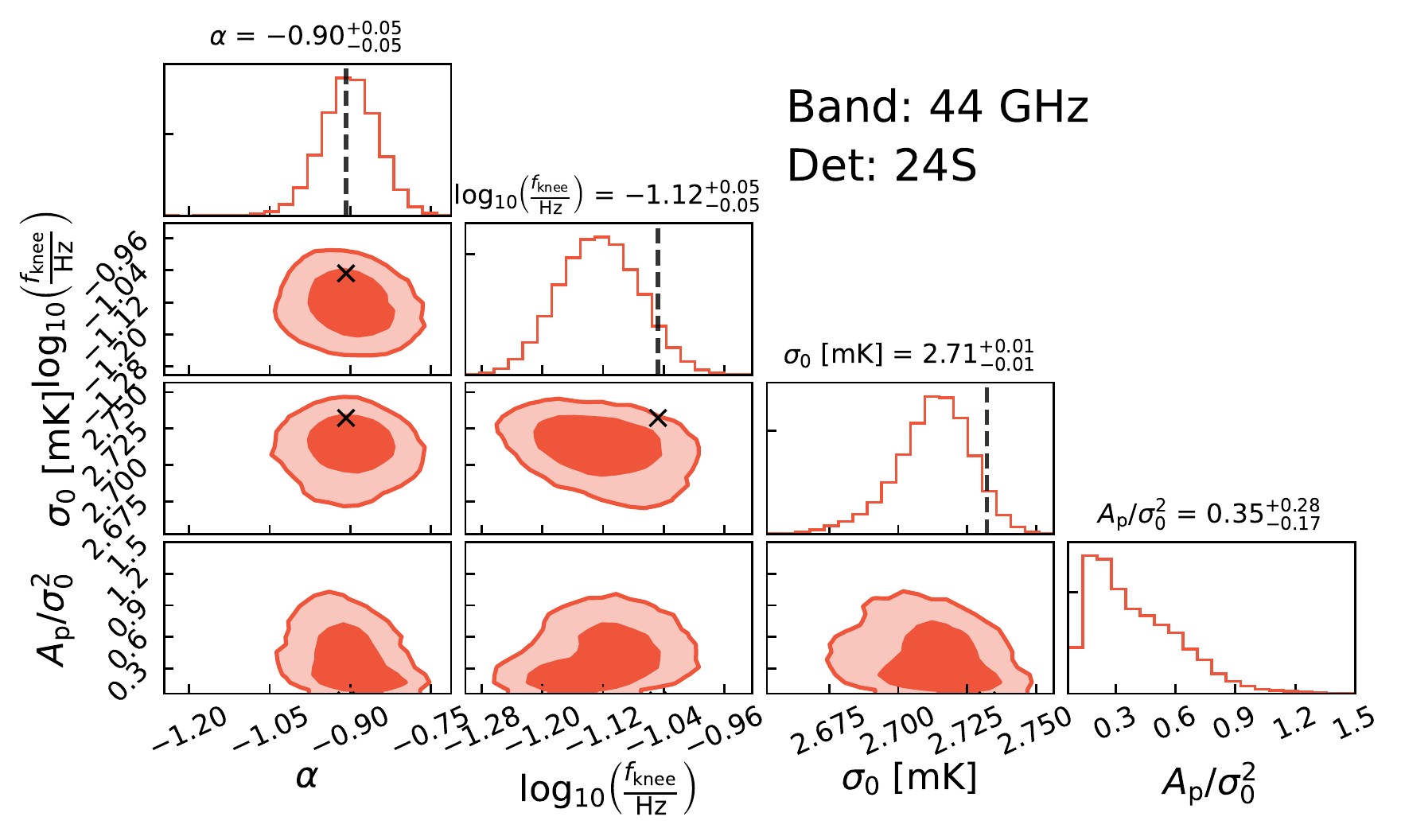}\\
		\includegraphics[width=0.495\linewidth]{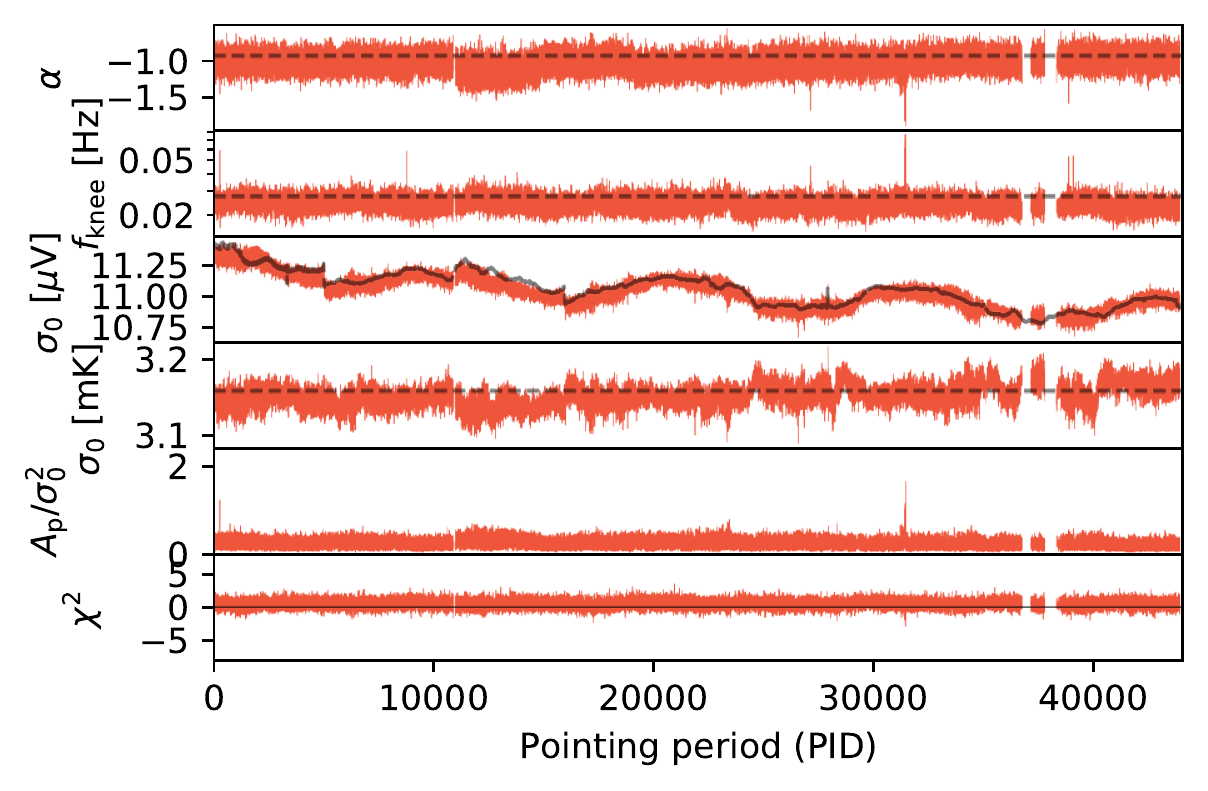}
		\includegraphics[width=0.495\linewidth]{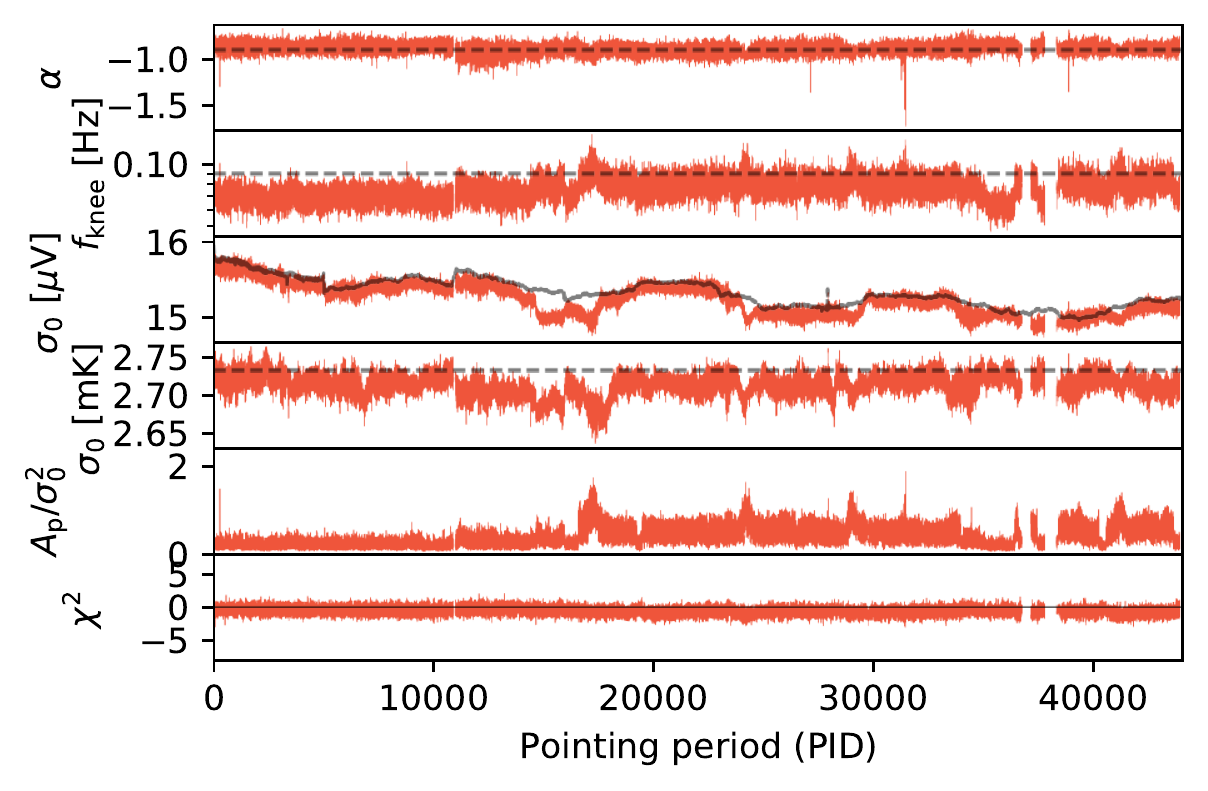}\\
		\vspace*{0.5mm}
		\includegraphics[width=0.495\linewidth]{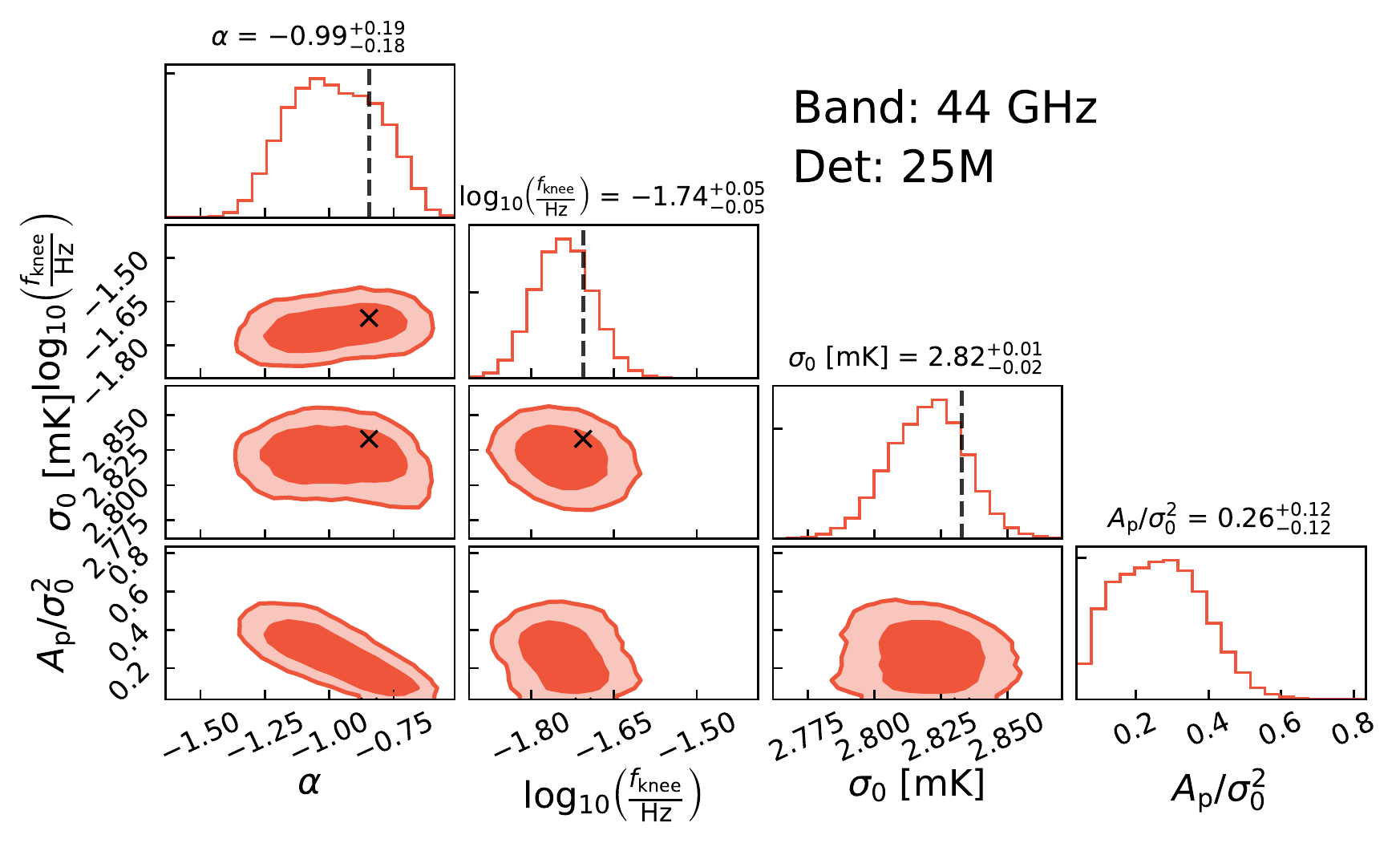}
		\includegraphics[width=0.495\linewidth]{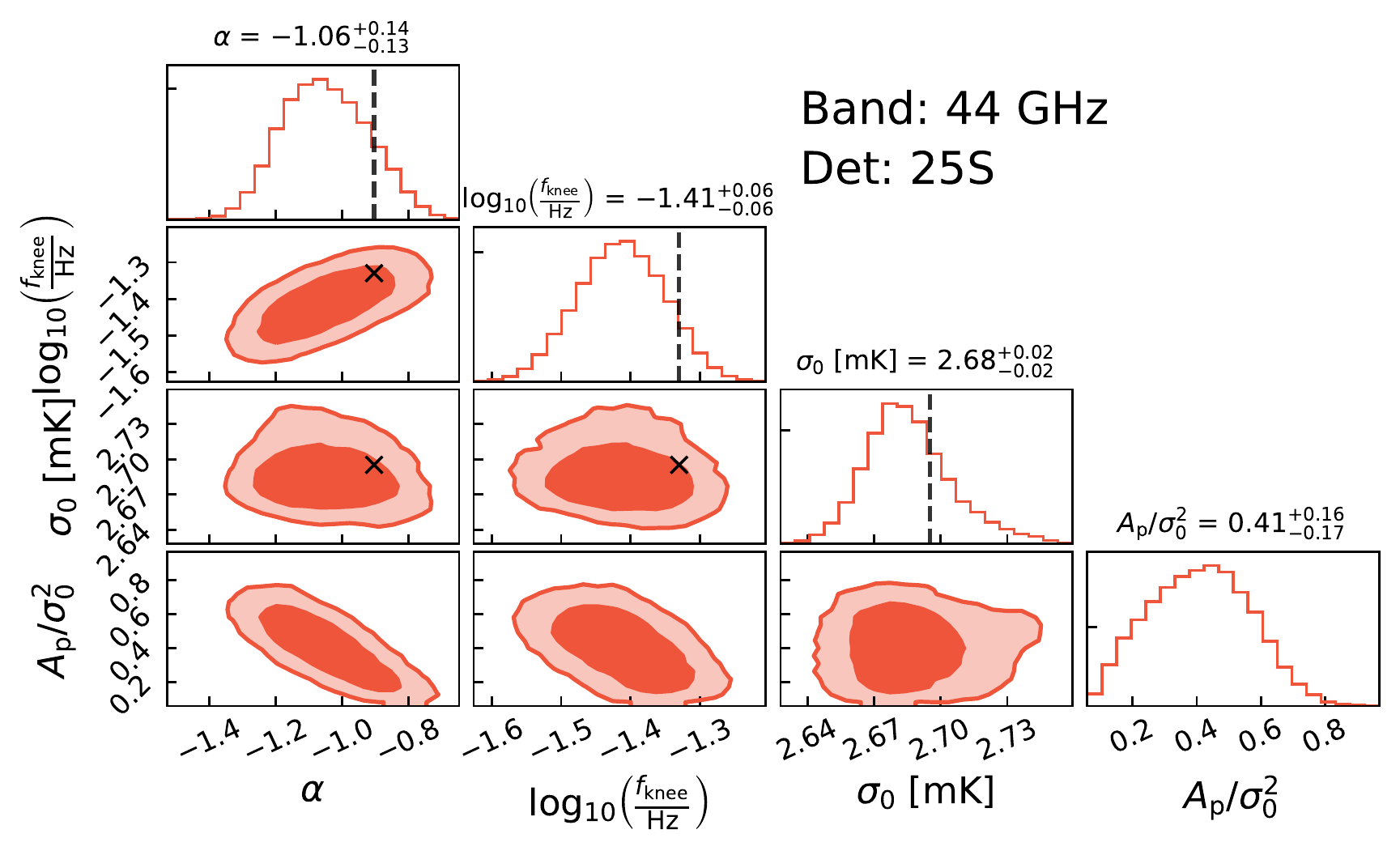}\\
		\includegraphics[width=0.495\linewidth]{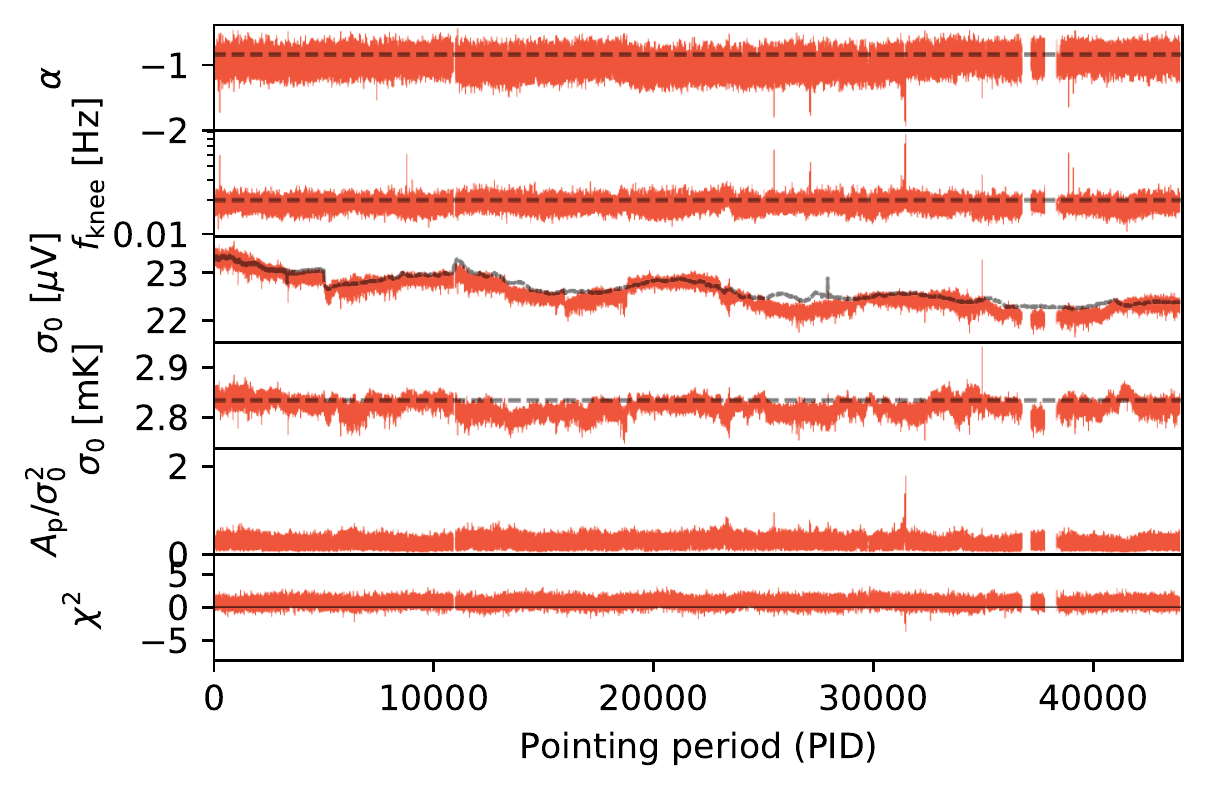}
		\includegraphics[width=0.495\linewidth]{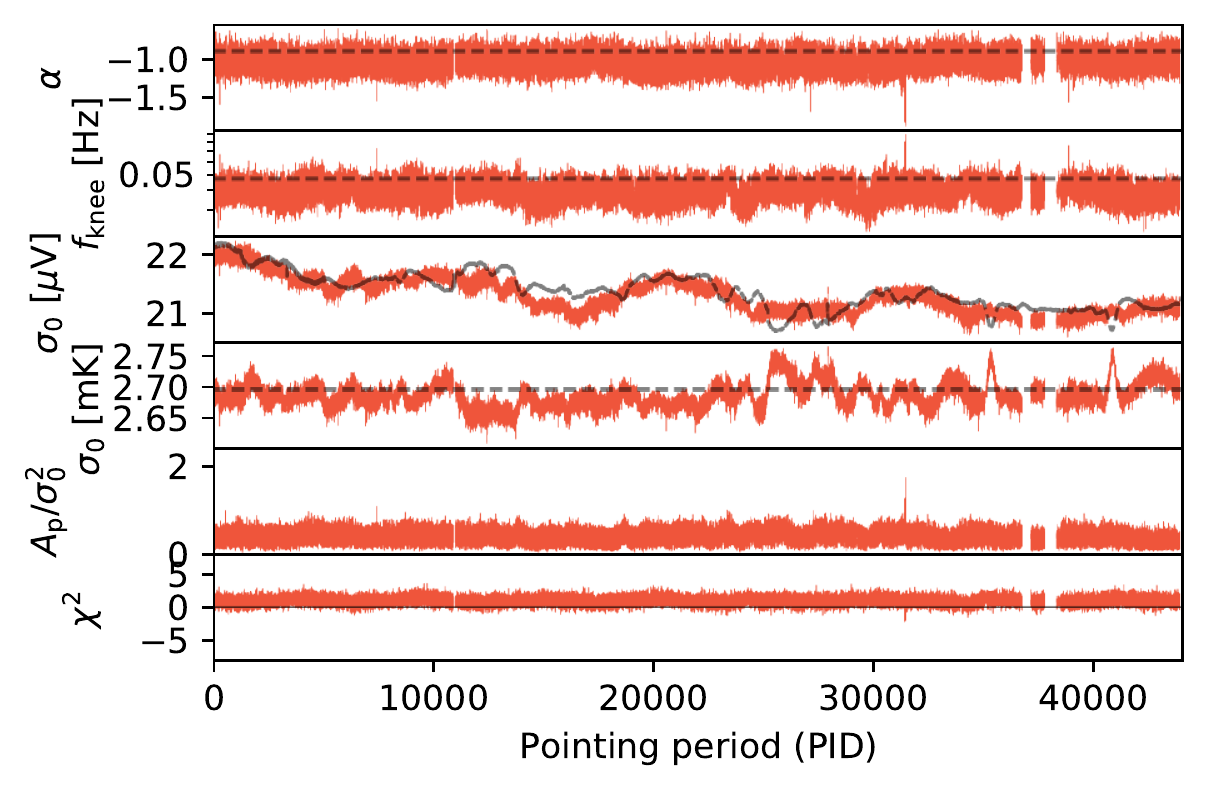}
		 \vspace*{-5.8mm}            
	\end{center}
	
	\caption{Noise characterization of the \Planck\ LFI 44\,GHz
		radiometers; 24M (\emph{top left}), 24S (\emph{top right}); 25M
		(\emph{bottom left}), and 25S (\emph{bottom right}). For each
		radiometer, the top figure shows distributions of noise parameters
		PSD, $\xi^n = \{\sigma_0, f_\mathrm{knee}, \alpha, A_\mathrm{p}\}$, averaged
		over all Gibbs samples for the full mission. The bottom figure
		shows the time evolution of the posterior mean of the noise
		parameters, and the bottom panel shows the evolution in reduced
		normalized $\chi^2$ in units of $\sigma$. Black dashed curves and crosses show corresponding values as derived by, and used in, the
		official \Planck\ LFI DPC pipeline.
		\label{fig:xi_prop_44}}
\end{figure*}

\begin{figure*}[t]
	\begin{center}
		\includegraphics[width=0.495\linewidth]{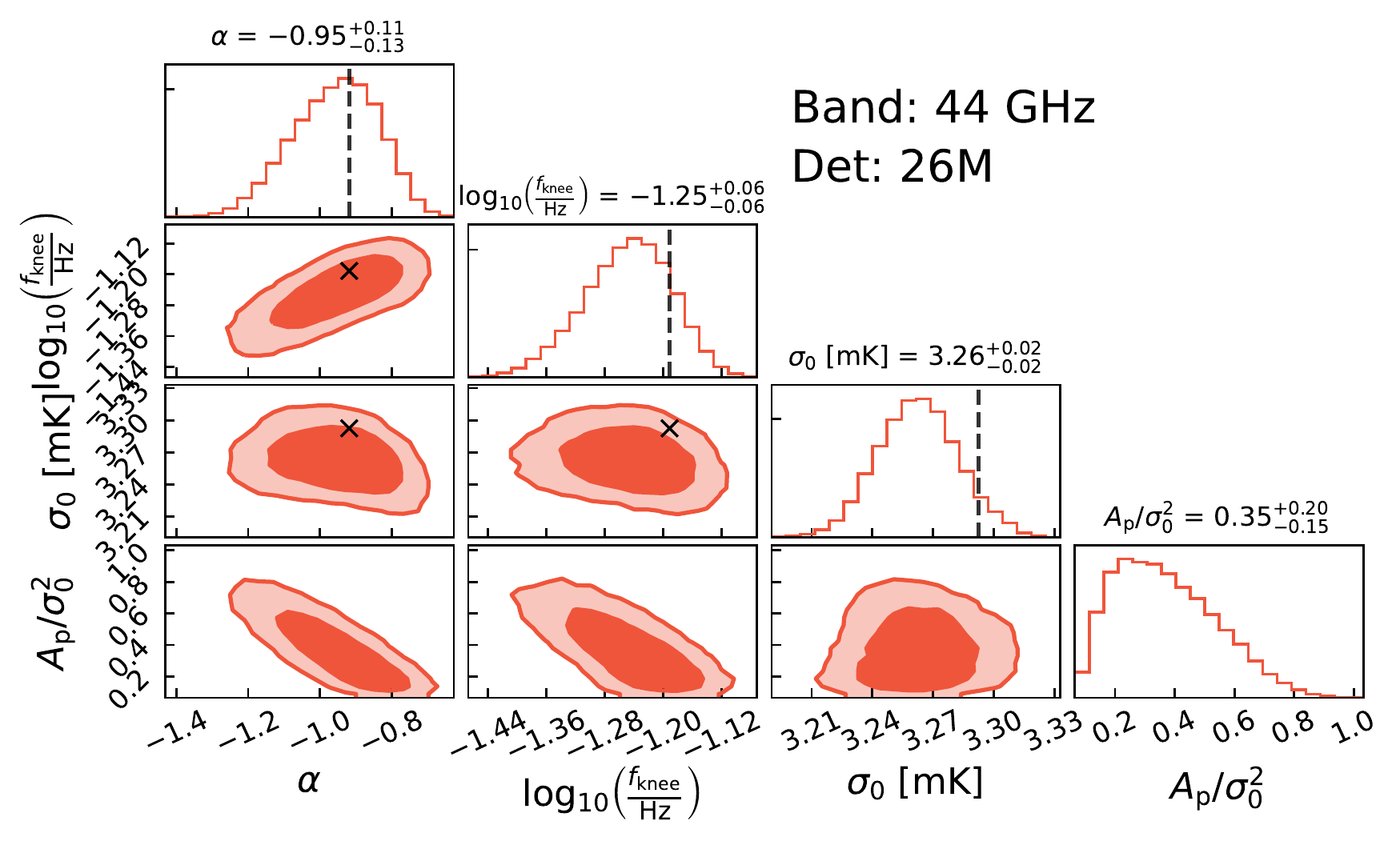}
		\includegraphics[width=0.495\linewidth]{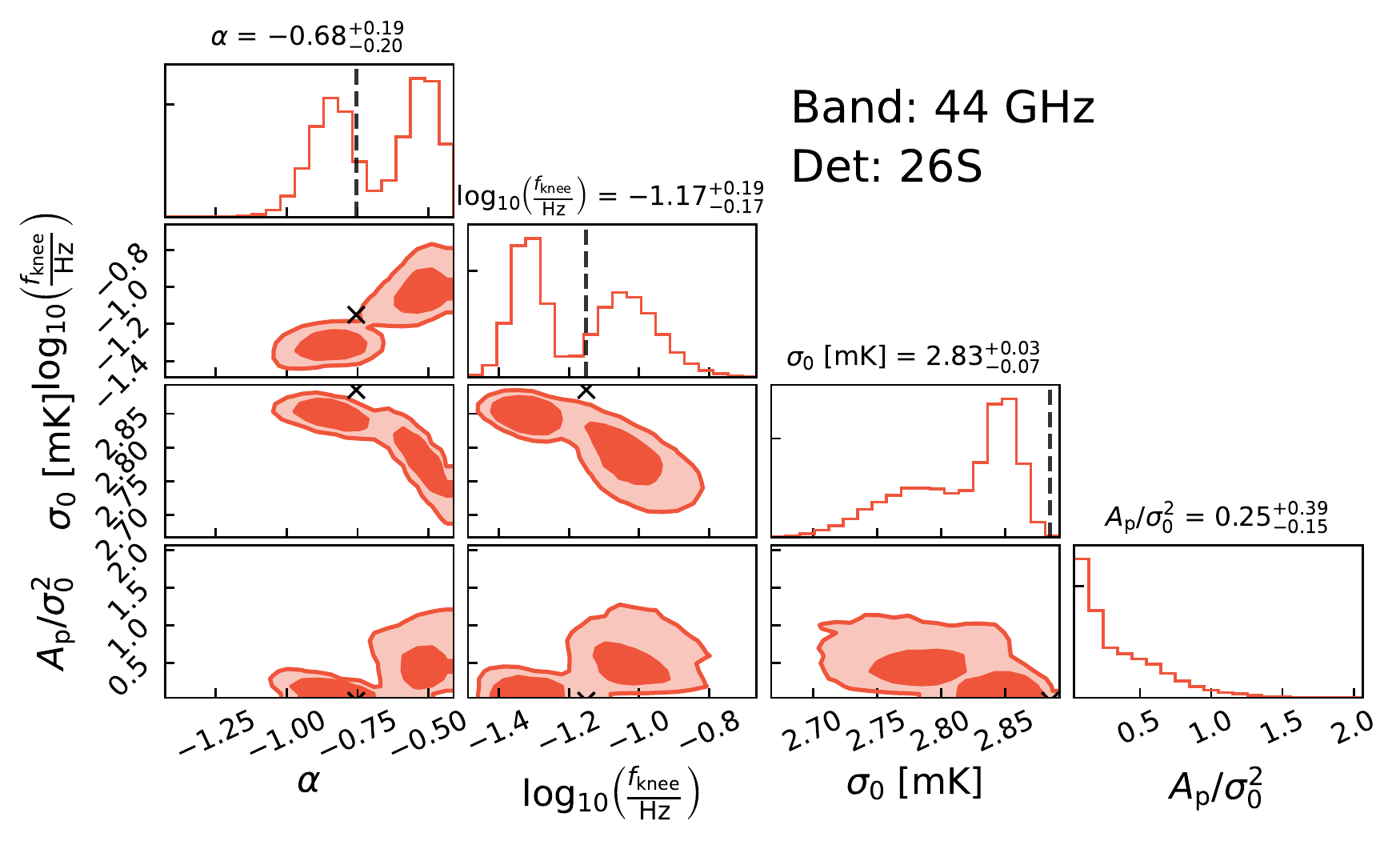}\\
		\includegraphics[width=0.495\linewidth]{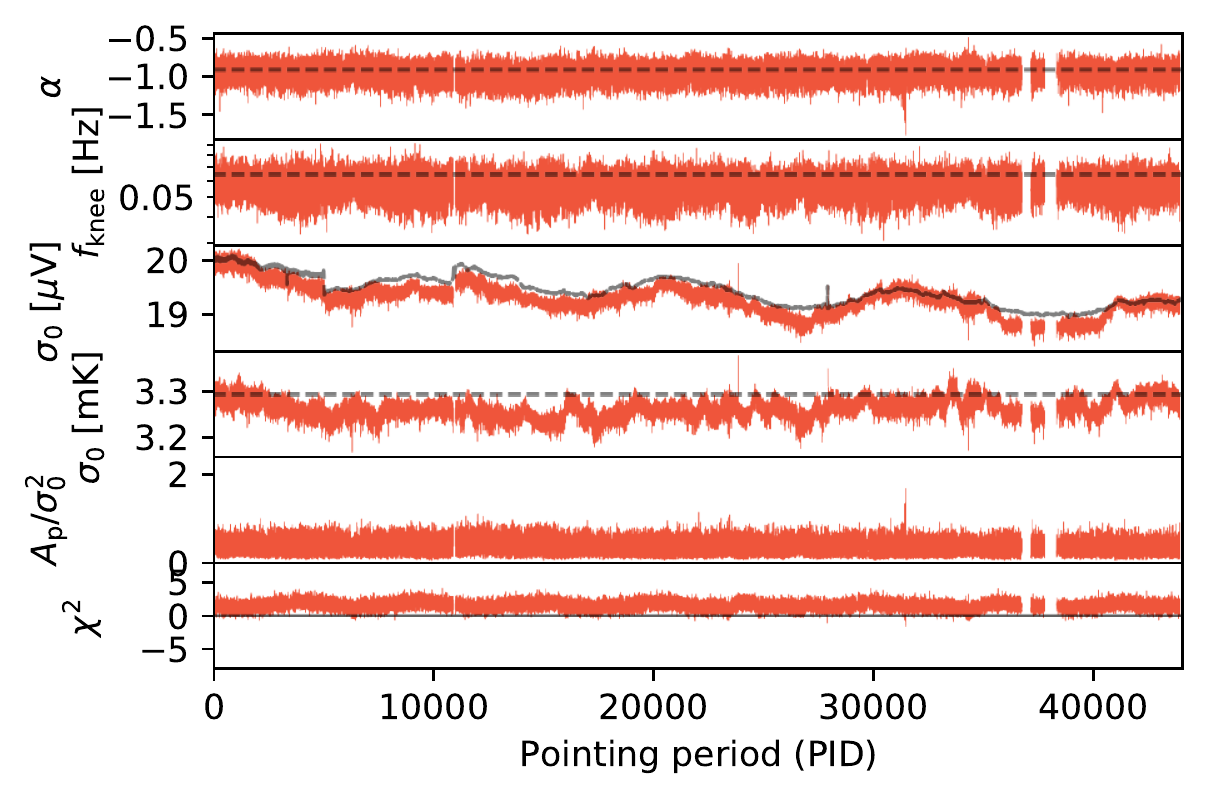}
		\includegraphics[width=0.495\linewidth]{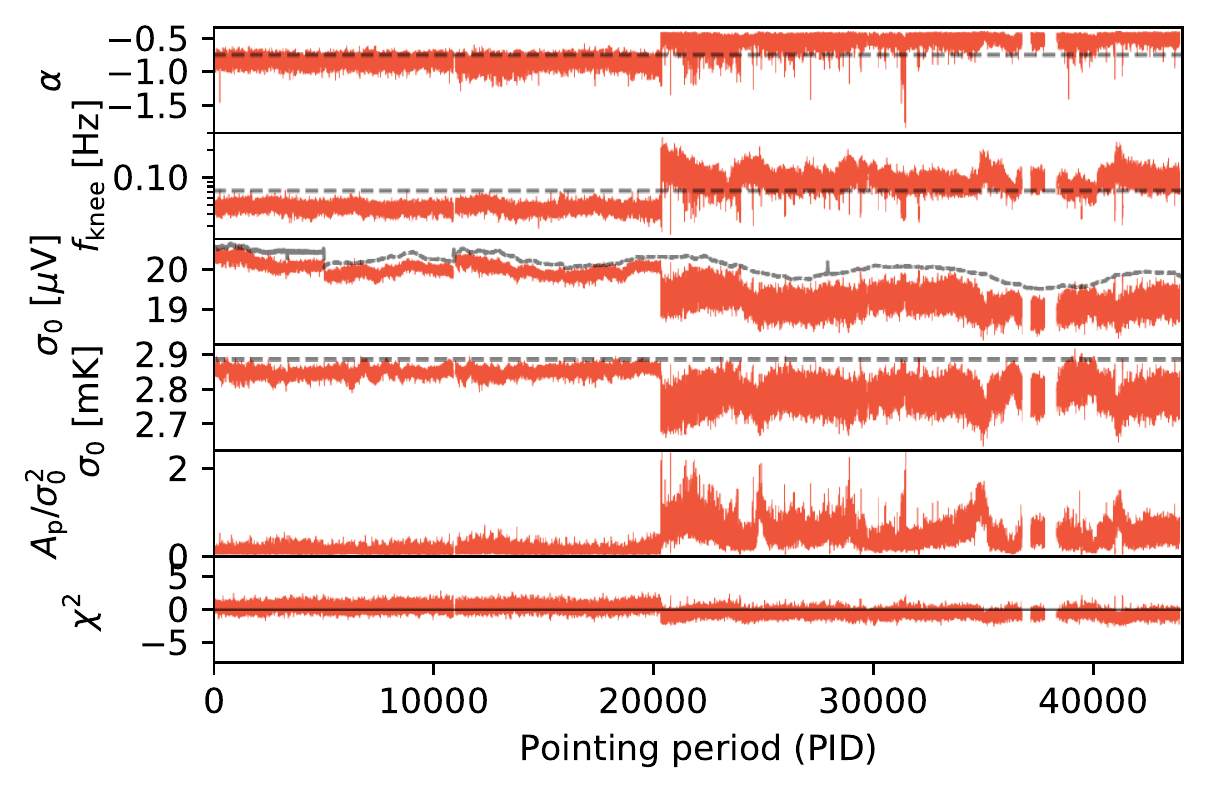}\\    
		\vspace*{-4.5mm} 
	\end{center}
	
	\caption{Noise characterization of the \Planck\ LFI 44\,GHz
		radiometers; 26M (\emph{left}), 26S (\emph{right}). For each
		radiometer, the top figure shows distributions of noise parameters
		PSD, $\xi^n = \{\sigma_0, f_\mathrm{knee}, \alpha, A_\mathrm{p}\}$, averaged
		over all Gibbs samples for the full mission. The bottom figure
		shows the time evolution of the posterior mean of the noise
		parameters, and the bottom panel shows the evolution in reduced
		normalized $\chi^2$ in units of $\sigma$. Black dashed curves and crosses show corresponding values as derived by, and used in, the
		official \Planck\ LFI DPC pipeline.
		\label{fig:xi_prop_44_2}}
\end{figure*}

\begin{figure*}[p]
	\begin{center}
		\includegraphics[width=0.495\linewidth]{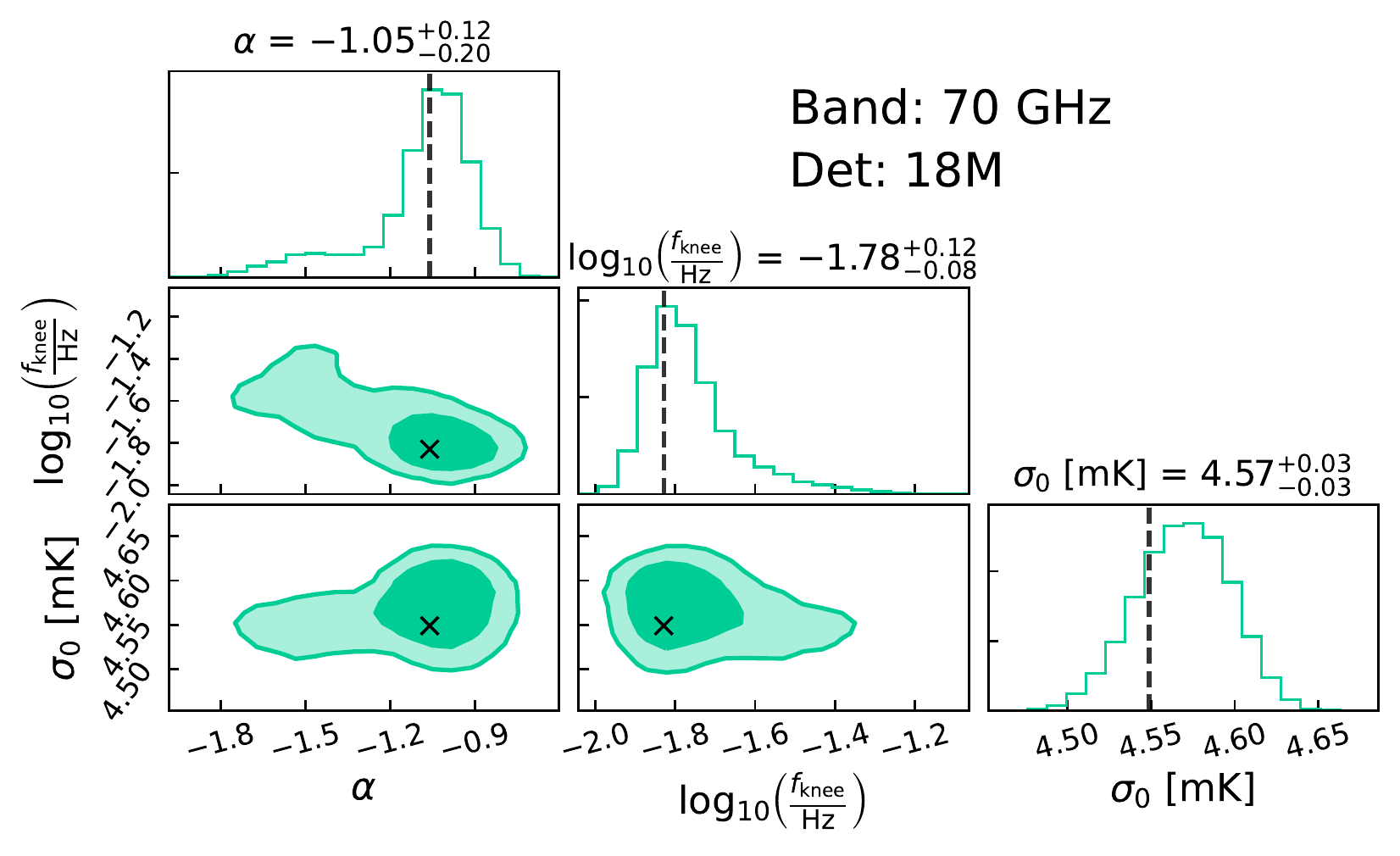}
		\includegraphics[width=0.495\linewidth]{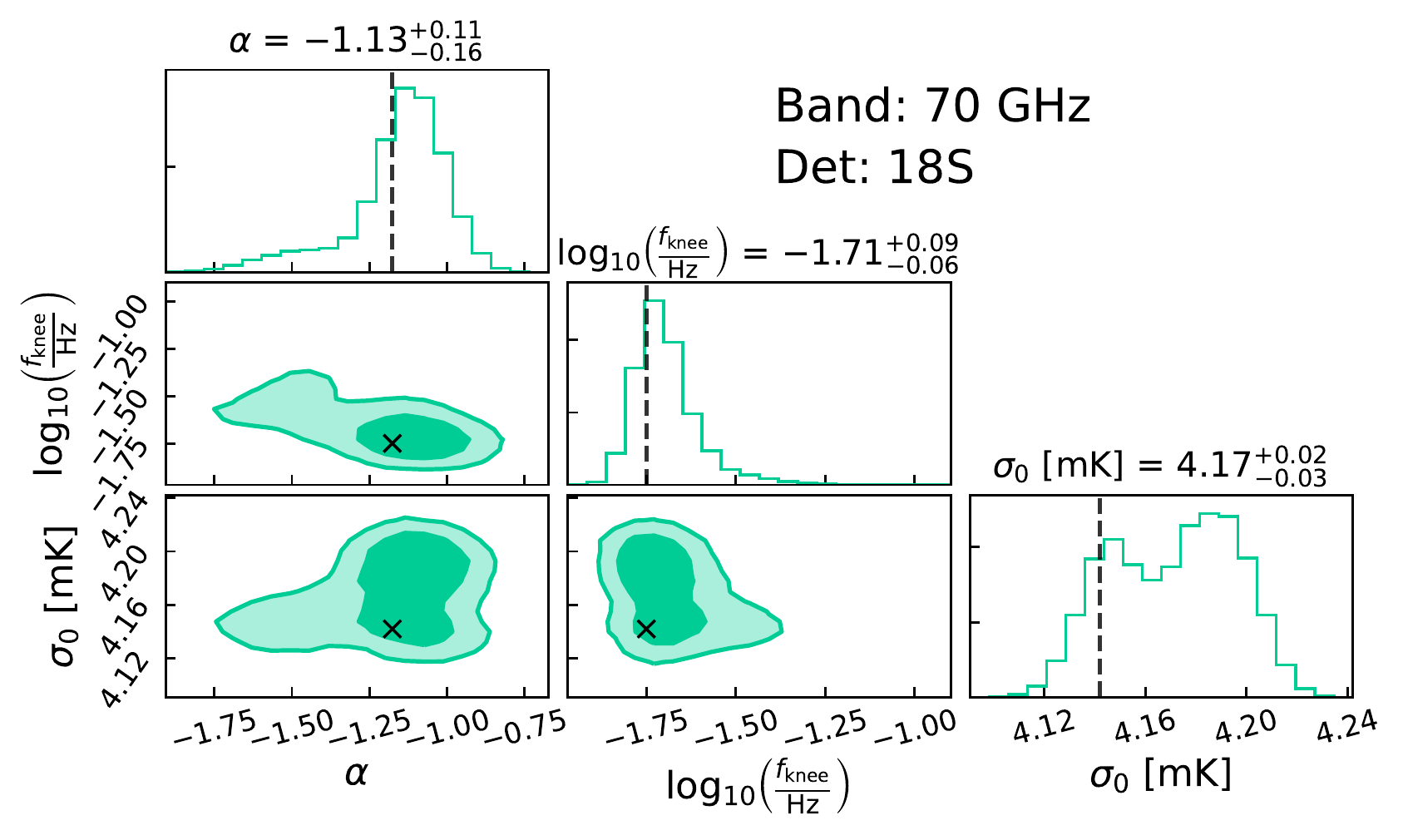}\\
		\includegraphics[width=0.495\linewidth]{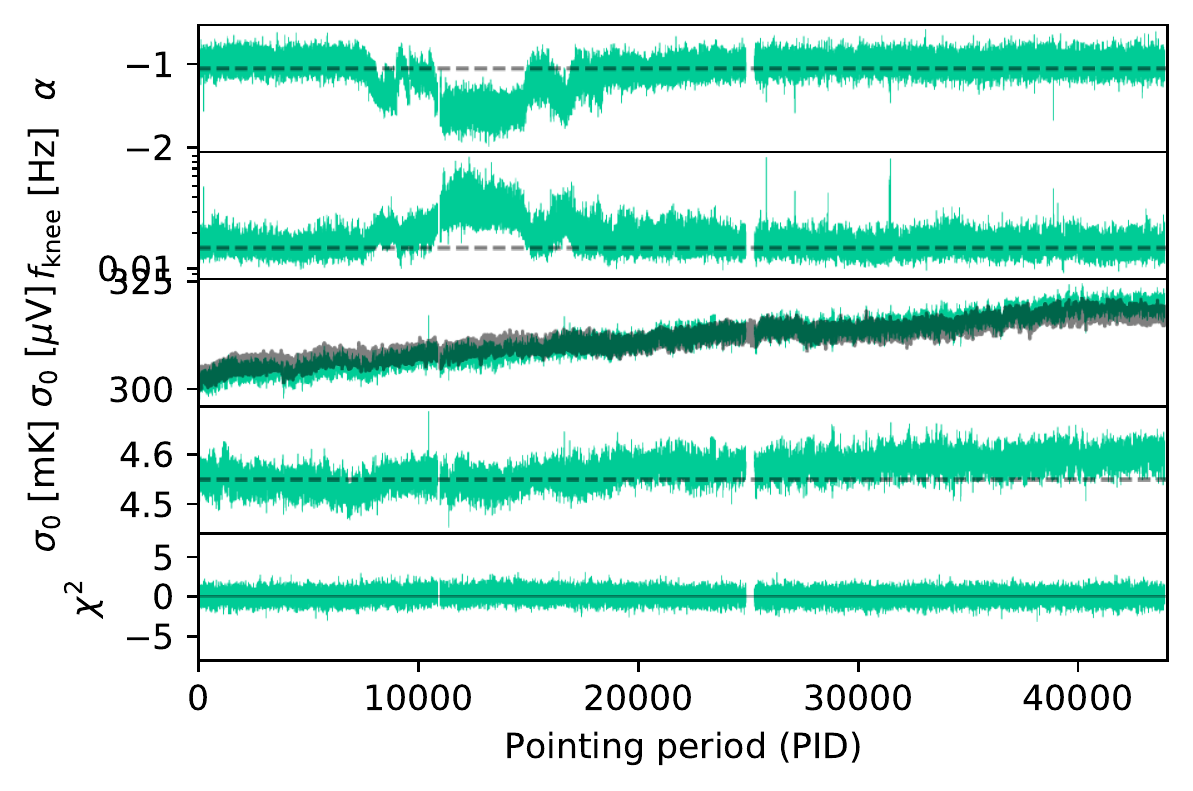}
		\includegraphics[width=0.495\linewidth]{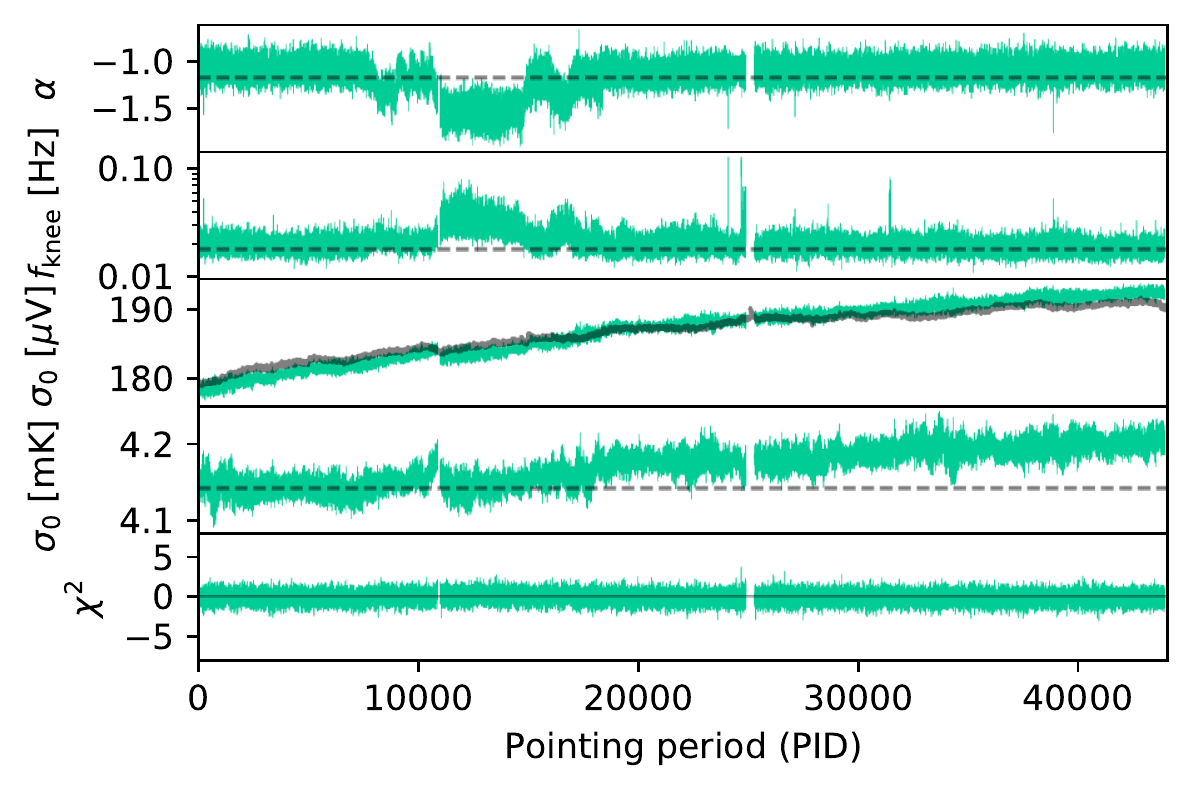}\\
		\vspace*{1mm}
		\includegraphics[width=0.495\linewidth]{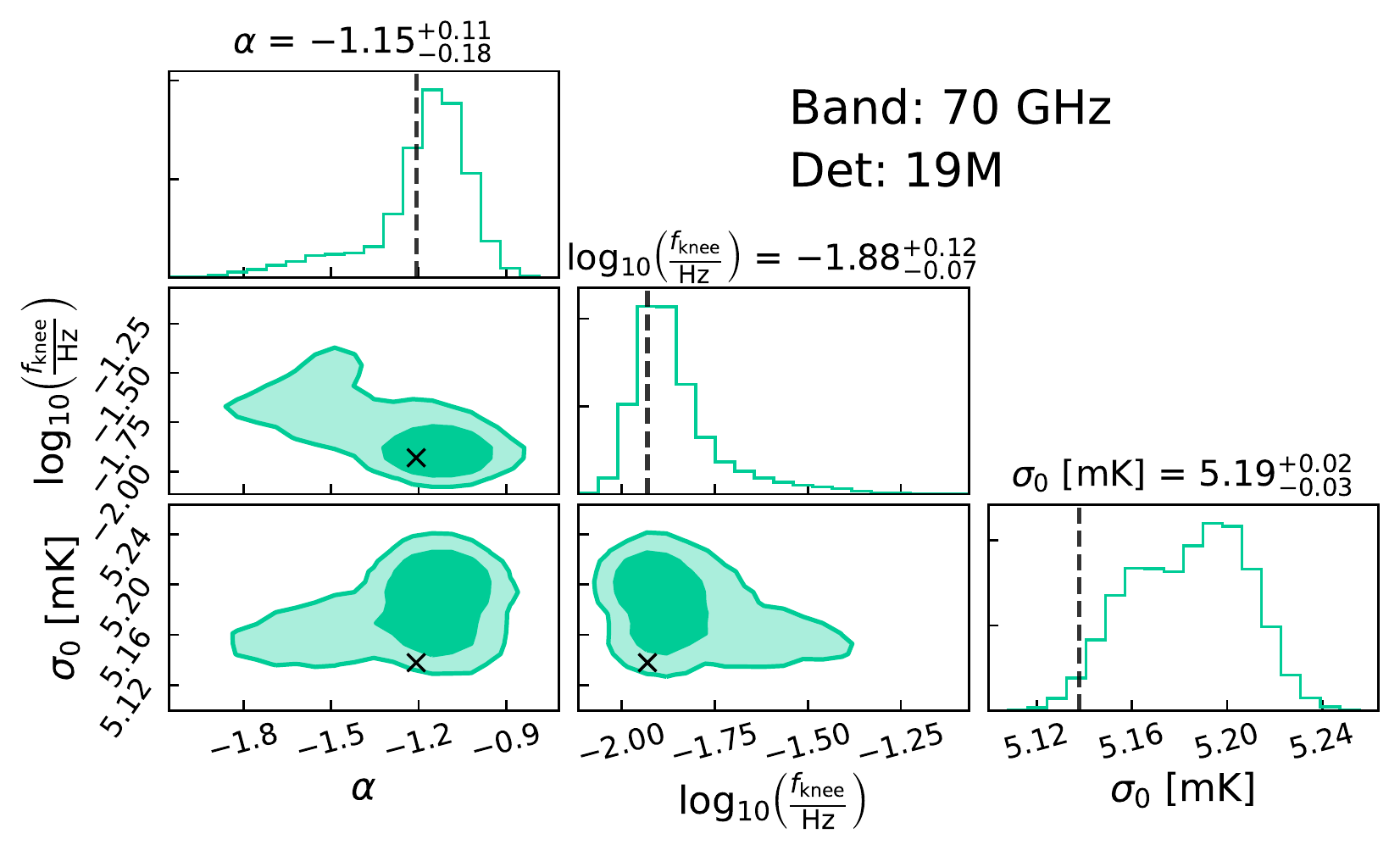}
		\includegraphics[width=0.495\linewidth]{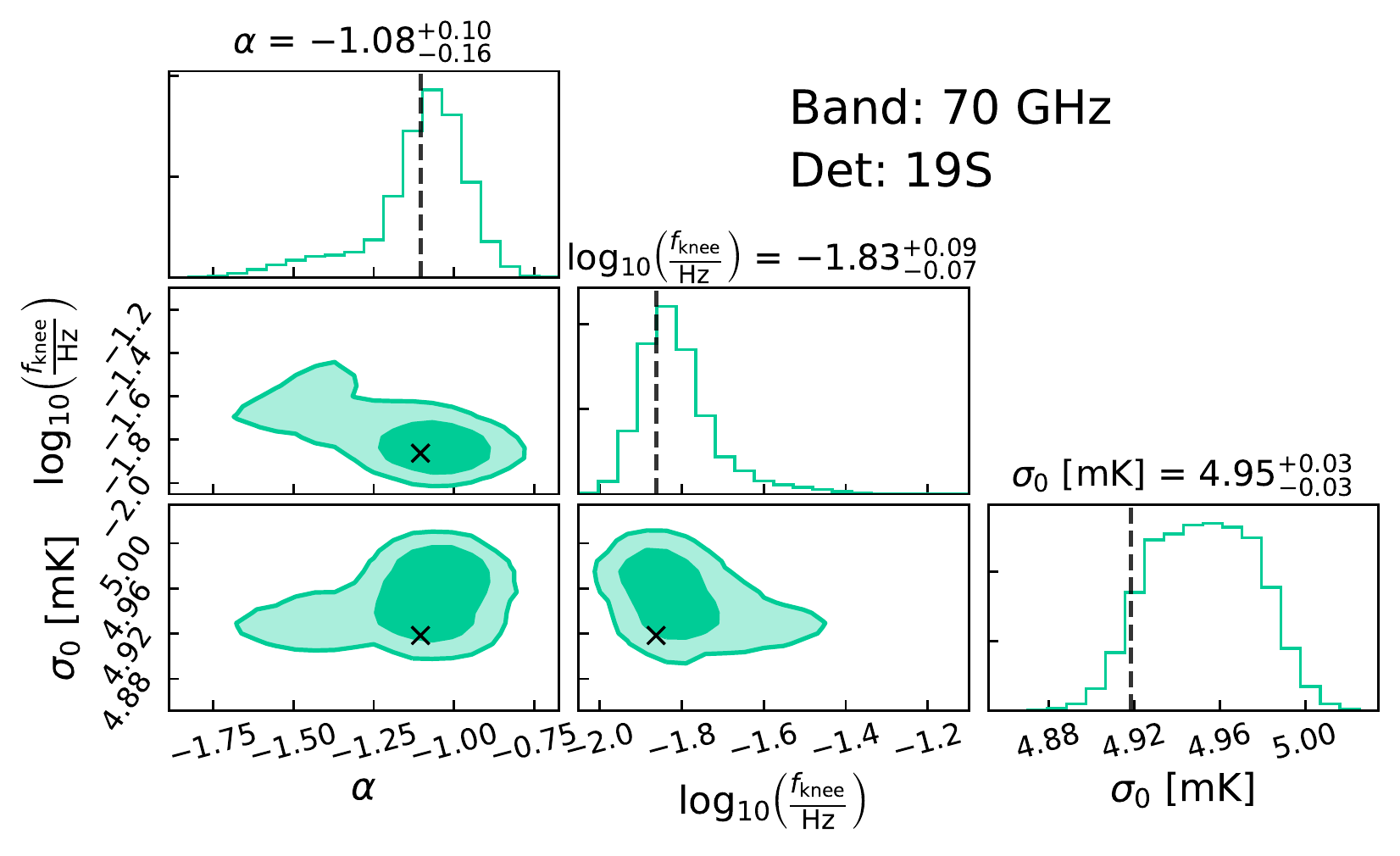}\\
		\includegraphics[width=0.495\linewidth]{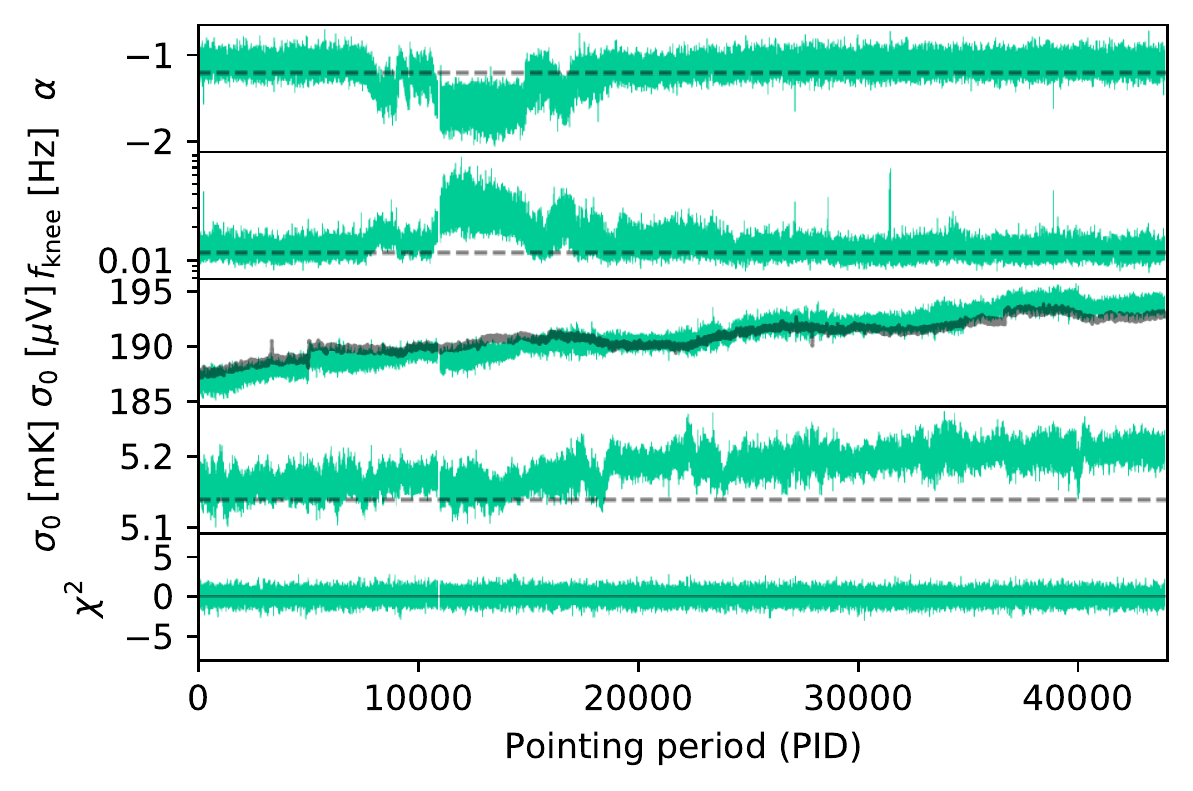}
		\includegraphics[width=0.495\linewidth]{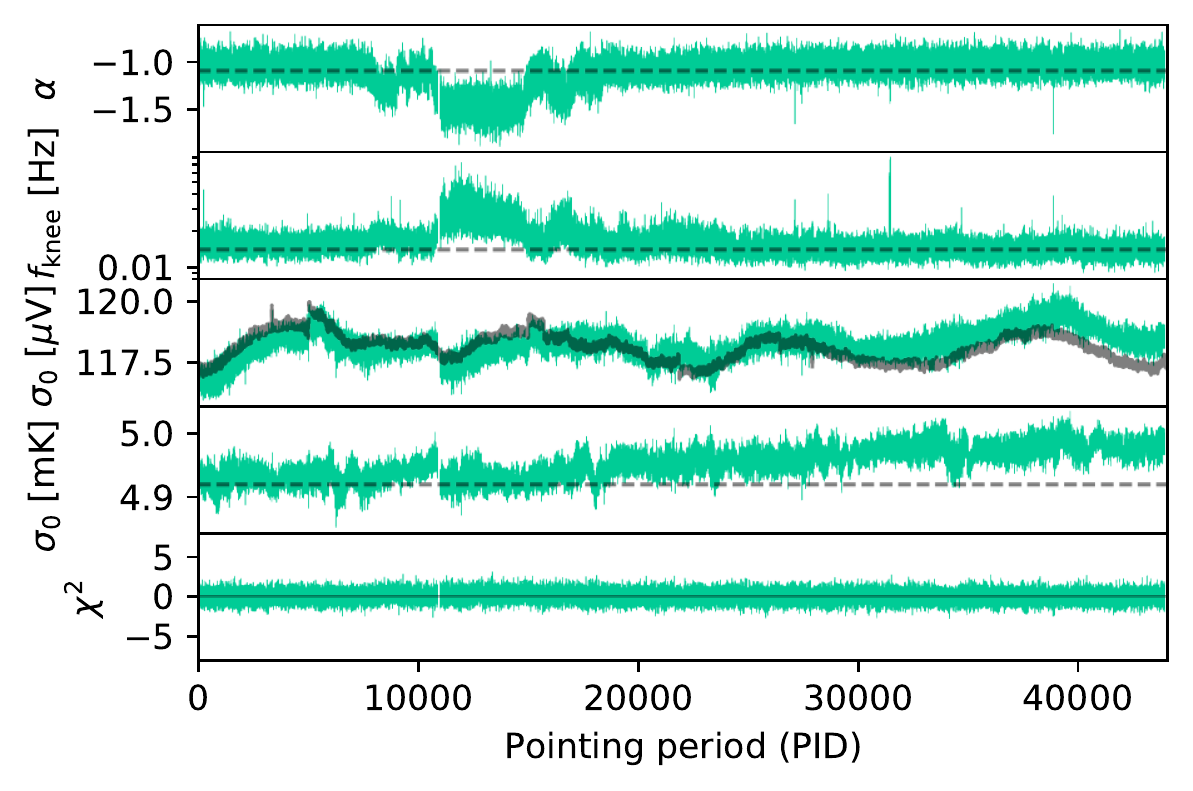}
		 \vspace*{-4.9mm}            
	\end{center}
	
	\caption{Noise characterization of the \Planck\ LFI 70\,GHz
		radiometers; 18M (\emph{top left}), 18S (\emph{top right}); 19M
		(\emph{bottom left}), and 19S (\emph{bottom right}). For each
		radiometer, the top figure shows distributions of noise parameters
		PSD, $\xi^n = \{\sigma_0, f_\mathrm{knee}, \alpha\}$, averaged
		over all Gibbs samples for the full mission. The bottom figure
		shows the time evolution of the posterior mean of the noise
		parameters, and the bottom panel shows the evolution in reduced
		normalized $\chi^2$ in units of $\sigma$. Black dashed curves and crosses show corresponding values as derived by, and used in, the
		official \Planck\ LFI DPC pipeline.
		\label{fig:xi_prop_70_1}}
\end{figure*}

\begin{figure*}[p]
	\begin{center}
		\includegraphics[width=0.495\linewidth]{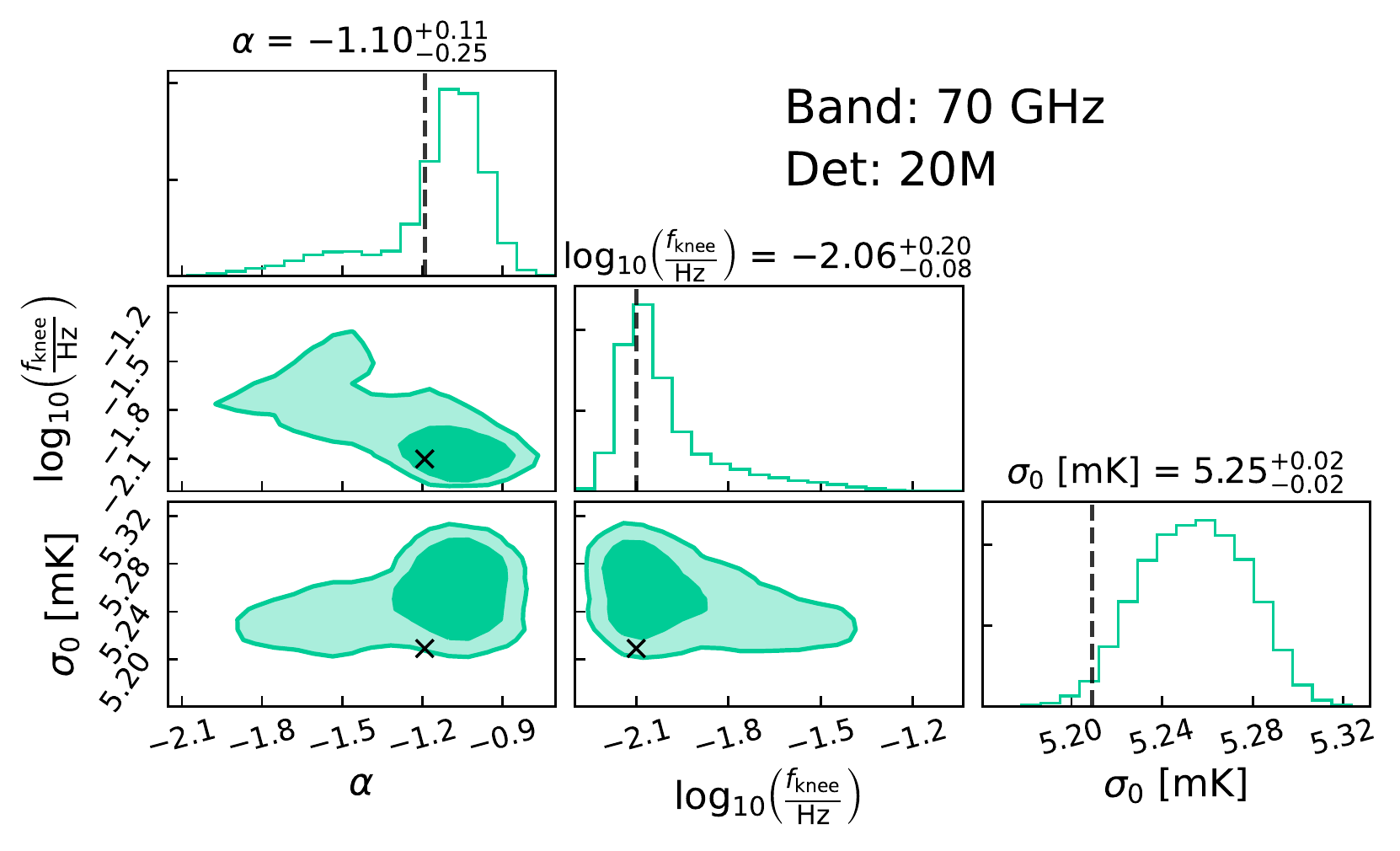}
		\includegraphics[width=0.495\linewidth]{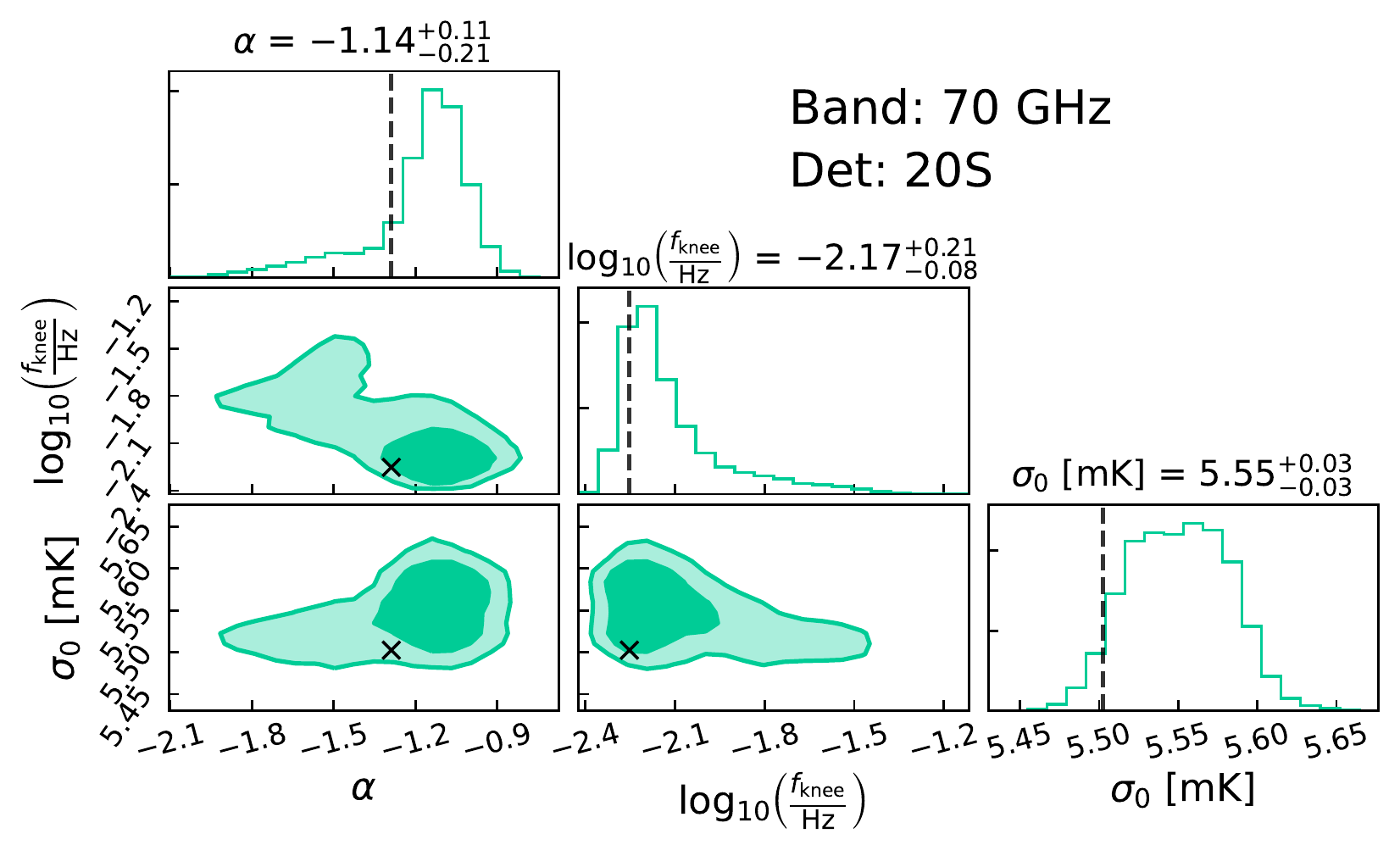}\\
		\includegraphics[width=0.495\linewidth]{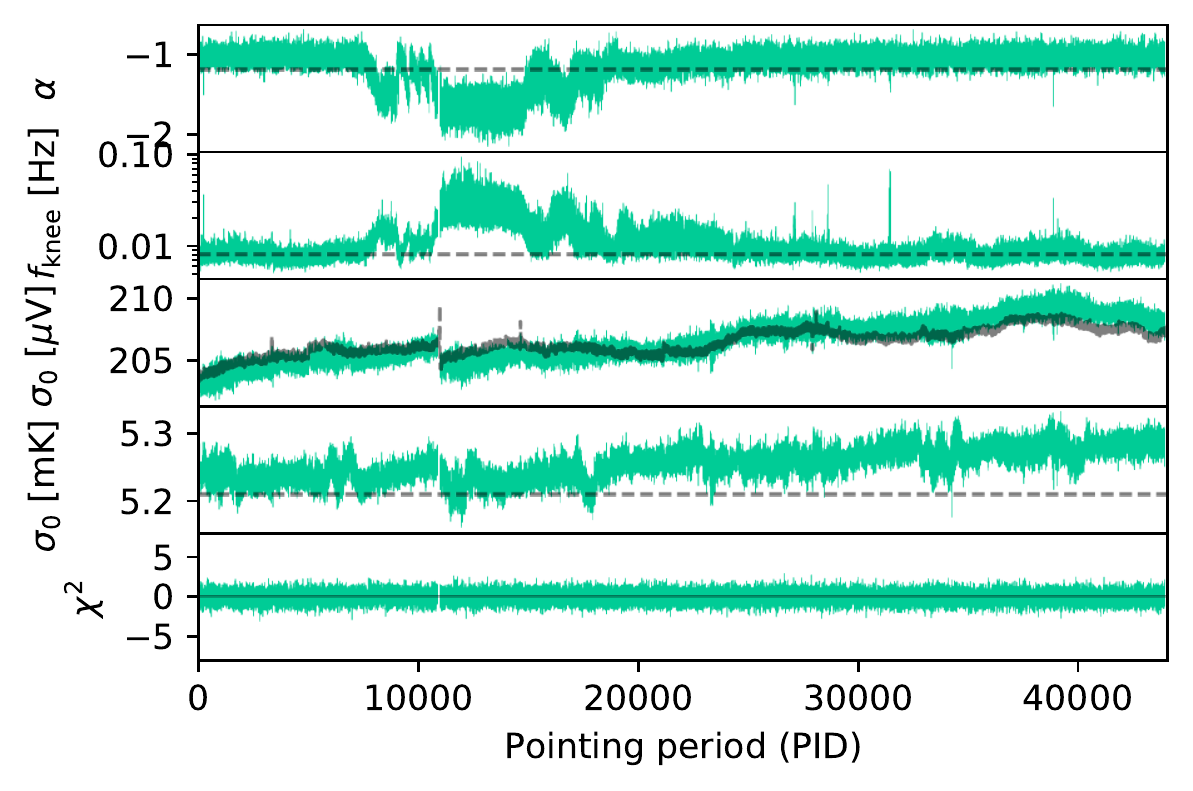}
		\includegraphics[width=0.495\linewidth]{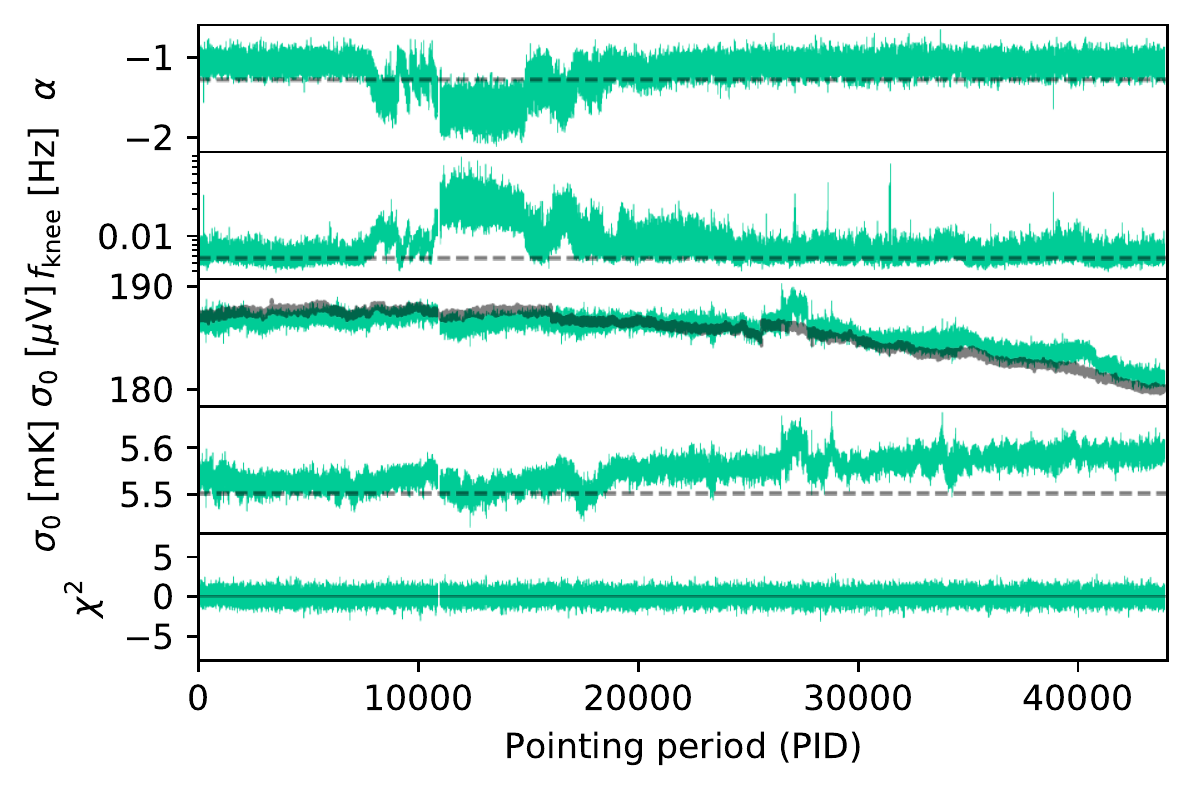}\\
		\vspace*{0.5mm}
		\includegraphics[width=0.495\linewidth]{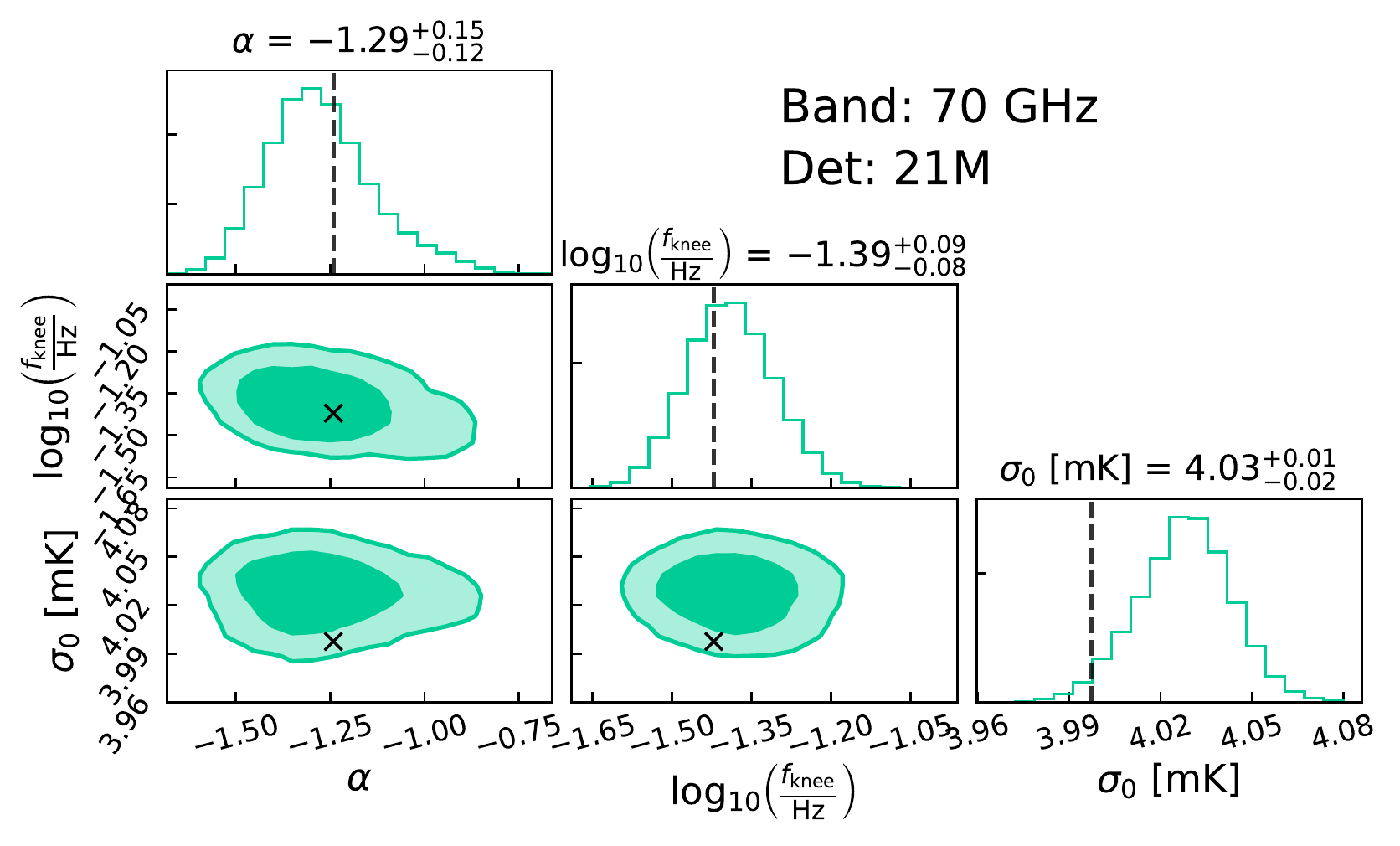}
		\includegraphics[width=0.495\linewidth]{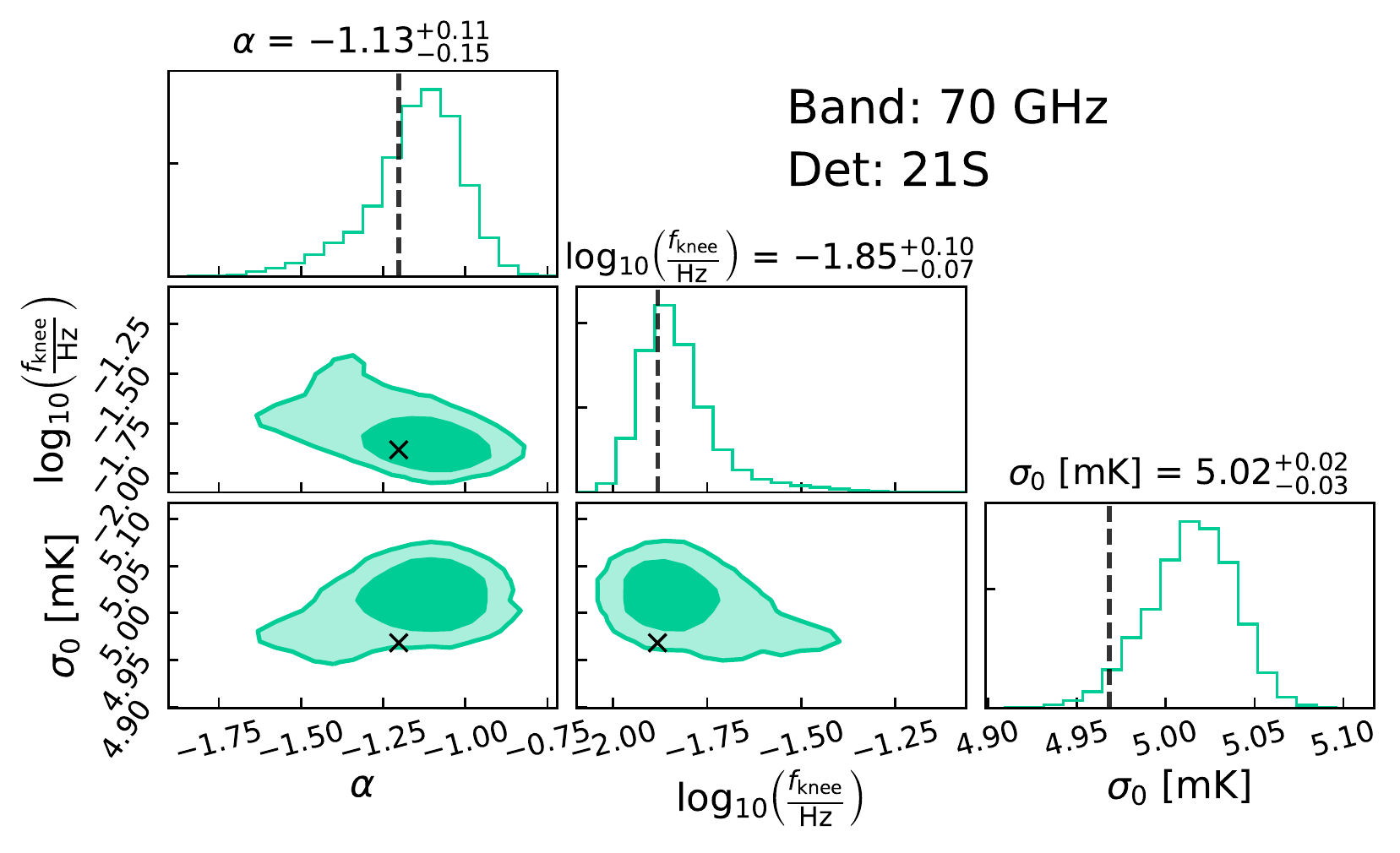}\\
		\includegraphics[width=0.495\linewidth]{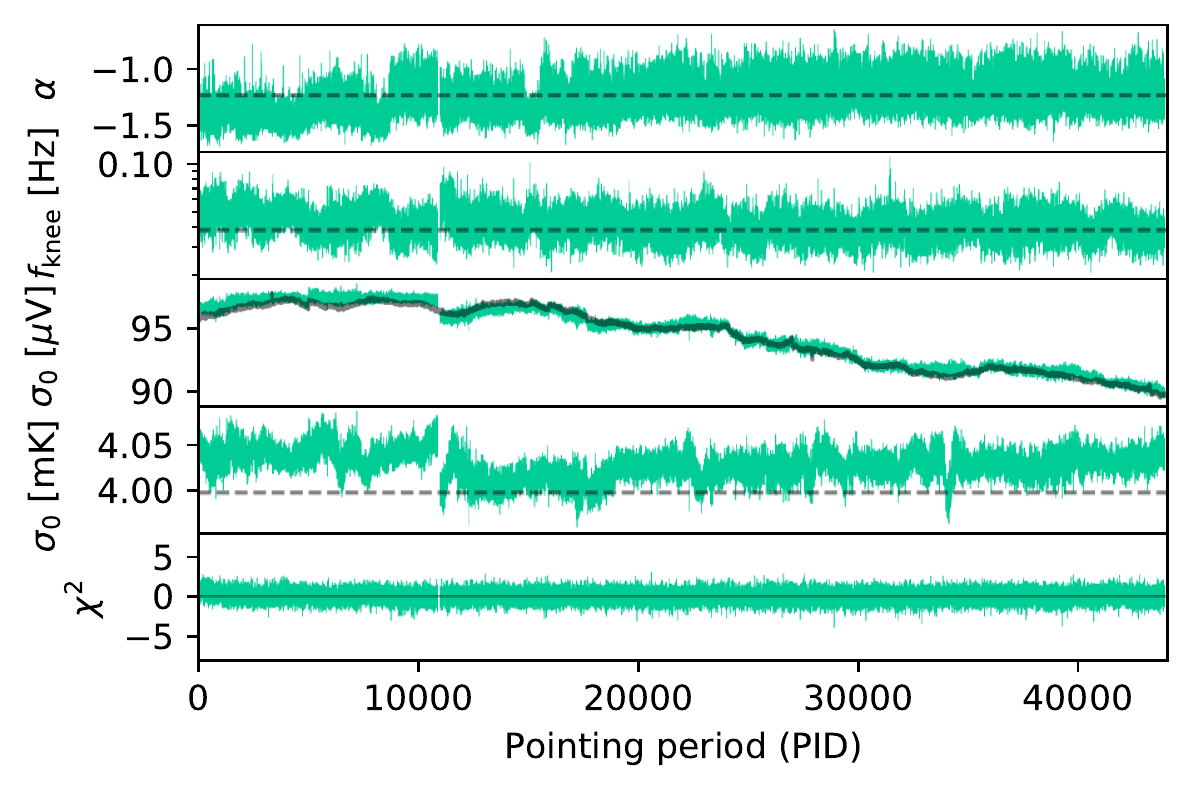}
		\includegraphics[width=0.495\linewidth]{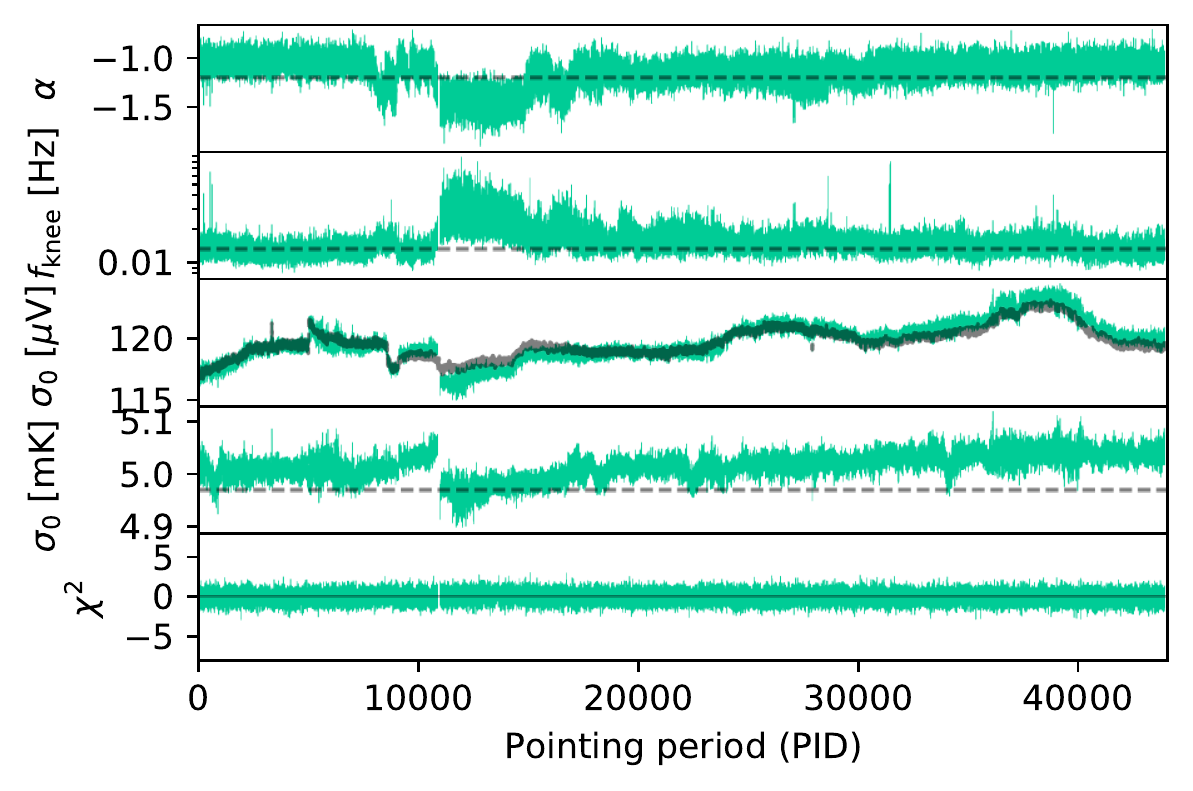}
		 \vspace*{-5.5mm}            
	\end{center}
	
	\caption{Noise characterization of the \Planck\ LFI 70\,GHz
		radiometers; 20M (\emph{top left}), 20S (\emph{top right}); 21M
		(\emph{bottom left}), and 21S (\emph{bottom right}). For each
		radiometer, the top figure shows distributions of noise parameters
		PSD, $\xi^n = \{\sigma_0, f_\mathrm{knee}, \alpha\}$, averaged
		over all Gibbs samples for the full mission. The bottom figure
		shows the time evolution of the posterior mean of the noise
		parameters, and the bottom panel shows the evolution in reduced
		normalized $\chi^2$ in units of $\sigma$. Black dashed curves and crosses show corresponding values as derived by, and used in, the
		official \Planck\ LFI DPC pipeline.
		\label{fig:xi_prop_70_2}}
\end{figure*}

\begin{figure*}[p]
	\begin{center}
		\includegraphics[width=0.495\linewidth]{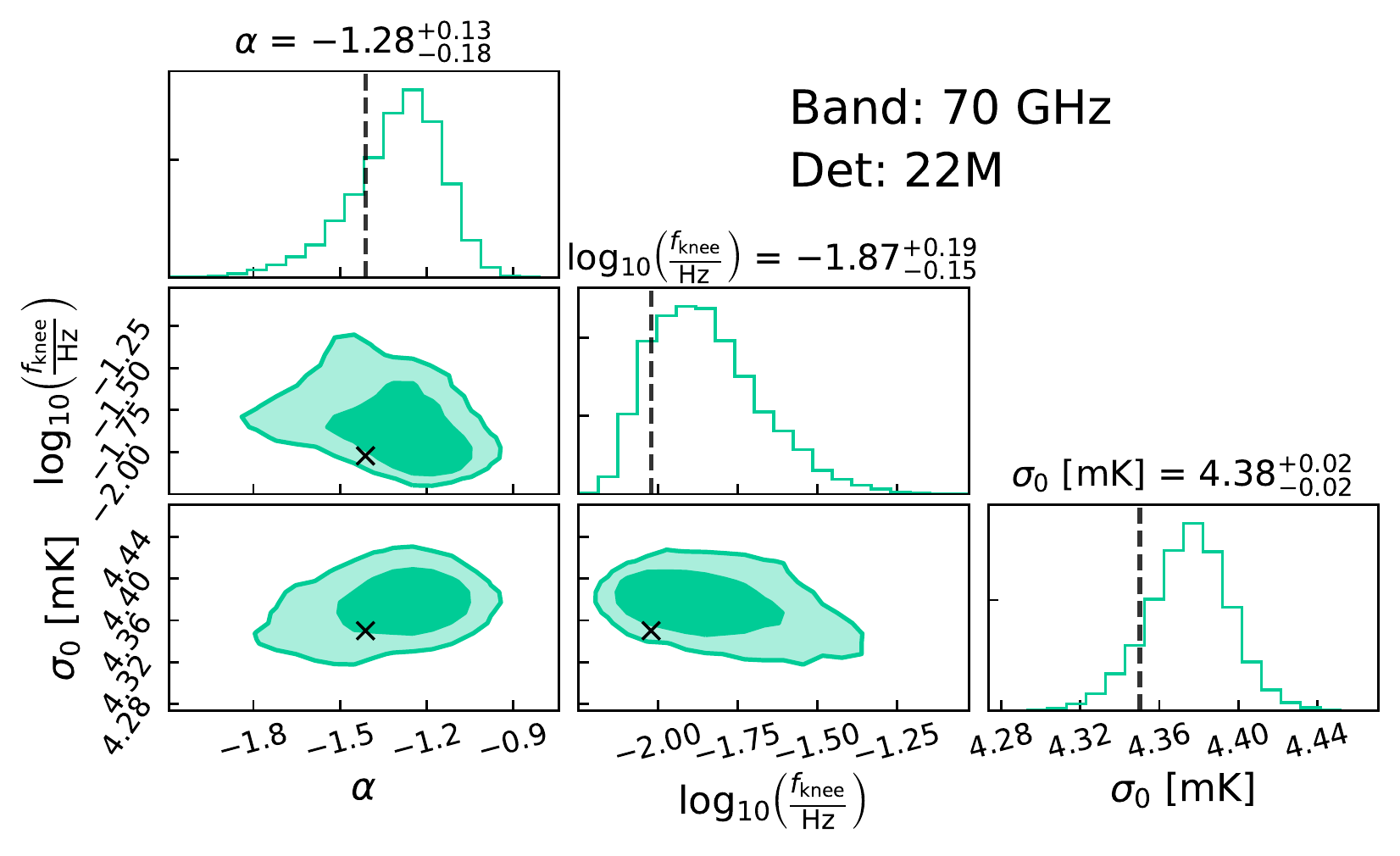}
		\includegraphics[width=0.495\linewidth]{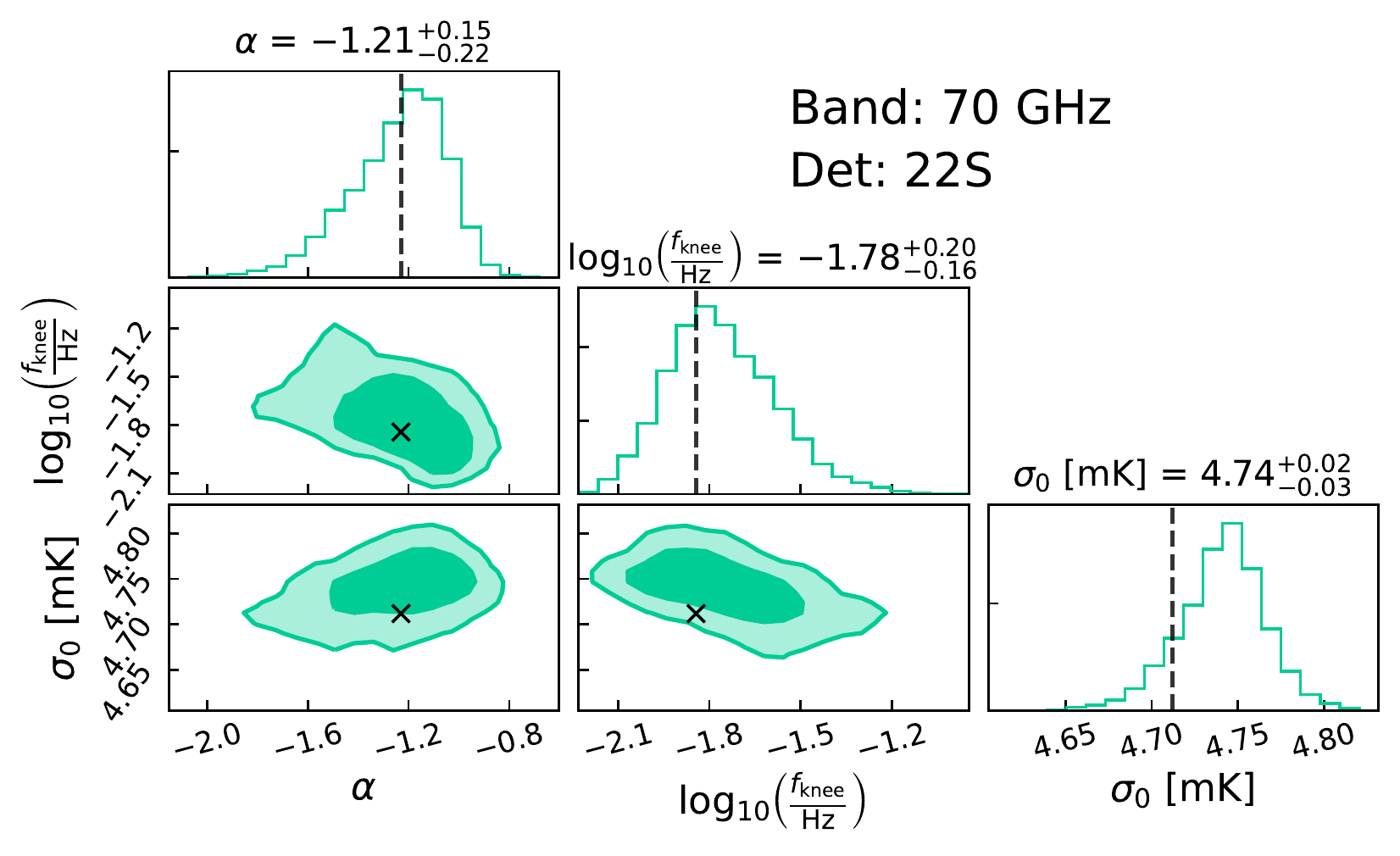}\\
		\includegraphics[width=0.495\linewidth]{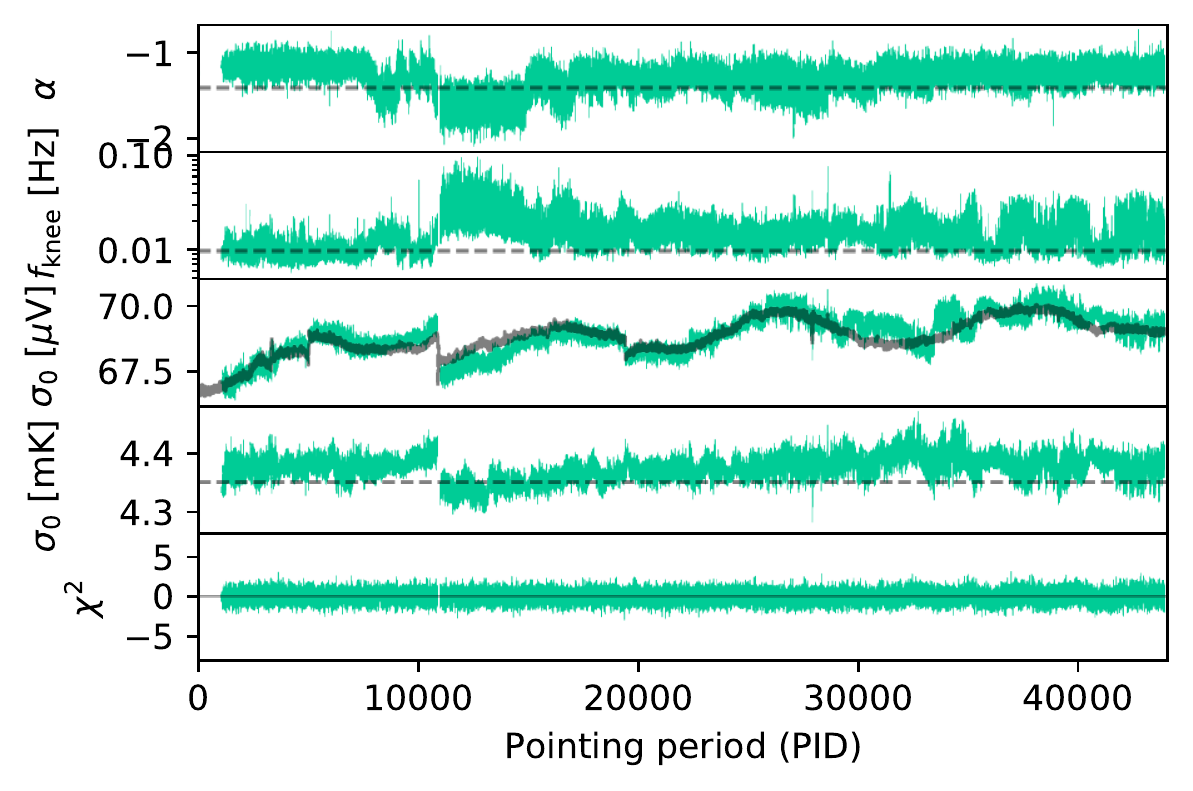}
		\includegraphics[width=0.495\linewidth]{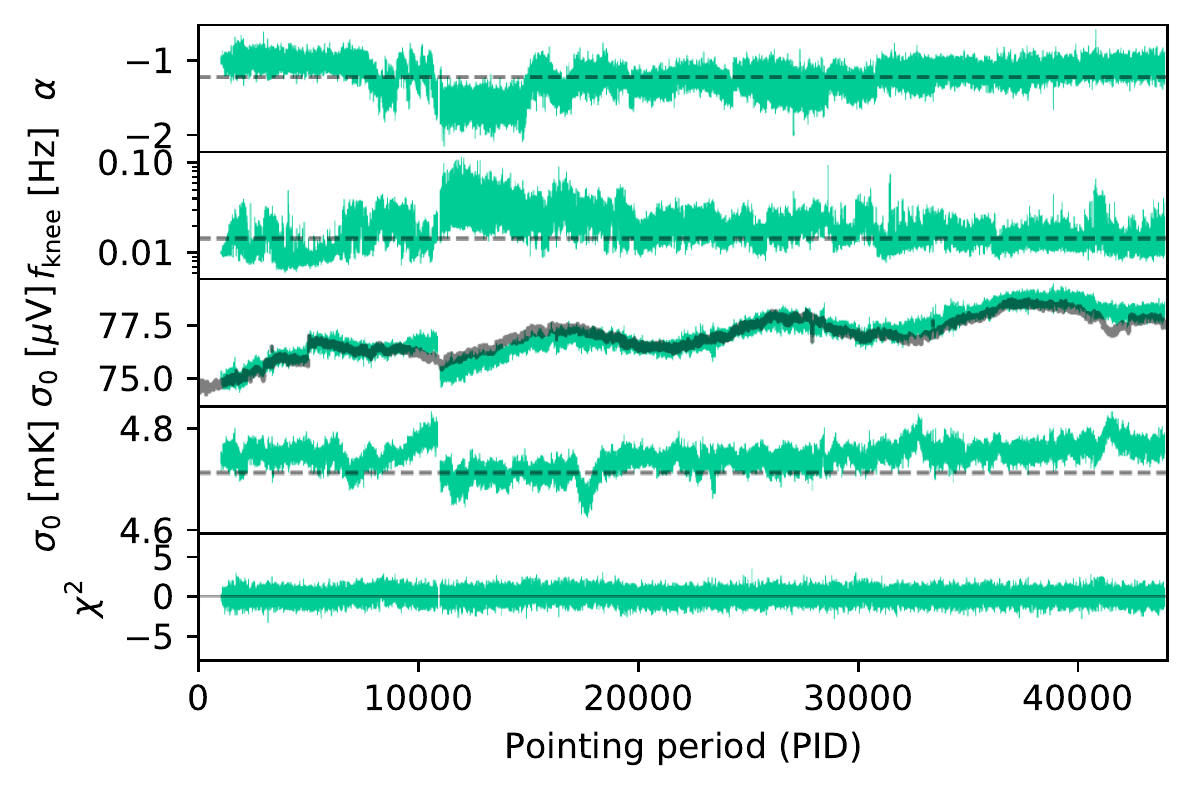}\\
		\vspace*{0.5mm}
		\includegraphics[width=0.495\linewidth]{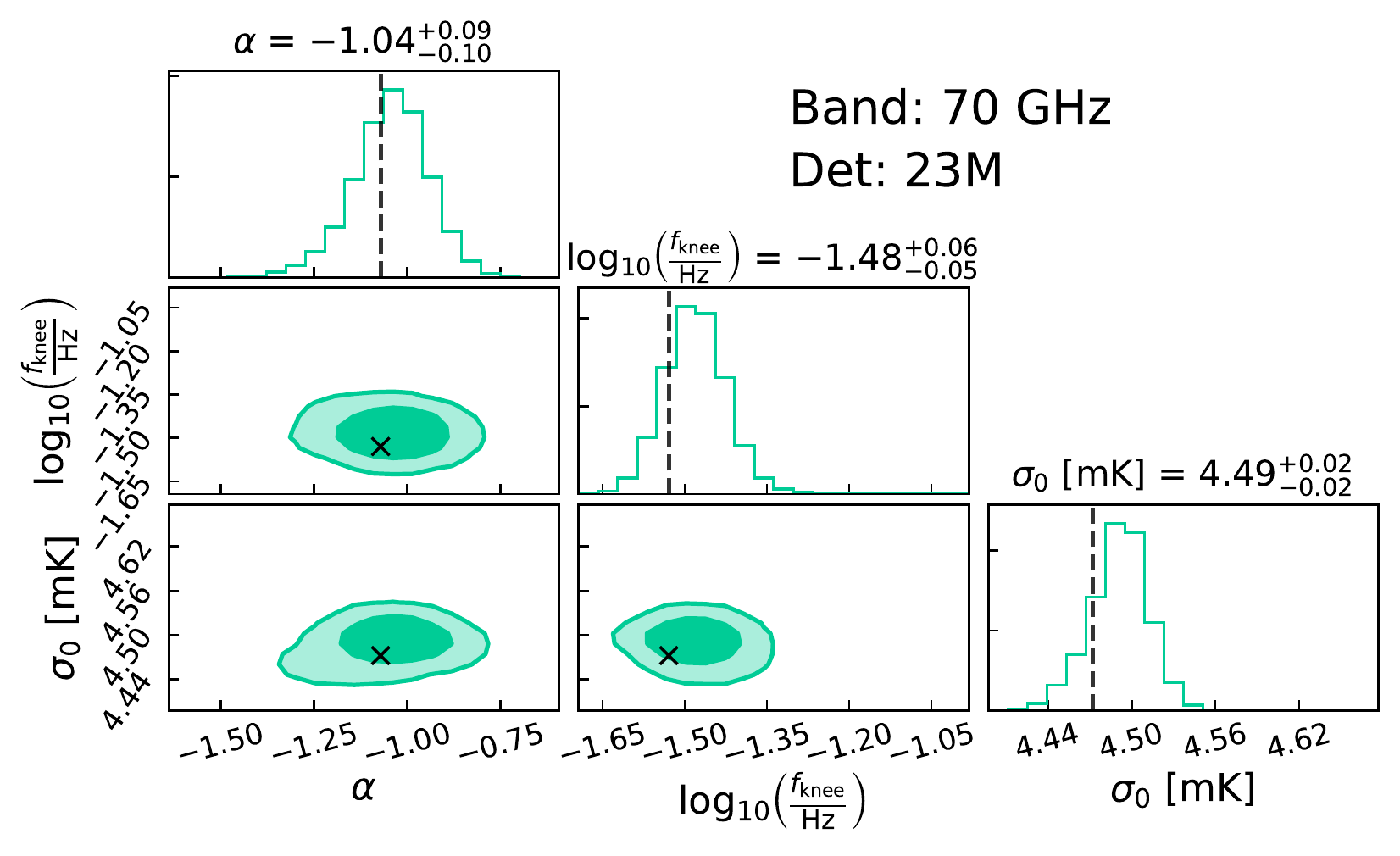}
		\includegraphics[width=0.495\linewidth]{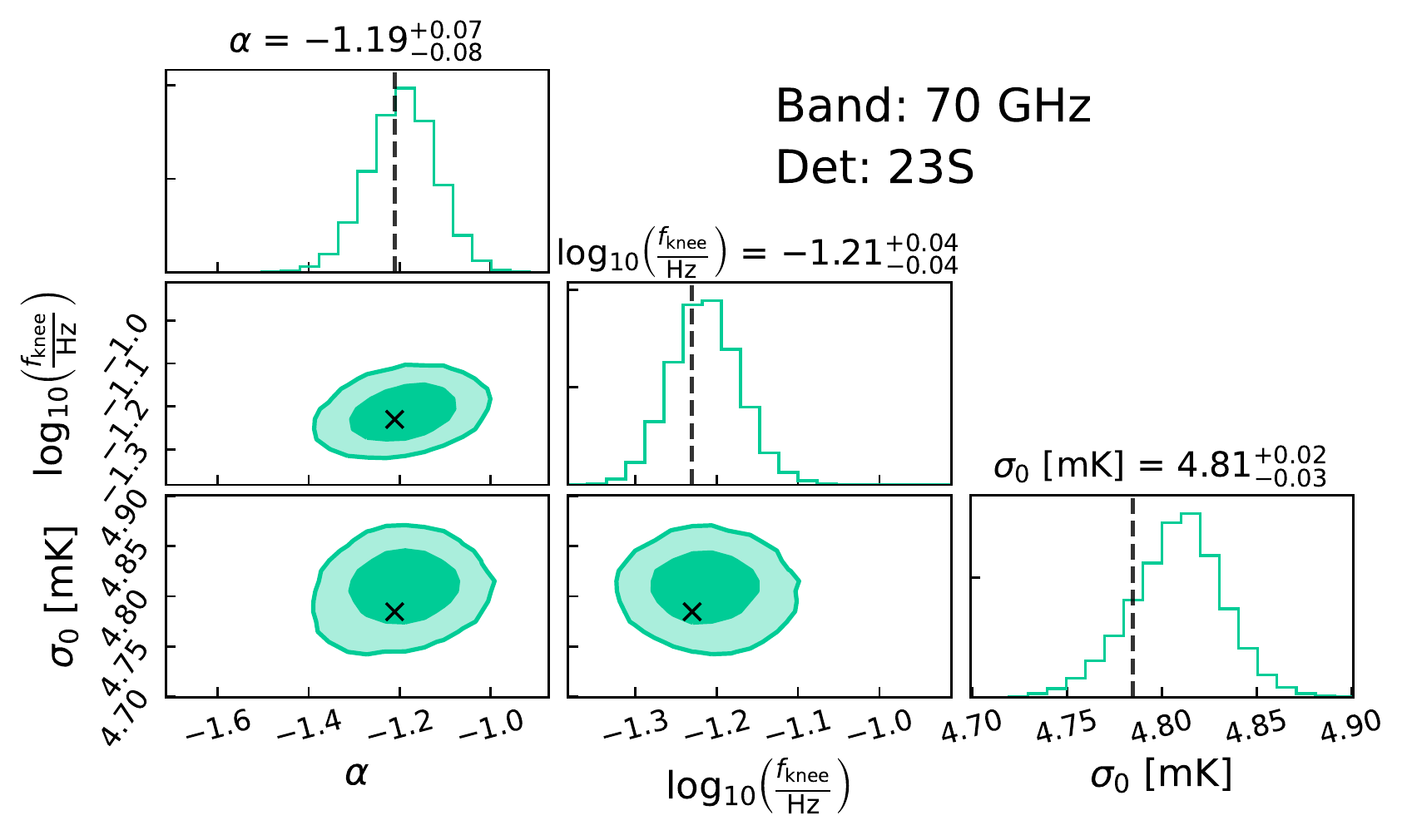}\\
		\includegraphics[width=0.495\linewidth]{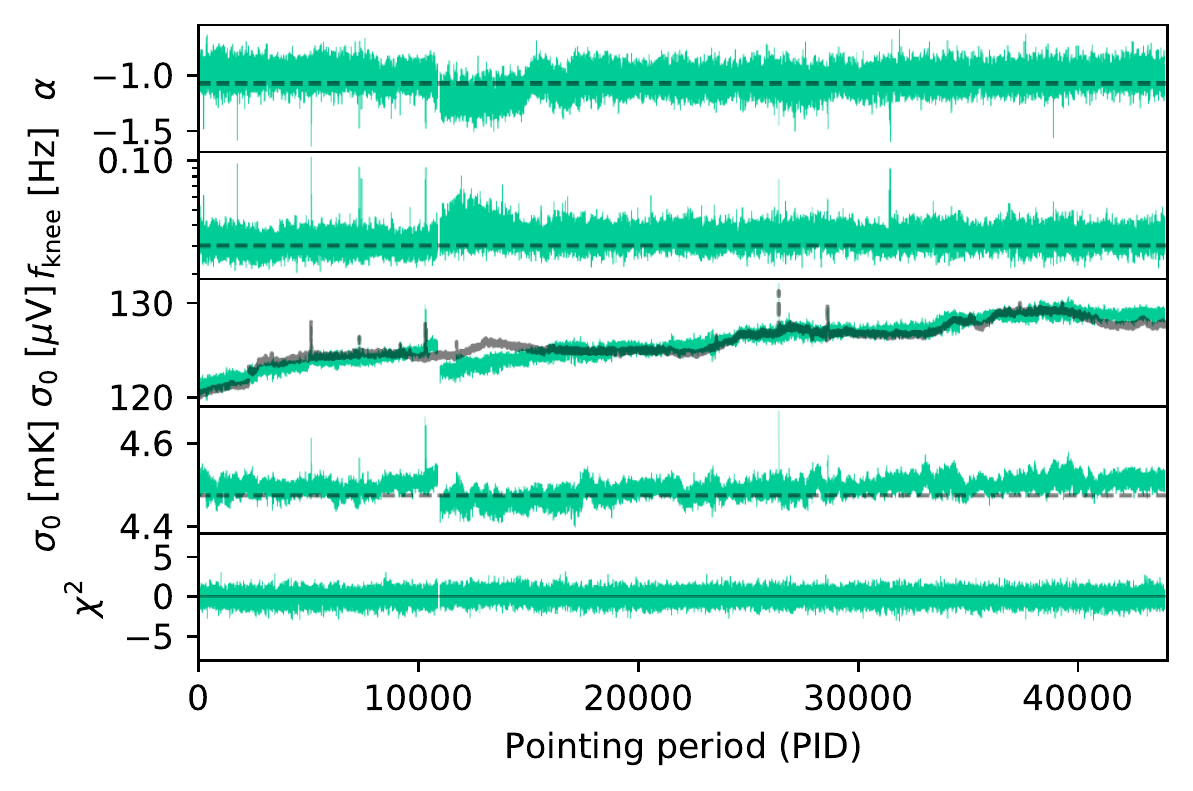}
		\includegraphics[width=0.495\linewidth]{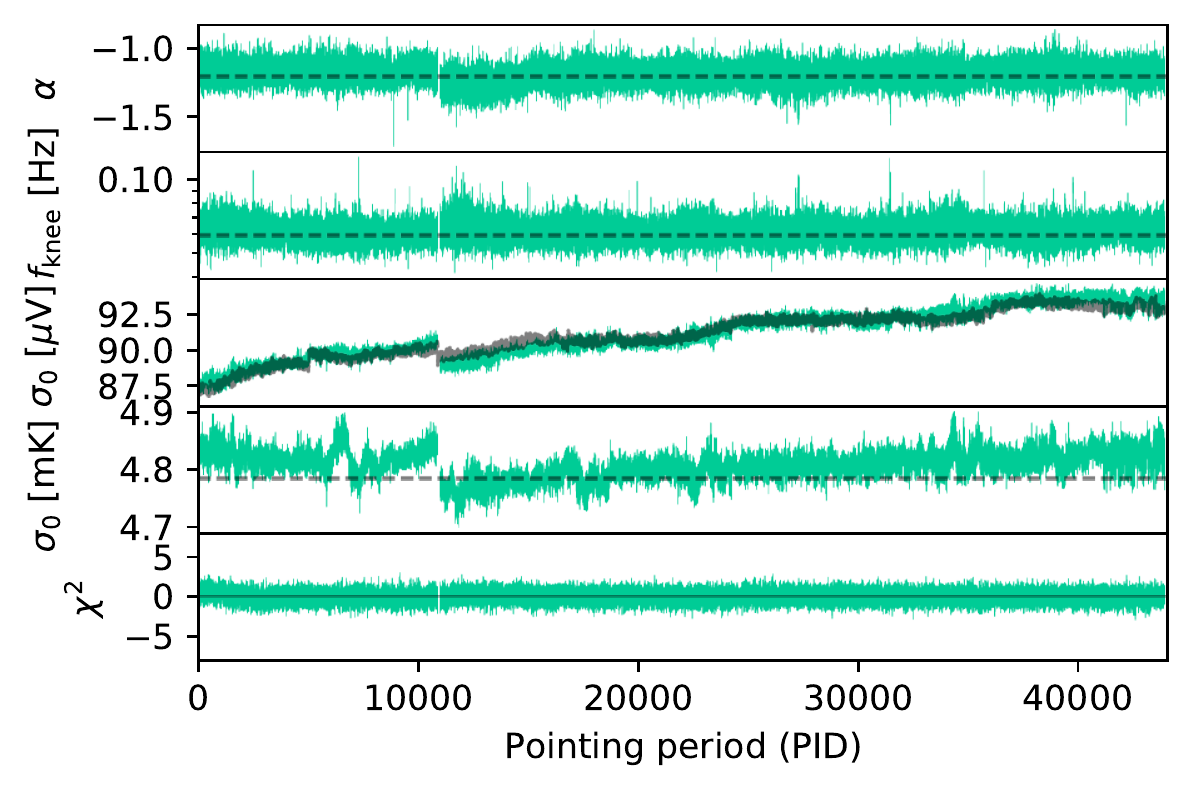}
		 \vspace*{-5.4mm}            
	\end{center}
	
	\caption{Noise characterization of the \Planck\ LFI 70\,GHz
		radiometers; 22M (\emph{top left}), 22S (\emph{top right}); 23M
		(\emph{bottom left}), and 23S (\emph{bottom right}). For each
		radiometer, the top figure shows distributions of noise parameters
		PSD, $\xi^n = \{\sigma_0, f_\mathrm{knee}, \alpha\}$, averaged
		over all Gibbs samples for the full mission. The bottom figure
		shows the time evolution of the posterior mean of the noise
		parameters, and the bottom panel shows the evolution in reduced
		normalized $\chi^2$ in units of $\sigma$. Black dashed curves and crosses show corresponding values as derived by, and used in, the
		official \Planck\ LFI DPC pipeline.
		\label{fig:xi_prop_70_3}}
\end{figure*}

We are now ready to present the main results obtained by applying the
methods described above to the \Planck\ LFI data within the \BP\ Gibbs
sampling framework \citep{bp01}, as summarized in terms of the
posterior distributions for each of the noise parameters. In total,
four independent Gibbs chains were produced in the main \BP\ analysis,
each chain including 500--1000 samples, for a total computational cost
of about 620\,000 CPU-hours \citep{bp01,bp03}.

\subsection{Posterior distributions and Gibbs chains}

First, we recall that at every step in the Gibbs chain, we sample the
correlated noise parameters for each pointing period and each
radiometer, both the time-domain realization $\n^{\mathrm{corr}}$ and
the PSD parameters $\xi^n$. To visually illustrate the resulting
variations from sample to sample in terms of PSDs,
Fig.~\ref{fig:ps_samples} shows three subsequent spectrum samples for
a single pointing period for the 23S radiometer. We see that the
correlated noise follows the data closely at low frequencies, while at
high frequencies the correlated noise is completely dominated by the
sampling terms, so at these frequencies the correlated noise PSD is 
effectively extrapolated based on the current noise PSD model. 
The scatter between the three colored curves shows the typical
level of variations allowed by the combination of white noise and
degeneracies with other parameters in the model.

Figure~\ref{fig:ncorr_map} shows the pixel-space correlated noise 
corresponding to a single Gibbs sample, obtained
after binning $\n^{\mathrm{corr}}$ for all radiometers and all PIDs
into an $IQU$ map. Columns show different frequency maps (30, 44, and
70\,GHz), and rows show different Stokes parameters ($I$, $Q$, and
$U$). Overall, we see that the morphology of each map is dominated by
stripes along the \Planck\ scanning strategy, as expected for
correlated $1/f$ noise, and we do not see any obvious signatures of
either residual foregrounds in the Galactic plane, nor CMB dipole
leakage at high latitudes. This suggests that the combination of the
data model and processing masks described above performs reasonably
well. We also note that the peak-to-peak values of the total
correlated noise maps are $\mathcal{O}(1\muK)$, which is of the same
order of magnitude as the predicted $E$-mode signal from cosmic reionization
\citep{planck2016-l04}. Thus, correlated noise estimation plays a
critical role for large-scale polarization reconstruction, while it is
negligible for CMB temperature analysis.

For both $\n^{\mathrm{corr}}$ and $\xi^n$, the main result of the
\BP\ pipeline are the full ensembles of Gibbs samples. These are
too large to visualize in their entirety here, and 
are instead provided digitally.\footnote{\url{http://cosmoglobe.uio.no}} In the
following, we will therefore focus on $\xi^n$, and as an example
Fig.~\ref{fig:samples} displays one of the full Gibbs chains for two
different PIDs for one radiometer from each LFI frequency band. We see
that the Gibbs chains appear both stable and well-behaved. Some chains
have longer Markov chain autocorrelation lengths than others, as
expected from their different levels of degeneracies both within the
noise model itself, and between the noise and the signal or
gain. In particular, the lognormal amplitude, $A_\mathrm{p}$, shows a 
long autocorrelation time, due to the degeneracy with $f_{\mathrm{knee}}$ 
and $\alpha$. However, while the long autocorrelation times are not ideal, 
all the chains are converging and seem to explore the full range of the 
distributions. In any case, moving power between the lognormal and the 
$1/f$ components has no
effect on the rest of the model. To account for burn-in we remove the first 50 samples from each
chain.

\begin{figure}
	\begin{center}
		\includegraphics[width=1.04\linewidth]{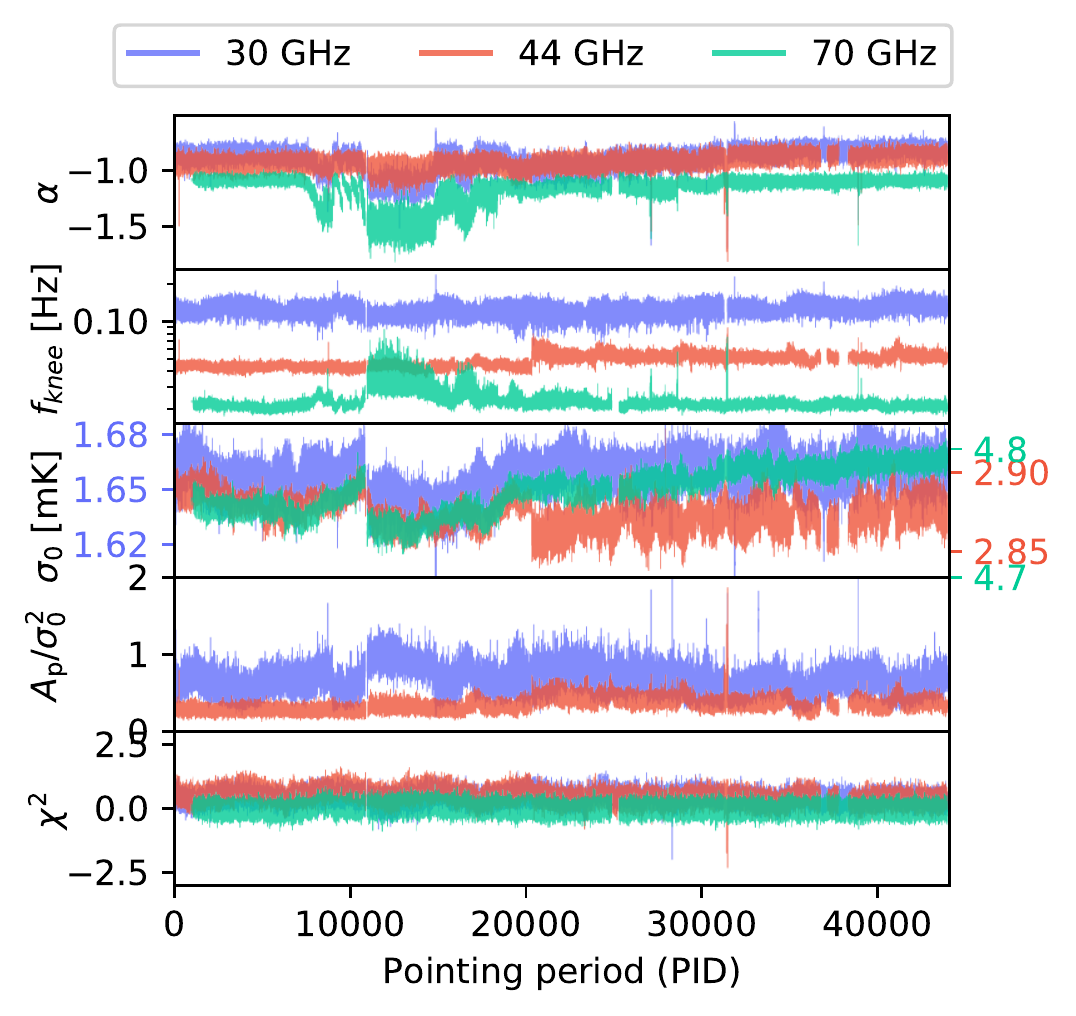}
	\end{center}
	\caption{Posterior mean noise parameters averaged over all radiometers in each band, for the full mission.
		\label{fig:xi_vs_pid_mean}}
\end{figure}

The main results are shown in
Figs.~\ref{fig:xi_prop_30}--\ref{fig:xi_prop_70_3}, which summarize
the noise PSD parameters for each LFI radiometer in terms of
distributions of posterior means (top section; histograms made from
the posterior means for all PIDs) and as average quantities as a function 
of PID (bottom section). The former are useful to obtain a quick overview of
the mean behavior of a given radiometer, while the latter is useful
to study its evolution in time. Blue, red, and green correspond
to 30, 44, and 70\,GHz radiometers, respectively. Mean $\xi^n$ values
are tabulated in Table~\ref{tab:mean_values}, while the average noise
properties of all radiometers in each band are plotted as a function
of time in Fig.~\ref{fig:xi_vs_pid_mean}.

Regarding mean values, we see that the 30\,GHz radiometers generally
have fairly high knee frequencies,
$f_\mathrm{knee}\sim100\,\textrm{mHz}$, and shallow power law slopes,
$\alpha\sim-0.85$. The 70\,GHz channels, on the other hand, have lower
knee frequencies, $f_\mathrm{knee}\sim20\,\textrm{mHz}$, and steeper
slopes, $\alpha\sim-1.2$. The 44\,GHz channels generally
fall between these two extremes. The (normalized) 30 GHz amplitudes of the 
lognormal noise PSD component, $A_\mathrm{p} / \sigma_0^2$, are 
typically a bit larger, fluctuating between $\sim 0-1.5$, than the 44 GHz ones, 
fluctuating between $\sim 0-0.7$. 

The values of $\alpha$ for the individual radiometers are, as expected,
typically around $-1$, with a spread that depends
on the detailed properties of each radiometer component. The
intrinsic correlated noise is dominated by fluctuations in the gain
and noise temperature of the FEM and BEM amplifiers, and on the
isolation of the phase switches. In addition, common mode correlated
noise is induced by temperature instabilities of the instrument
interfaces (mainly in the 20\,K and 300\,K stages), with different amplitudes for
different radiometers depending on their individual thermal
susceptibility coefficient. All these noise contributions are then
processed by the LFI pseudo-correlation scheme, which largely
suppresses their effect in the scientific data by differencing the sky
signal with the signal from the internal 4\,K reference load, whose
instabilities may in principle contribute to the correlated noise. A
prediction of the combined effects of all these sources in terms of
$\alpha$ and $f_{\mathrm{knee}}$ of the differenced data for the individual
radiometers is beyond the reach of our instrument model, and those
values can only be obtained by measurement.

The dashed lines in Figs.~\ref{fig:xi_prop_30}--\ref{fig:xi_prop_70_3}
show the \Planck\ LFI DPC values for each parameter
\citep{planck2016-l02}, which are assumed to be constant throughout
the mission. In most cases, these agree well with the results
presented here. The main exception is the 30\,GHz white noise level,
$\sigma_0$, for which we on average find about 2\,\% lower values. It is
difficult to precisely pinpoint the origin of these differences, but
we do note that Galactic foregrounds are particularly bright at
30\,GHz. One possible hypothesis is therefore that these are fitted
slightly better in the joint and iterative \BP\ approach, as compared
to the linear pipeline DPC approach.

\subsection{Time variability and goodness-of-fit}
\label{sec:time_variability}
Perhaps the single most important and visually immediate conclusion to be drawn
from these plots is the fact that the noise properties of the LFI
instrument vary significantly in time. This is evident in all three
frequency channels and all radiometers. Furthermore, by comparing the
time evolution between different radiometers, we observe many common
features, both between frequencies and, in particular, among
radiometers within the same frequency band. Many of these may be associated
with specific and known changes in the thermal environment of the
satellite, and can be traced using thermometer housekeeping data; this will
be a main topic for the next section.

The bottom panels in
Figs.~\ref{fig:xi_prop_30}--\ref{fig:xi_prop_70_3} show a $\chi^2$ per
PID of the following form,
\begin{equation}
  \chi^2 \equiv
  \frac{\sum_{i=1}^{n_\mathrm{samp}}\left(\frac{r_i}{\sigma_0}\right)^2 -
    n_\mathrm{samp}}{\sqrt{2n_\mathrm{samp}}}\
  \label{eq:chisq}
\end{equation}
where $n_\mathrm{samp}$ is the number of samples, and $r_i$ is the
residual for sample $i$ as defined by Eq.~(\ref{eq:residual}). Thus,
this quantity measures the normalized mean-subtracted $\chi^2$ for
each PID, which should, for ideal data and $n_\mathrm{samp}\gg1$, be
distributed according to a standard Gaussian distribution. 

Starting with the 70\,GHz channel, which generally is the most
well-behaved, we see that the $\chi^2$ fluctuates around zero for most
channels, with a standard deviation of roughly unity. 
In general, the 30 and 44\,GHz channels appear less stable than
the 70\,GHz channels in terms of overall $\chi^2$, with several detectors
showing a positive bias of 1--$2\,\sigma$ per PID, with
internal temporal variations at the $1\,\sigma$ level. This suggests that,
while the addition of the lognormal noise PSD component certainly improves
the fit of the noise PSD model, we still do not 
always get a perfect fit. This is not very surprising, since the 
shape parameters of the lognormal component are not chosen separately for each
radiometer, only the overall amplitude is fit. 
Sampling the lognormal shape parameters independently for each 
radiometer is a main goal for future LFI analysis. 

\begin{figure}
	\begin{center}
		\includegraphics[width=\linewidth]{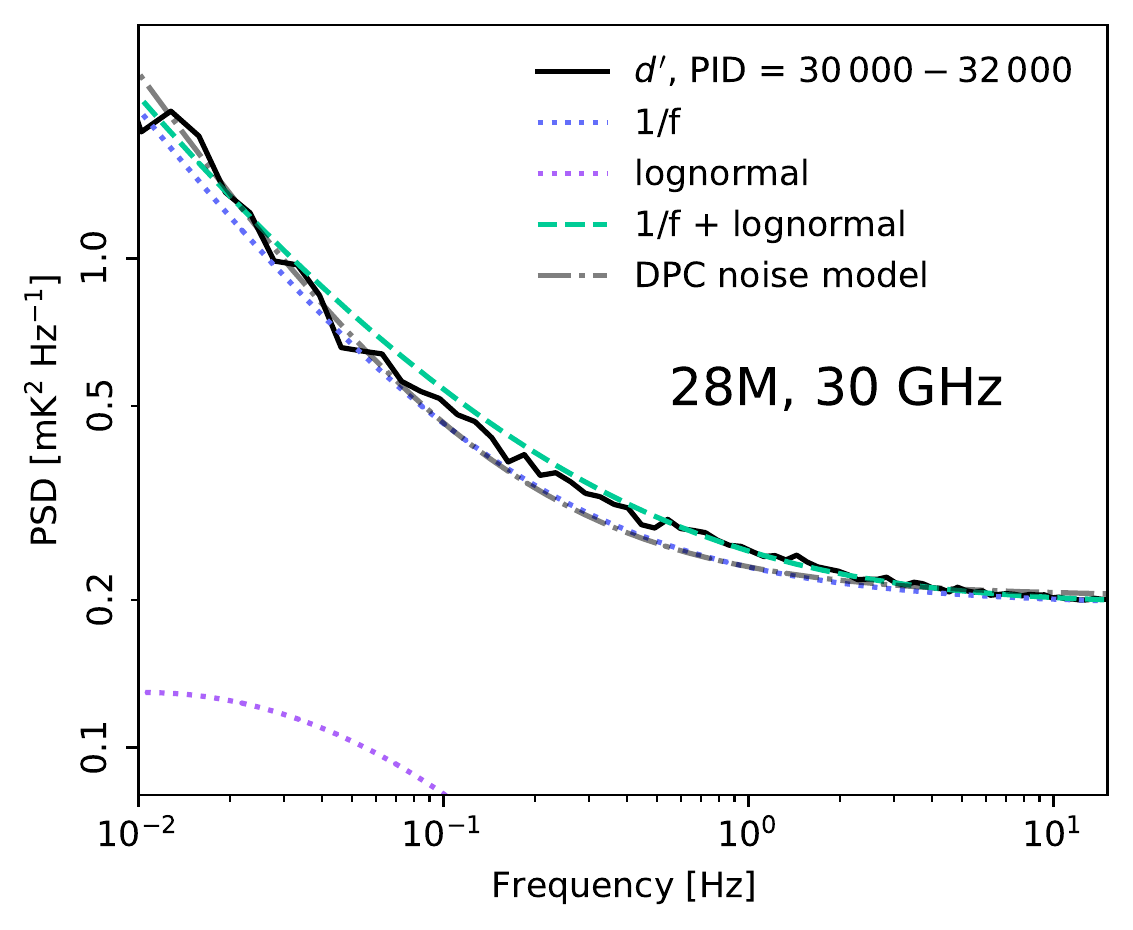}
	\end{center}
	\caption{PSD of signal-subtracted data from radiometer 28M,
          averaged over 18 PIDs (at intervals of 100 PIDs) in the range 30\,000--32\,000
          (black). The dashed lines show the mean \BP\ (dashed green)
          and LFI DPC (dashed and dotted gray) noise models for the same data. Dotted lines show the $1/f$ (blue) and lognormal (purple) components of the \BP\ noise model. 
		\label{fig:ps_compare_28M}}
\end{figure}

As a typical illustration of noise PSD model fits, Fig.~\ref{fig:ps_compare_28M} 
shows the PSD for a range of 18 PIDs for
the 28M 30\,GHz radiometer. Here the $1/f$ model is
not able to fit the correlated noise to sufficient statistical
accuracy at intermediate temporal frequencies, between 0.1 and 10\,Hz,
but rather shows a generally flatter trend. The addition of the lognormal
component greatly improves the fit between about 0.3--3\,Hz, but we 
still see significant deviations from the model at slightly lower frequencies
this is an example of features that could be better fit if the shape parameters
were fit individually for each radiometer. 
Similar behavior is seen in many 30 and 44\,GHz radiometers, 
while the 70\,GHz radiometers are better behaved.

Turning our attention to the $\xi^n$ parameters, we see much larger
variability than in the $\chi^2$. First, we note a period of
significant instability in most channels between PIDs 8000--20\,000, but
most strikingly in the 70\,GHz $\alpha$ estimates. This feature will be
discussed in more detail in Sect.~\ref{sec:systematics}, where it is
explicitly shown to be correlated with thermal variations. We note,
however, that the noise model seems flexible enough to adjust to
these particular changes, as no associated excess $\chi^2$ is observed
in the same range.

Next, when considering the white noise level, $\sigma_0$, given in
units of volts or kelvins, we see the pattern anticipated
in the previous section. The uncalibrated white noise in units of volts
follows the slow drifts of the gain, which typically manifests itself
in slow annual gain oscillations. In contrast, the calibrated noise in
units of $\mathrm{K}_{\mathrm{CMB}}$ is far more stable. 

Other significant features include sharp jumps in noise properties,
most notably seen around PID 20\,000 for radiometer 26S (see Figure~\ref{fig:xi_prop_44_2}), but also 
seen in most radiometers in the period before and after PID 11\,000. Both will be discussed in Sect.~\ref{sec:systematics}.

\section{Systematic effects}
\label{sec:systematics}
\begin{figure}
  \begin{center}
    \includegraphics[width=\linewidth]{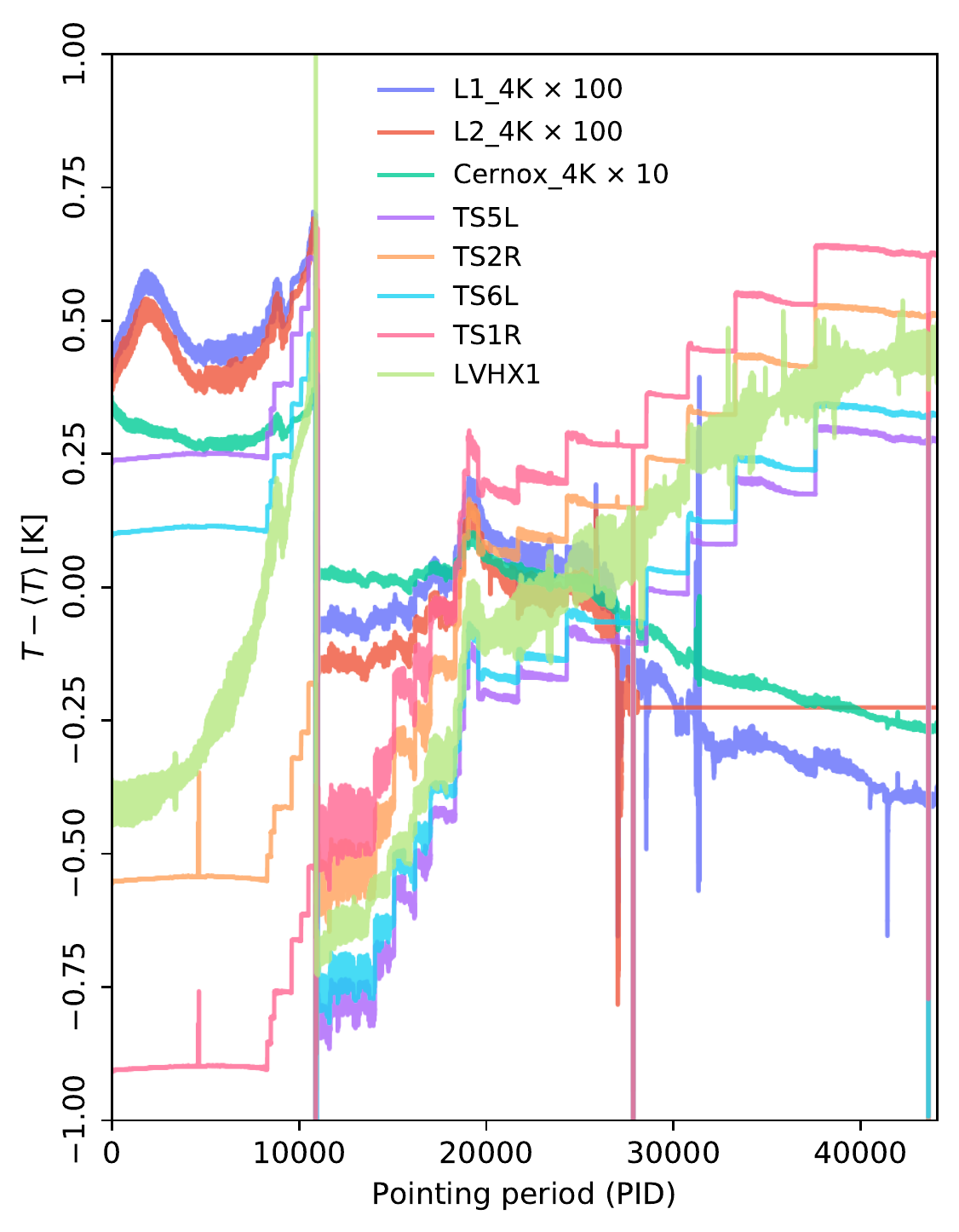}
  \end{center}
  \caption{High-level comparison of measurements from eight
    temperature sensors that are particularly relevant for
    LFI. Sensors TS5L, TS6L, TS1R, TS2R are installed in the 20\,K
    focal plane frame, while LVHX1 is the liquid-vapor heat exchanger
    providing 18\,K to HFI; sensors L1\_4K, L2\_4K and Cernox\_4K are on
    the HFI 4\,K stage supporting the LFI 4\,K reference loads. The
    step-like increases in the 20\,K stage are visible both before and
    after the sorption cooler switchover event (near PID 11\,000).
    For
    details on the locations of the various temperature sensors, see Fig.~21
    of \citet{bersanelli2010} and Fig.~18 of \citet{lamarre2010}. For
    visualization purposes, the mean value has been subtracted from
    each data set, and some have been scaled by one or two orders of
    magnitude, as indicated in the legend. 
    \label{fig:hk_all}}
\end{figure}

Previous LFI analyses have assumed a stationary noise model with three
fixed parameters ($\sigma_0 \,\mathrm{[K]}$, $f_{\rm knee}$, and
$\alpha$) for each of the 22 radiometers. In contrast, each of these
parameters is estimated for every PID in \BP\, increasing the total
number of PSD noise parameters from 66 to more than 3 million. This
increase of information allows us to capture the effects of evolution
in the radiometer responses and local thermal environment, as well as
subtle interactions between them. In this section, we will use this
new information to characterize potential residual systematic effects
in the data, and, as far as possible, associate these with independent
housekeeping data or known satellite events. An overview of the
measurements from eight temperature sensors that are particularly
important for LFI is provided in Fig.~\ref{fig:hk_all}. For details on
the locations of the various temperature sensors, see Fig.~21 of
\citet{bersanelli2010} and Fig.~18 of \citet{lamarre2010}.

\subsection{Temperature changes in the 20\,K stage}
A key element for the LFI thermal environment was the
\Planck\ sorption cooler system (SCS), which provided the 20\,K stage
to the LFI front-end and the 18\,K pre-cooling stage to HFI. The SCS
included a nominal and a redundant unit \citep{planck2011-1.3}. In
August 2010 (around PID 11\,000), a heat switch of the nominal cooler
unit reached its end-of-life, and the SCS was therefore switched over
to the redundant cooler.\footnote{This operation took place at PID
  10911, corresponding to Operation Day (OD) 454.}  This ``switchover''
event implied a major redistribution of the temperatures in the LFI
focal plane, with variations at $\sim$1\,K level, for two main
reasons. First, the efficiency of the newly active redundant cooler
led to an overall decrease of the absolute temperature. Second,
because of the different location of the interface between the focal
plane structure and the cold-end for the redundant cooler, a change of
temperature gradients appeared across the focal plane.

Since the SCS dissipated significant power, changes in its
configuration produced measurable thermal effects in the entire \Planck\
spacecraft, and most directly in the 20\,K stage. In the period
preceding the switchover, starting around PID 8000, a series of power
input adjustments were commanded to reduce thermal fluctuations in the
20\,K stage while optimizing the sorption cooler lifetime, which
generated a number of step-like increases in the LFI focal plane
temperature. These are measured by all of the LFI temperature 
sensors located in the 20\,K focal plane unit, as shown in Fig.~\ref{fig:hk_all}.

Following the switchover, in the period with PIDs 11\,000--15\,000, a
significant increase of 20\,K temperature fluctuations was
observed. These excess fluctuations were understood as due to residual
liquid hydrogen sloshing in the inactive cooler and affecting the
cold-end temperature. The issue was resolved by heating the unit and
letting the residual hydrogen evaporate. Afterwards, to optimize the
performance and lifetime of the operating cooler, several periodic,
step-like adjustments were again introduced in the operational
parameters of the cooler. This resulted in a semi-gradual, monotonic
increase of the LFI focal plane temperature from switchover to end of
mission of $\sim$1.3\,K.

\begin{figure}
	\begin{center}
		\includegraphics[width=\linewidth]{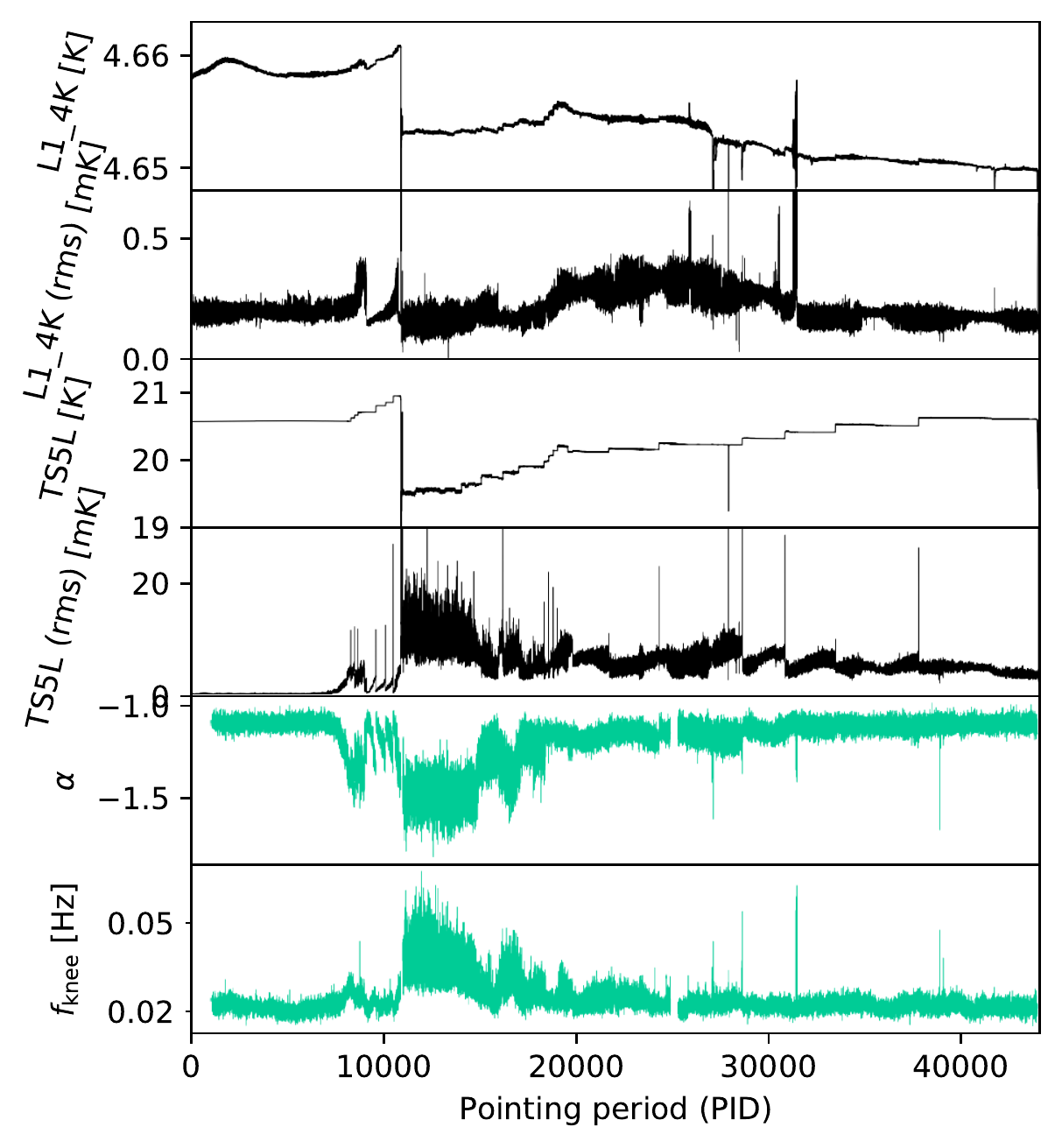}
	\end{center}
	\caption{Average correlated noise properties of the 70\,GHz radiometers
          (bottom two panels) compared with 4\,K and 20\,K temperature sensor
          read-outs (top four panels) for the full mission.
		\label{fig:hk_TSL_70_full}}
\end{figure}

\begin{figure}
	\begin{center}
		\includegraphics[width=\linewidth]{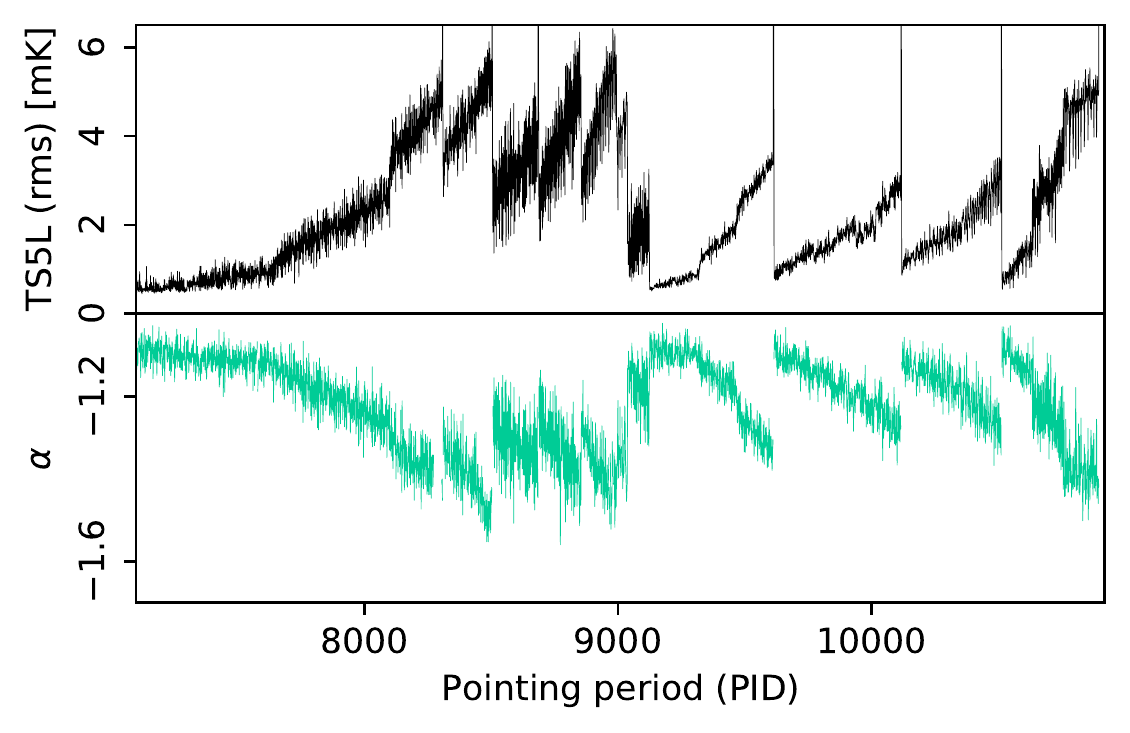}
	\end{center}
	\caption{Same as Fig.~\ref{fig:hk_TSL_70_full}, but zoomed in
          on PIDs 7000--11\,000.
		\label{fig:hk_TSL_70_zoom}}
\end{figure}

In Fig.~\ref{fig:hk_all} the sudden discontinuity at switchover (PID 11\,000) is visible 
for all temperature sensors, and the stepwise up-ward trend driven by 
SCS operational adjustments can be seen in all 20\,K sensors.
These temperature variations directly affected the
LFI noise performance for most radiometers, as observed in the lower
panels of Figs.~\ref{fig:xi_prop_30}--\ref{fig:xi_prop_70_3}. To see
this, it may be useful to concentrate on a well-behaved case (e.g.,
radiometers 22 or 23, Fig.~\ref{fig:xi_prop_70_3}) and then recognize
the same features in other radiometers.

The effect of the SCS switchover shows up as a sharp discontinuity
also in the white noise levels near PID 11\,000. The sudden decrease of
the focal plane temperature of about 1\,K implies a change in
radiometer gain, as well as a genuine reduction in radiometer
noise. This leads to a decrease not only of $\sigma_0 \,\mathrm{[V]}$ but also
of $\sigma_0 \,\mathrm{[K]}$. Furthermore, due to the change in cold-end
interface, the temperature drop at switchover was larger on the
top-right-hand side of the focal plane (as defined by the view in
Fig.~6 of \citealp{bp01}) than in other regions. In particular, we
see in Figs.~\ref{fig:xi_prop_30}--\ref{fig:xi_prop_70_3} that the
drop in $\sigma_0 \,\mathrm{[K]}$ is particularly pronounced for radiometers 21,
22, 23, 27 (both M and S), which are all located in that portion of
the focal plane.

Using again Fig.~\ref{fig:xi_prop_70_3} as a guide, we can also
recognize the effect of the incremental increase of focal plane
temperature due to sorption cooler adjustments, both before and after
switchover. The increasing physical temperature of the focal plane
drives a corresponding increase of $\sigma_0 \,\mathrm{[K]}$, which is visible
for most of the radiometers in
Figs.~\ref{fig:xi_prop_30}--\ref{fig:xi_prop_70_3}. However, we cannot
exclude that part of the observed slow increase of white noise is due
to aging effects degrading the intrinsic noise performance of the
front-end amplifiers. To disentangle these two components would
require a more detailed thermal and radiometric model.

\subsection{Temperature fluctuations and $1/f$ parameters}
\label{sec:temp_fluct_1f}

In Fig.~\ref{fig:hk_TSL_70_full} (top four panels) we report the value and rms of 
representative temperature sensor of the 4\,K and 20\,K stages 
(L1\_4K and TS5L). During the thermal instability period 
that followed the switchover, the noise
properties of essentially all the 70\,GHz radiometers markedly changed
their $1/f$ noise behavior (with the only notable exception of
21M). This is highlighted in the lower two panels of
Fig.~\ref{fig:hk_TSL_70_full}, which show the averaged values of $\alpha$ and
$f_{\rm knee}$ for all the 70\,GHz radiometers. The correlation between $1/f$ noise parameters and
temperature fluctuations is very strong, with higher fluctuations producing an
increase in $f_{\mathrm{knee}}$ and a steepening (i.e., more negative)
slope $\alpha$. The latter is a typical behavior of thermally driven
instabilities, which tend to transfer more power to low frequencies,
and thus steepen the $1/f$ tail. This behavior shows up
also in the individual 70\,GHz radiometers
(Figs.~\ref{fig:xi_prop_70_1}--\ref{fig:xi_prop_70_3}).

Figure~\ref{fig:hk_TSL_70_zoom} is a zoom into the pre-switchover period
(PID 7000--11\,000) of the upper plot. Here we see the effect of some
of the step-wise adjustments in the sorption cooler operation, whose
main effect is to temporarily reduce the temperature fluctuations. The
observed tight correlation with the steepening of the slope is
striking.

\begin{figure}
	\begin{center}
		\includegraphics[width=\linewidth]{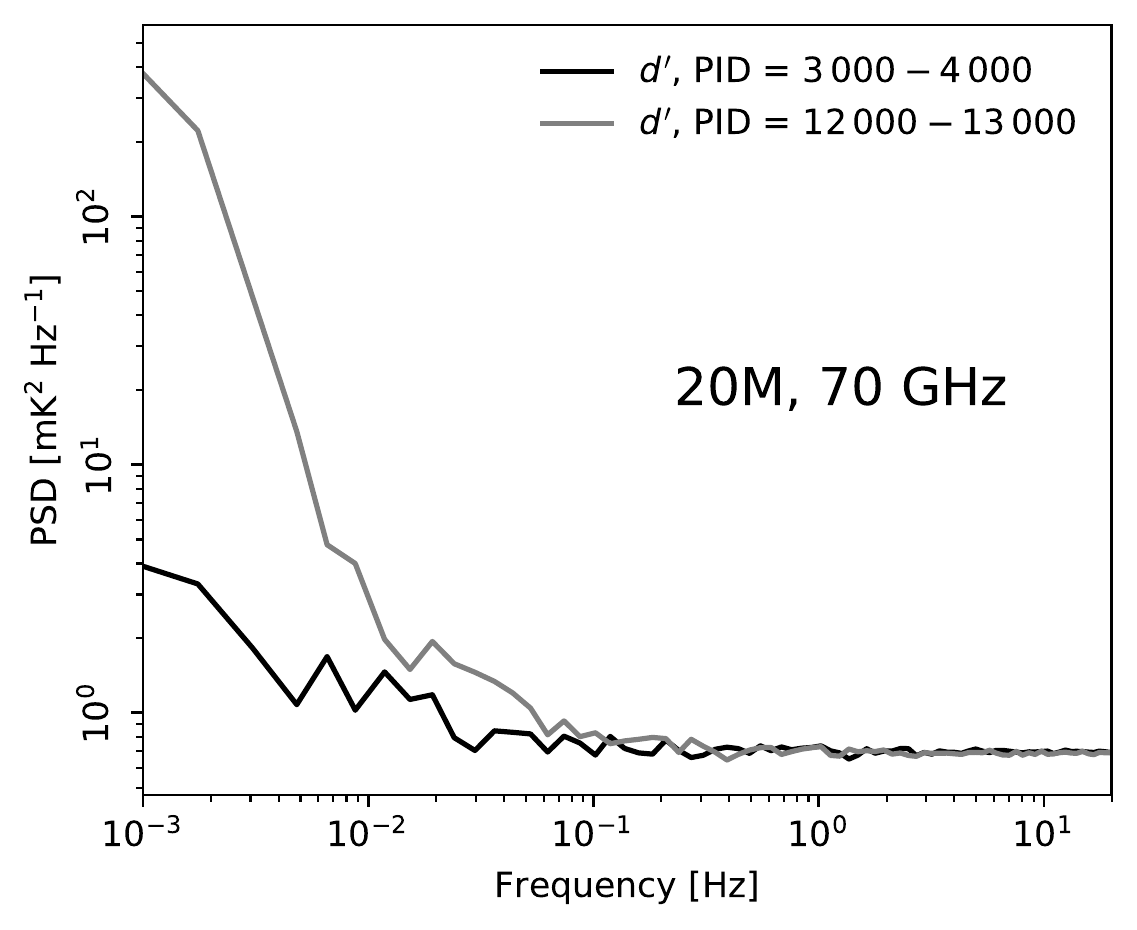}
	\end{center}
	\caption{PSD of signal subtracted data from radiometer 20M, averaged over 10
		PIDs in the ranges 3000--4000 (black) and 12\,000--13\,000 (grey). 
		\label{fig:ps_compare_sorption}}
\end{figure}

For a specific example of how noise property variations modify the
noise PSD, Fig.~\ref{fig:ps_compare_sorption} shows the average PSD
for 10 PIDs between 3000--4000 (black) compared to 10 PIDs between
12\,000--13\,000 (grey) for the 70\,GHz 20M radiometer. We see large
increases in power at low frequencies, and a shift in the knee
frequency.

These correlations appear more weakly in the 30
and 44\,GHz radiometers (see Fig.~\ref{fig:xi_vs_pid_mean}). In
particular, there is no correlation with the knee frequency. This
behavior could be partly explained by the fact that, by mechanical
design, the front-end modules (FEMs) of the 30 and 44\,GHz are less
thermally coupled to the frame and cooler front-end; or it could be
indicative of an additional source of non-thermal correlated noise
that dominates the slope and knee frequency of these channels. This
could be the case also for the 70\,GHz radiometer 21M, for which the
lack of correlation cannot be explained in terms of poor thermal
coupling.

These hypotheses are supported by Fig.~\ref{fig:ps_compare_28M},
which compared the PSD of the 30\,GHz 28M signal-subtracted data,
averaged over 18 PIDs in a typical stable period, with both the
\BP\ and LFI DPC noise models for the same period. We see that the $1/f$ 
model is not able to properly describe the observed data. The deviation
indicates that there is an excess power in the frequency range between
0.1 and 5\,Hz. This and similar excesses in many of the other 30 and 44\,GHz channels are the motivation for adding the lognormal component 
to the noise model for these bands. 

\subsection{Seasonal effects and slow drifts}
\label{sec:seasonal}
The changing Sun-satellite distance during the yearly \Planck\ orbit
around the Sun produced a seasonal modulation of the solar power
absorbed by the spacecraft. The corresponding effect on the LFI
thermal environment was negligible for the
actively-controlled front-end, as demonstrated by the lack of yearly
modulation in the 20\,K temperature sensors (see Fig.~\ref{fig:hk_all} and
upper panel of Fig.~\ref{fig:hk_TSL_70_full}). However, the 300\,K
environment and the passive cooling elements (V-groove radiators) were
affected by a $\sim$1\,\% seasonal modulation (see Fig.~6 of
\citealp{planck2013-p01}).

Since the radiometer back-end modules (BEMs) provided a major
contribution to the radiometer gain $g$, and these are located in the
300\,K service module (SVM), the thermal susceptibility of the BEMs
coupled with local thermal changes is expected to induce radiometer
gain variations. On the other hand, since the BEMs are downstream
relative to the $>$30\,dB amplification from the FEMs, their
contribution to the noise temperature, $T_{\rm sys}$, is
negligible. Therefore we may expect the LFI uncalibrated signal (and
the uncalibrated noise $\sigma_0 \,\mathrm{[V]}$) to show a seasonal
modulation due to thermally-driven BEM gain variations, with
essentially no degeneracy with $T_{\rm sys}$.

Figures \ref{fig:xi_prop_30}--\ref{fig:xi_prop_70_3} show that
several LFI radiometers exhibit such modulation in the uncalibrated
white noise, $\sigma_0 \,\mathrm{[V]}$, throughout the four year survey. For all of
these, the modulation disappears in $\sigma_0 \,\mathrm{[K]}$, indicating that
our gain model properly captures this effect. We also observe that the
sign of the modulation is opposite for the 70\,GHz and the 30/44\,GHz
radiometers. Furthermore, all radiometers that exhibit seasonal
modulation also show a systematic slow drift of $\sigma_0 \,\mathrm{[V]}$
throughout the mission with the same sign as the initial modulation
(which corresponds to increasing physical temperatures in the
SVM). Since the spacecraft housekeeping recorded a slow overall
increase in temperature throughout the mission ($\Delta T \approx
5$\,K), the observed drift of $\sigma_0 \,\mathrm{[V]}$ is qualitatively consistent
with the hypothesis of BEM susceptibility as the origin of the effect.

For each radiometer, the amplitude of the modulation depends on the
details of the thermal susceptibility of the LFI elements down-stream
relative to the third V-groove, including waveguide losses, BEM
components, particularly low-noise amplifiers (LNAs), detector diodes,
data acquisition electronics (gain and offset), etc.
The dominant element is the BEM, whose
thermal susceptibility was measured in the LFI pre-launch test
campaign for the 30 and 44\,GHz radiometers \citep{villa2010}. The
change in BEM output voltage, $\Delta V_{\rm out}$, as a function of
the variation in BEM physical temperature, $\Delta T_{\rm BEM}$, can
be written as
\begin{equation}
\Delta V_{\rm out} \propto \phi_{\rm BEM} \Delta T_{\rm BEM}
\left(T_{\rm sys} + T_{\rm in} \right),
\end{equation}
where $T_{\rm in}$ is the input signal temperature (either sky or
reference load) and $\phi_{\rm BEM}$ is a transfer function
quantifying the BEM thermal susceptibility. The measured values of
$\phi_{\rm BEM}$ \citep{villa2010} were slightly negative for all the
30 and 44\,GHz radiometers, ranging from $-0.01$ to $-0.02$, and this
is consistent with both the observed overall drift and the seasonal
effect. No such ground tests could be done for the 70\,GHz instrument.
However, in-flight tests during commissioning \citep{Cuttaia2011}
revealed that the sign of $\phi_{\rm BEM}$ for the 70\,GHz radiometers
was opposite to those of 30 and 44\,GHz, which is consistent with our
interpretation.
\begin{figure}
	\begin{center}
		\includegraphics[width=\linewidth]{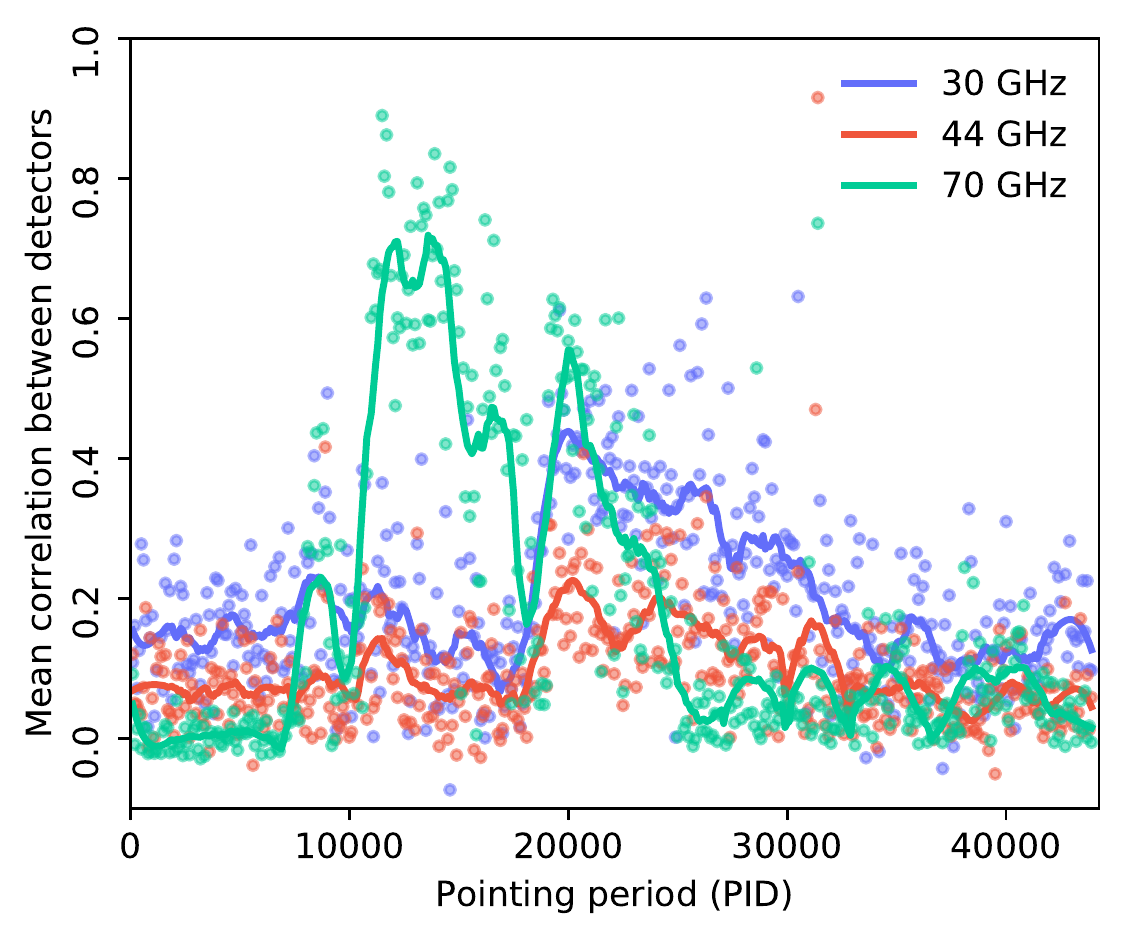}
	\end{center}
	\caption{Average correlation between the correlated noise in the different radiometers 
		in each frequency band for the full mission. Colored dots indicate correlations found 
		in a single PID, while the colored curves are a "running mean" of the corresponding 
		points. We use every hundreth PID, so 1\,\% of all PIDs are included in the plot. 
		\label{fig:corr_v_pid}}
\end{figure}

\begin{figure}
	\begin{center}
		\includegraphics[width=\linewidth]{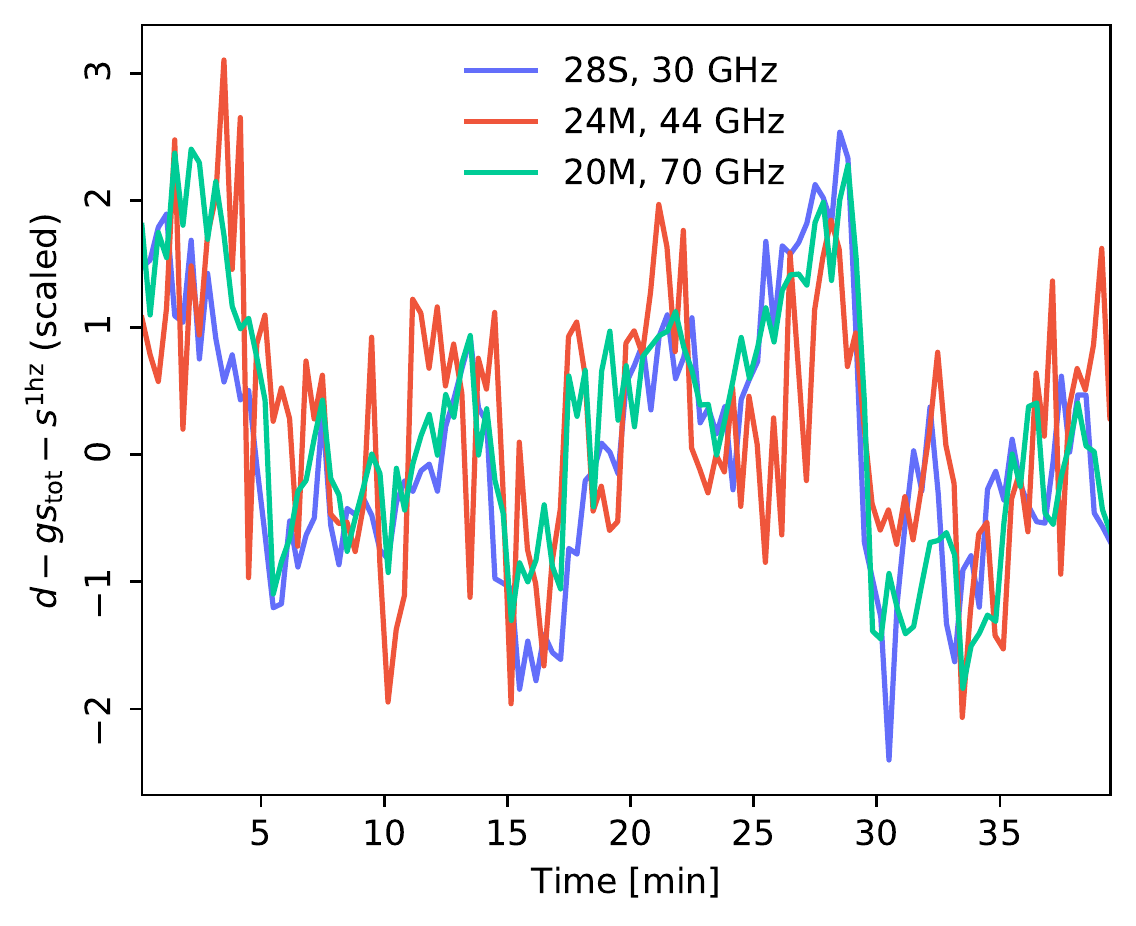}
	\end{center}
	\caption{Signal subtracted data from radiometers from all three bands for PID 12\,301. The data is averaged over a 20 second timescale and scaled to fit in the same plot. 
		\label{fig:data_compare_corr}}
\end{figure}

\begin{figure}
	\begin{center}
		\includegraphics[width=\linewidth]{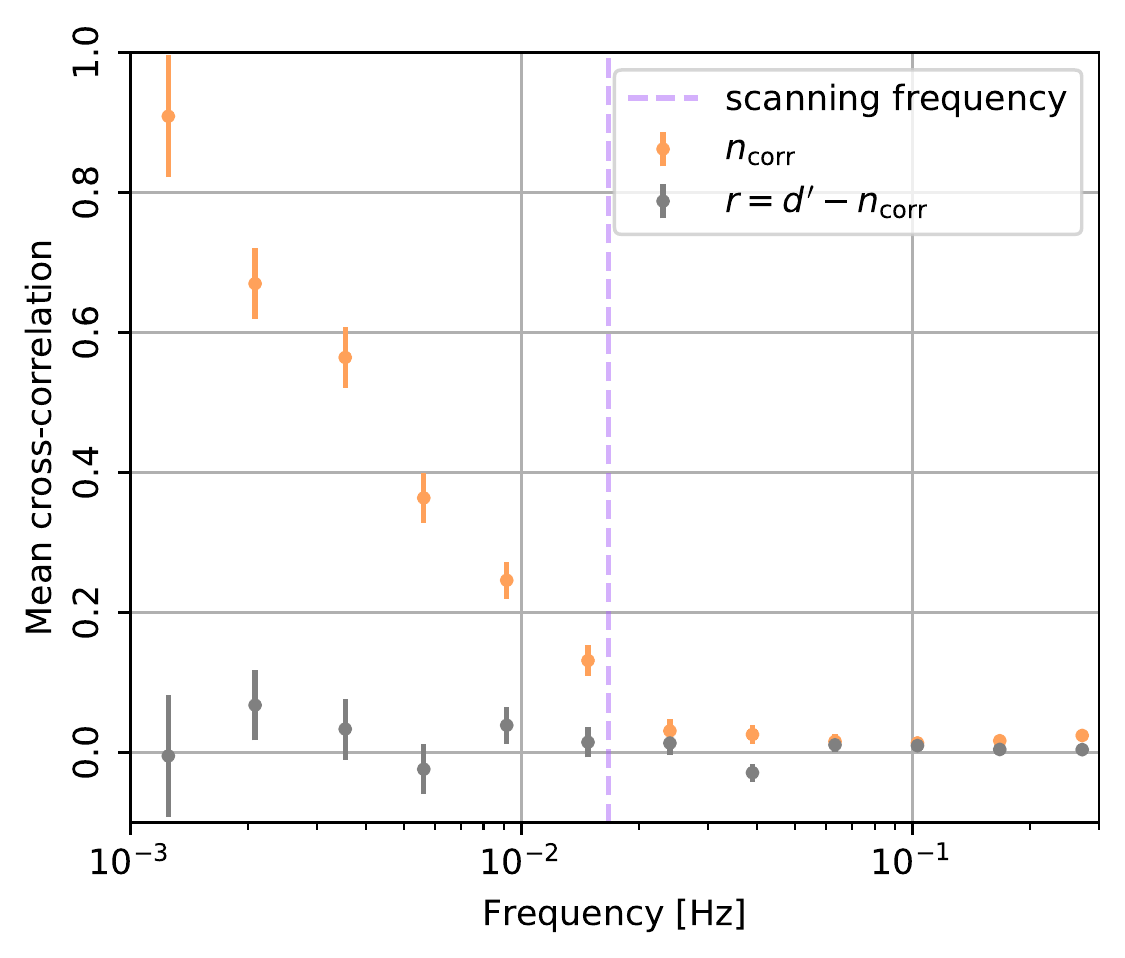}
	\end{center}
	\caption{Average cross correlation between timestreams of all 70 GHz
    radiometers for PID 12301. Orange points show average correlation between
    the correlated noise components, while the grey points shows average
    correlation between the residuals. We see that even though the correlation
    between the correlated noise components is large, the residuals are completely uncorrelated, indicating that this correlated signal does not leak into the rest of the pipeline, but is all incorporated into the correlated noise.
		\label{fig:mean_xcorr_70}}
\end{figure}

\subsection{Inter-radiometer correlations}

\begin{figure*}
	\begin{center}
		\includegraphics[width=1.0\linewidth]{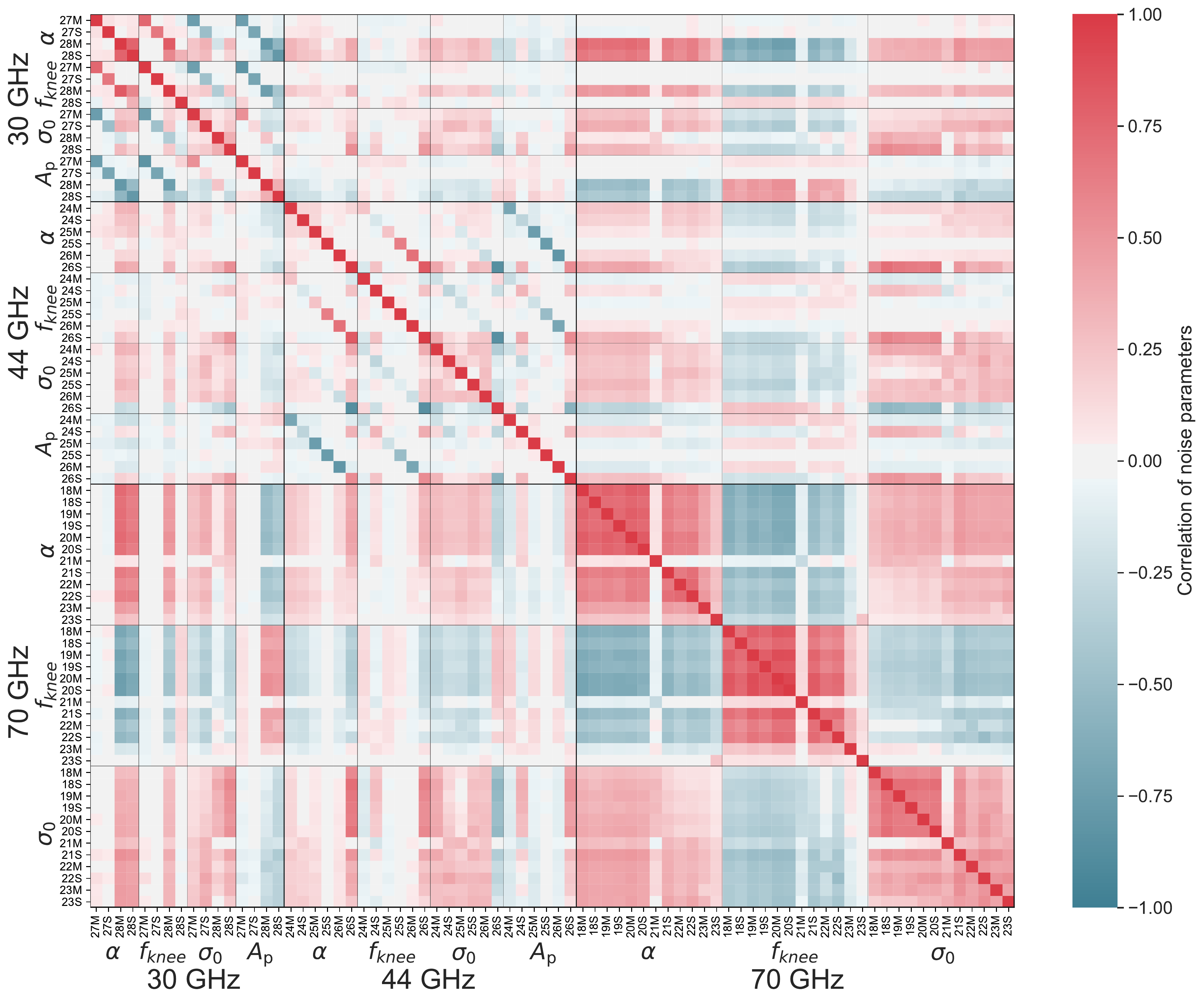}
	\end{center}
	\caption{Noise parameter correlation matrix. We average over all Gibbs samples of the noise parameters, $\xi^n = \{\sigma_0, f_\mathrm{knee}, \alpha, A_\mathrm{p}\}$, for each PID. We then find the correlation in time between these averages for the different bands and detectors. The results here are for the calibrated white noise level, $\sigma_0$\,[K].
		\label{fig:xi_corr}}
\end{figure*}

So far, we have mostly considered noise properties as measured
separately for each radiometer. However, given the significant
sensitivity to external environment parameters discussed above, it is
also interesting to quantify correlations between detectors. As a
first measure of this, we plot in Fig.~\ref{fig:corr_v_pid} the
correlation of $\n^{\mathrm{corr}}$ averaged over all pairs of
radiometers within each frequency band as a function of PID. As
expected from the previous discussion, we find a large common
correlation for the 70\,GHz channel that peaks in the post-switchover
period. Similar coherent patterns are seen in the 30 and 44\,GHz
channels, but at somewhat lower levels.

As a specific example of such common mode noise,
Fig.~\ref{fig:data_compare_corr} shows the signal subtracted
timestreams for one radiometer from each band for PID 12\,301, which
is representative for the period of maximum correlation. Here we
see that the same large scale fluctuations are present in all
three bands, though at different levels of amplitude. In 
Fig.~\ref{fig:mean_xcorr_70} we show the average cross-correlation 
between time streams of all 70 GHz radiometers for the
same PID. We compare the average correlation between the correlated
noise components, $\n_{\rm corr}$, with the correlation between the
residuals, $\d'-\n_{\rm corr}$. We see that even though the
correlations between the $\n_{\rm corr}$ components are large, the
residuals are highly uncorrelated. This is an indication that the
common mode signal is efficiently described by $\n^{\mathrm{corr}}$,
and it therefore does not leak into the rest of the \BP\ pipeline.

Figure~\ref{fig:xi_corr} shows a global correlation matrix of all the noise
parameters for all the LFI radiometers throughout the mission.
A number of interesting features can be recognized in this diagram:
\begin{enumerate}
	\item
	We note that all 70\,GHz radiometers exhibit an internally coherent trend, where $f_{\rm knee}$ and $\alpha$ behave essentially as a common mode for the entire 70\,GHz array, with the only exception being 21M. This coherent behavior reflects the common thermal origin of the $1/f$ noise of the 70\,GHz radiometers, as discussed in Sect.~\ref{sec:temp_fluct_1f}. We also see that $\sigma_0 \,\mathrm{[K]}$ shows a similar common mode behavior for the 70\,GHz radiometers and, to a lesser extent, it also correlates with the $\sigma_0 \,\mathrm{[K]}$ of the 30 and 44\,GHz radiometers. This is indicative of the fact that changes in the LFI radiometers’ sensitivities are driven by the global LFI thermal environment, most importantly by the slow increase in temperature at the 20\,K temperature stage.
	
	\item
	For 30 and 44\,GHz we do not observe the same common mode behavior for
        $f_{\rm knee}$ and $\alpha$ as for the 70\,GHz. Rather, we see positive
        correlation (red pixels in Fig.~\ref{fig:xi_corr}) between $f_{\rm
        knee}$ and $\alpha$ within each single radiometer. This suggests that
        (a) the dominant source of $1/f$ noise is independent for each 30 and
        44\,GHz radiometer, and (b) for a given radiometer, as $f_{\rm knee}$
        increases, the slope becomes flatter (i.e., $\alpha$ becomes less negative). This behavior further supports the hypothesis that the dominant source of correlated noise in the 30 and 44\,GHz is not of thermal origin.
	
	\item 
	We see that the amplitude of the lognormal noise component, $A_\mathrm{p}$, is negatively correlated with both $\alpha$ and $f_{\rm knee}$. This makes sense, since, with the lognormal component added, $\alpha$ and $f_{\rm knee}$ no longer have to adjust to the intermediate frequency noise signal, which means that $f_{\rm knee}$ is lower and $\alpha$ gets more steep. 
	
	\item
	Finally, we observe an anti-correlation between $f_{\rm knee}$ and $\sigma_0$ (as a common mode at 70\,GHz and individually for 30 and 44\,GHz). Slightly larger values of $f_{\rm knee}$ for lower $\sigma_0$ can be understood in terms of the correlated fluctuations becoming subdominant near $f_{\rm knee}$ when the white noise increases during the mission time.
	
\end{enumerate}

\subsection{Correlation with housekeeping data}
\begin{figure}
	\begin{center}
		\includegraphics[width=\linewidth]{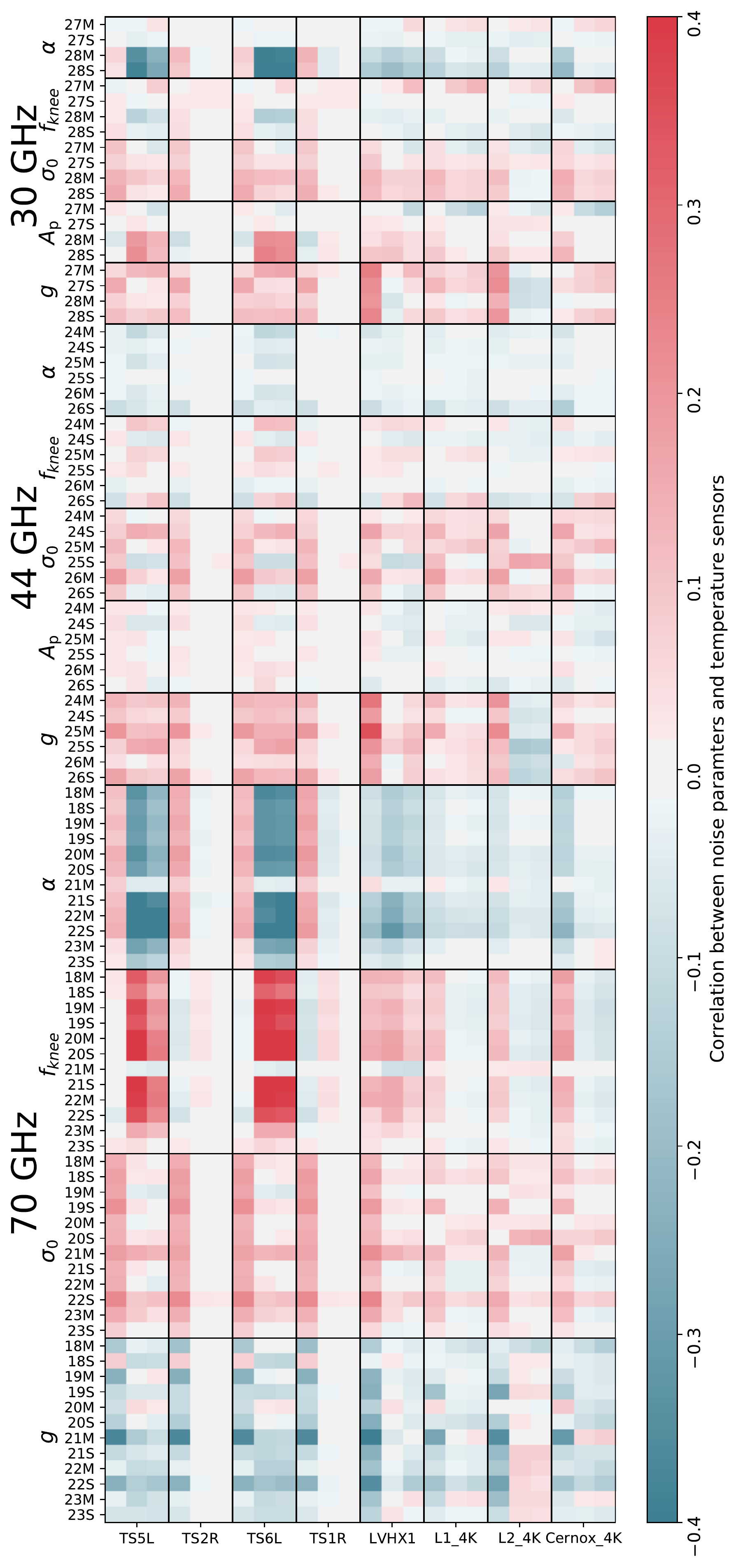}
	\end{center}
	\caption{Correlation in time, for the complete mission, between noise parameters and the temperature sensors. For each sensor we show the results (from left to right) from the mean temperature, the temperature rms and the peak-to-peak temperature of each sensor within each pointing period. The results here are for the calibrated white noise level, $\sigma_0$\,[K]. We have imposed a mild highpass-filter in time of the different datasets in order to avoid random correlations on the very longest timescales.
		\label{fig:hk_corr}}
\end{figure}

Next, we correlate the LFI noise parameters with housekeeping data, and
in particular with temperature sensors that are relevant for LFI. This
is summarized in Fig.~\ref{fig:hk_corr}, showing the correlation
coefficients with respect to several sensors that monitor the 20\,K
stage (TS5L, TS2R, TS6L, TS1R, LVHX1) and the 4\,K stage (L1\_4K,
L2\_4K, Cernox\_4K).
Some significant patterns appear that can be interpreted in terms of
the general instrument behavior:
\begin{enumerate}
	\item
	For the 70\,GHz radiometers, both the rms and the peak-to-peak
        variations of the 20\,K temperature sensor fluctuations 
        correlate with $f_{\rm knee}$ and anti-correlate with $\alpha$
        (i.e., they prefer a steeper power-law slope).  This indicates that the $1/f$
        noise of the 70\,GHz radiometers is dominated by residual
        thermal fluctuations in the 20\,K stage. A similar trend can
        be seen also at 30\,GHz in the two horn-coupled receivers 28M
        and 28S. However, the 44\,GHz channels show no
        sign of this behavior. 
        Combined with the lack of correlation with the 4\,K
        sensors, this is consistent with the hypothesis that the $1/f$ noise
        of the 44\,GHz (and partly the 30\,GHz) radiometers is dominated by
        non-thermal fluctuations.
	\item
	Weaker correlations are seen between the various noise parameters and the
        4\,K temperature sensors. The lack of significant correlation of the rms and peak-to-peak variations of 4\,K sensors with any of the $1/f$ parameters, $f_{\rm knee}$ and  $\alpha$, is an indication that the 4\,K reference loads do not contribute significantly to the radiometers correlated noise.
	\item
	A strong anti-correlation (correlation) of the gain $g$ with
        the absolute value of the 20\,K sensors for the 70\,GHz
        (30/44\,GHz) radiometers is observed. Based on the discussion
        in Sect.~\ref{sec:seasonal}, this pattern can be understood by
        noting that the 20\,K stage temperature systematically
        increased throughout the mission, driven by sorption cooler
        adjustments. The same monotonic trend was also on-going in the
        300\,K stage, which controls the BEM amplifiers. This is thus a
        spurious correlation, for which the increasing back-end
        temperature actually leads to lower (higher) values of $g$ for
        the 70\,GHz (30/44\,GHz) radiometers.
\end{enumerate}

\subsection{Issues with individual radiometers}
In addition to the overall behavior and correlations that are common to many or most radiometers,
there are issues that only seem to affect individual radiometers. Here we point out two special cases, namely 26S and 21M. 

\begin{figure}
	\begin{center}
		\includegraphics[width=\linewidth]{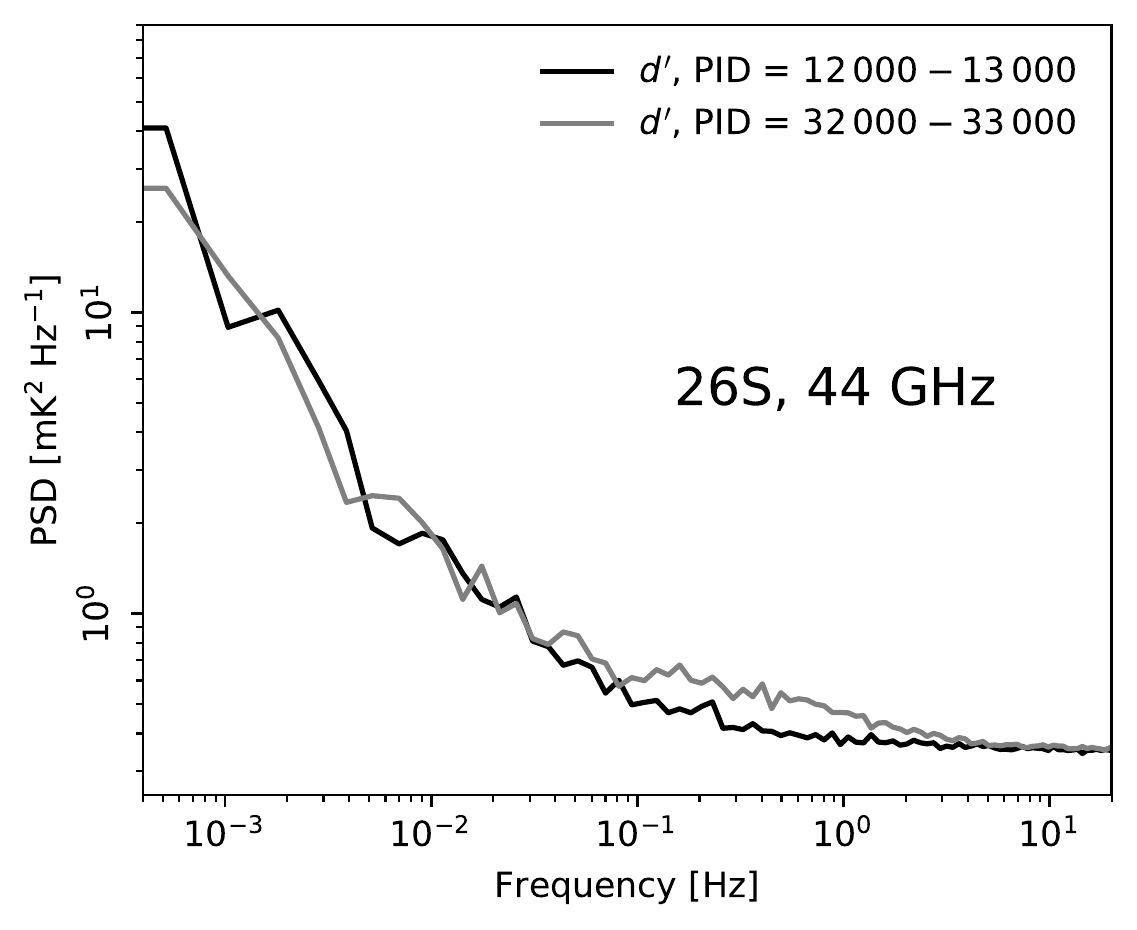}
	\end{center}
	\caption{PSD of signal subtracted data from radiometer 26S, averaged over 10 PIDs (at intervals of 100 PIDs) in the ranges 
    12\,000--13\,000 (black) and 32\,000--33\,000 (grey). We see that there is
    significantly more power in the frequency range 0.1--10\,Hz in the later period. 
		\label{fig:ps_compare_26S}}
\end{figure}

First, as discussed above, we often find excess noise power in the 30
and 44\,GHz channels in the signal-subtracted data at intermediate
frequencies, $\sim$0.01--5\,Hz, which cannot be described with a
$1/f$ noise model. The most extreme
example of this is the 44\,GHz 26S radiometer, as shown in the bottom
panel of Fig.~\ref{fig:xi_prop_44_2}. Here we see a jump in $f_\mathrm{knee}$ around PID 20\,800,
after which the noise parameters change abruptly. This is elucidated in Fig.~\ref{fig:ps_compare_26S},
which compares the noise PSD averaged over 10 PIDs in the
12\,000--13\,000 range with a corresponding average evaluated in the
32\,000--33\,000 range. We see that the signal from the early period is consistent with a $1/f$ spectrum, while for the later period we find an excess in power at intermediate
frequencies that is not possible to fit with the $1/f$ noise
model. Considering that the \Planck\ spin period is 60\,s, temporal
frequencies of 0.1--1\,Hz correspond to angular scales of
6--$60^{\circ}$ on the sky. This noise excess therefore represents
a significant contaminant with respect to large-scale CMB polarization
reconstruction, which is one of the main scientific targets for the
current \BP\ analysis. This is why the lognormal component was added, to prevent this excess noise from leaking into any of the sky components or other parts of our model.

The sudden degradation of 26S at around PID 21\,000 has no
simultaneous counterpart in any other LFI radiometer, including the
coupled 26M which exhibits a normal behavior (see
Fig.~\ref{fig:xi_prop_44_2}).  This suggests a singular event within the
26S itself, or in the bias circuits serving its RF components. Since
we do not observe significant changes in the radiometer output signal
level and no anomalies are seen in the LNAs currents, it is unlikely
that the problem resides in the HEMT amplifiers. A more plausible
cause would be a degradation of the phase switch performance, possibly
due to aging, instability of the input currents, or loss of internal
tuning balance \citep{mennella2010,cuttaia2009}. Indeed sub-optimal
operation of the phase switches would not significantly change the
signal output level, but is known to introduce excess $1/f$ noise, as
verified during the ground testing and in-flight commissioning phase.

\begin{figure}
	\begin{center}
		\includegraphics[width=\linewidth]{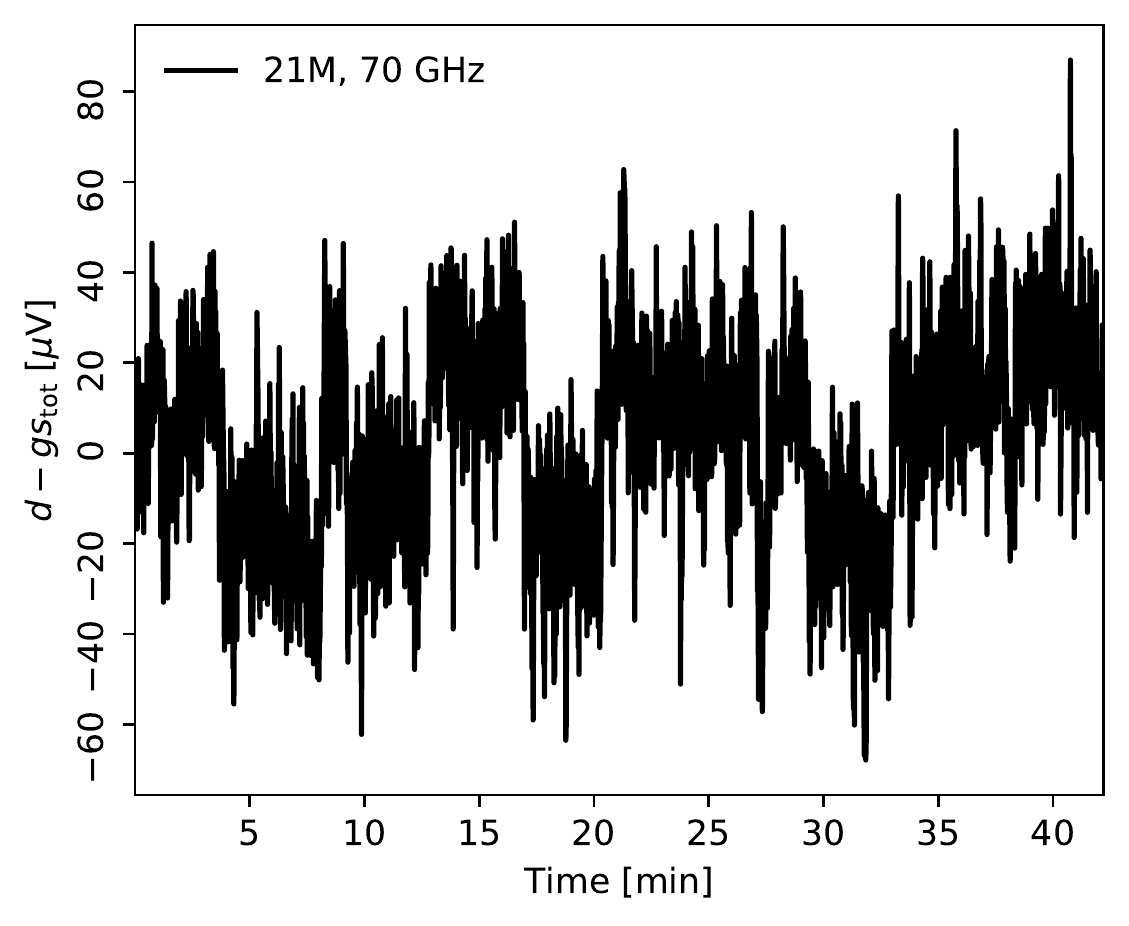}
	\end{center}
	\caption{Signal subtracted data from radiometer 21M for PID 2201. The data are averaged over a one second timescale. 
		\label{fig:popcorn_21M}}
\end{figure}

The second anomalous case is the 70\,GHz 21M radiometer.  While the
noise properties of the other 70\,GHz channels are internally
significantly correlated, this particular channel does not show
similar correlations. The reason for the different behavior of 21M is
still not fully understood. However, as shown for PID 2201 in
Fig.~\ref{fig:popcorn_21M}, this particular radiometer exhibits a
typical “popcorn” or “random telegraph” noise, i.e., a white noise
jumping between two different offset states.  During ground testing
this behavior was noted in the undifferenced data of LFI21 and LFI23
and ascribed to bimodal instability in the detector diodes. The effect
was then recognized in-flight and this prevented proper correction of
ADC nonlinearity effect \citep{planck2013-p02a}. However, because the
timescale of diode jumps between states (typically a few minutes) is
longer than the differencing between sky and reference load
($0.25$\,ms, corresponding to the phase switch frequency of 4\,kHz),
the effect is efficiently removed in the differenced data. In the
current analysis, we actually observe popcorn behavior in the
differenced data, suggesting either an increased instability of the
affected diode in 21M (possibly due to aging), or a different origin
of the effect.  Popcorn noise has been also found in some HFI channels
\citep{planck2011-1.5}. We have not seen any sign of popcorn noise in
any of the other LFI channels besides 21M, but we have also not
performed a deep dedicated search for it. However, the fact that the
$\chi^2$ distribution for channel 21M appears acceptable suggests that
this effect, even if surviving in the differenced data stream, happens
at a sufficiently long timescale that $\n_{\rm corr}$ is able to
absorb it, preventing it from leaking into other astrophysical
components.

\section{Conclusions}
\label{sec:conclusions}

This paper has two main goals. First, it aims to describe Bayesian
noise estimation within a global CMB analysis framework
\citep{bp01}. As such, this work represents the first real-world
application and demonstration of methods originally introduced by
\citet{wehus:2012}, while at the same time taking advantage of
important numerical improvements introduced by \citet{bp02}. The
second main goal is to apply this method to the \Planck\ LFI
measurements to characterize their noise properties at a more
fine-grained level than done previously \citep{planck2016-l02}.

An important question regarding the original work of
\citet{wehus:2012} was whether the method would be practical for
real-world observations, or whether it was too computationally
expensive to be useful in a real pipeline. We are now in a position to
quantitatively answer this question. As summarized by \citep{bp01,bp03},
the noise estimation step in the \BP\ pipeline accounts for 19\,\% of
the total runtime, or 27 CPU-hours per sample for the 70\,GHz channel,
most of which is spent Fourier transforming the raw time-ordered
data. As such, exact Bayesian noise estimation certainly is an
expensive pipeline component---but it is by no means
prohibitive. Additionally, it is important to note that Bayesian
correlated noise sampling replaces both traditional mapmaking and
noise covariance matrix evaluations \citep{bp02,bp10}, which are two
of the most expensive procedures in a traditional CMB analysis
pipeline \citep{planck2016-l01}, and, in fact, this leads to lower
computational requirements overall. As an example, we note that a full
\BP\ Gibbs sample (which includes both low- and high-level processing
with all LFI channels) costs 243\,CPU-hours, while producing a single
component of the \Planck\ Full Focal Plane (FFP) simulation of
the 70\,GHz channel costs 9360\,CPU-hours \citep{planck2014-a14}. Likewise, we also note that
the current \BP\ analysis was run on an in-house cluster with 256
cores and 4\,TB of RAM, while the \Planck\ simulations were produced
on a large national computing center with $\mathcal{O}(10^5)$ cores
\citep{planck2016-l01,planck2020-LVII}. We believe that the computational speed
of this method alone should make it an attractive option for other CMB
experiments, not to mention the possibilities of performing joint
exact Bayesian analysis. 

As far as LFI-specific analysis is concerned, the current results
point toward generally complex noise behavior with subtle
contributions from origins that have not yet been fully accounted
for. Most notably, the noise properties of each LFI radiometer vary
significantly in time, and depend sensitively on the thermal
environment of the instrument. For the 70\,GHz channel, for which the
correlated noise amplitude (and knee frequency) is generally low, most
of these variations may be described in terms of a simple $1/f$ model
with time-dependent parameters. With very few exceptions, the
time-domain $\chi^2$ of this channel is statistically acceptable
throughout the mission.

However, for the 30 and 44\,GHz channels a more complex picture has
emerged. Multiple observations suggest a yet unexlained source of
non-thermal correlated noise in the 44\,GHz (and at a lower level in
the 30 GHz radiometers) that is responsible for mild, and possibly
time-varying, deviations from the simple $1/f$ model. Inspection of
individual PIDs indicates the presence of excess power between 0.1 and
5\,Hz, well above the \Planck\ scanning frequency of 0.017\,Hz,
thereby affecting the angular scales that are relevant for large-scale
CMB polarization science. This excess motivated the addition of a lognormal noise component, at intermediate frequencies, in addition to the $1/f$ component, for these bands. 

Our analysis suggests that these effects are
not due to temperature fluctuations, but rather associated with other
effects, such as electrical instabilities or other environmental
issues. We have carried out a preliminary investigation by correlating
the LFI radiometers whose LNA bias were supplied by common electronics
groups,\footnote{There were four such power groups in LFI, feeding the
  radiometers associated with the following horn sets: (19-20-28),
  (18-26), (21-22-24-27), (23-25).}  but we have found no compelling
evidence of correlations or anomalies.  Many other electrical effects
must be studied by exploiting all the available housekeeping
information.  Most of the spikes in the rms of the temperature sensors
(see, e.g., Fig.~\ref{fig:hk_TSL_70_full}) are readily understood as
due to commanded cooler adjustments, but a few of them deserve further
investigation. The influence of transient perturbations should also be
systematically investigated, including the possible effect of cosmic
rays and solar flares.

A complete and quantitative analysis will require a detailed thermal
model of the full instrument that includes the back-end unit and
interfaces with the V-grooves, coupled with thermal susceptibility
parameters of the relevant components (LNAs, OMTs, waveguides, BEMs,
detector diodes, data acquisition electronics). Such a detailed study is beyond the
scope of this work, but this is now made possible through the present
study.

It is important to note that while we have mainly been discussing the application to the LFI time-ordered-data, the methods presented here are much more general. As the addition of the lognormal noise component shows, the methods can accomodate any parametric noise PSD model, as long as the data can constrain it. It can also be straightforwardly extended to cases where we have correlated noise components that are common to several detectors, which for instance is typically the case for atmospheric fluctuations.

In general, the detailed \BP\ modelling approach allows us to
highlight a number of subtle systematic patterns in the LFI
radiometers that were already noted and reported in previous analyses,
but only now, for the first time, have been possible to elucidate and
understand in greater detail. Examples are a detailed characterization
of the nature of seasonal modulations and long term drifts,
correlations between instrument noise parameters and temperature
sensor read-out information and deviations from the $1/f$ noise model. 
These methods are likely to play a
central role in the analysis of future high-sensitivity CMB $B$-mode
experiments, for instance \textit{LiteBIRD} \citep{litebird2020,litebird2022}.

\begin{acknowledgements}
  We thank Prof.\ Pedro Ferreira and Dr.\ Charles Lawrence for useful suggestions, comments and 
  discussions. We also thank the entire \Planck\ and \WMAP\ teams for
  invaluable support and discussions, and for their dedicated efforts
  through several decades without which this work would not be
  possible. The current work has received funding from the European
  Union’s Horizon 2020 research and innovation programme under grant
  agreement numbers 776282 (COMPET-4; \BP), 772253 (ERC;
  \textsc{bits2cosmology}), and 819478 (ERC; \textsc{Cosmoglobe}). In
  addition, the collaboration acknowledges support from ESA; ASI and
  INAF (Italy); NASA and DoE (USA); Tekes, Academy of Finland (grant
   no.\ 295113), CSC, and Magnus Ehrnrooth foundation (Finland); RCN
  (Norway; grant nos.\ 263011, 274990); and PRACE (EU).
\end{acknowledgements}

\bibliographystyle{aa}

\bibliography{Planck_bib,BP_bibliography}

\end{document}